\documentclass[12pt,a4paper]{article}
\usepackage{jheppub}
\usepackage[T1]{fontenc} 
\usepackage{epsfig} 
\usepackage{amscd}
\usepackage{verbatim}
\usepackage{latexsym}
\usepackage{amsfonts,amsthm,amsmath,mathtools}
\usepackage{amssymb,stmaryrd,textcomp,enumerate,bbm}
\usepackage{color}
\usepackage{xcolor}
\usepackage{booktabs}
\usepackage{float}
\usepackage{graphicx}
\usepackage{lscape}
\usepackage{slashed}

\notoc 

\title{\centering{c-extremization from toric geometry}}
\author[a]{Antonio Amariti}
\author[b]{\!\!,~Luca Cassia}
\author[b]{\!\!,~Silvia Penati}
\affiliation[a]{Albert Einstein Center for Fundamental Physics, Institute for Theoretical Physics, University of Bern, Sidlerstrasse 5, Bern, ch-3012, Switzerland}
\affiliation[b]{Universit\`a degli studi di Milano Bicocca and INFN, Sezione di Milano--Bicocca, Piazza della Scienza 3, 20161, Milano, Italy}
\emailAdd{amariti@itp.unibe.ch,luca.cassia@mib.infn.it,silvia.penati@mib.infn.it}

\newcommand{\nocomma}{}

\newcommand{\tmop}[1]{\ensuremath{\operatorname{#1}}}

\newcommand{\nocontentsline}[3]{}
\newcommand{\tocless}[2]{\bgroup\let\addcontentsline=\nocontentsline#1{#2}\egroup}

\abstract{
We derive a geometric formulation of the 2d central charge $c_r$ from  
infinite families of 4d $\mathcal{N}=1$ superconformal field theories topologically twisted on 
constant curvature Riemann surfaces.
They correspond to toric quiver gauge theories and are associated to
D3 branes probing five dimensional Sasaki-Einstein
geometries in the AdS/CFT correspondence.
We show that $c_r$ can be expressed
in terms of the areas of the toric diagram describing the moduli space
of the 4d theory, both for toric geometries with smooth and singular horizons.
We also study the relation between a-maximization in 4d and c-extremization in 2d, giving further evidences of the mixing of the baryonic 
symmetries with the exact R-current in two dimensions.
}

\begin{document}

\maketitle

\newpage
\tableofcontents
%
%
\section{Introduction}
%
%
Anomalies play a crucial role in the analysis of conformal field theories (CFTs).
They provide consistency checks and impose several constraints on the 
existence of the IR fixed points and on the behavior of the RG flows.
A well studied anomaly in 4d is the coefficient of the Euler density 
of $T_\mu^\mu$, referred as the central charge $a$.
This quantity satisfies a c-theorem, decreasing between the endpoints of RG flows 
\cite{Cardy:1988cwa,Komargodski:2011vj}.
When considering $\mathcal{N}=1$ supersymmetric CFTs (SCFTs), 
the central charge $a$, non perturbatively obtained in \cite{Anselmi:1997am},
is maximized
by the exact R--current of the superconformal algebra \cite{Intriligator:2003jj}.
The exact R--current is a linear combination of the trial UV R-current and 
the other global symmetries. By maximizing the central charge these
mixing coefficients can be exactly computed.

The right--moving
central charge $c_r$ of 2d $\mathcal{N}=(0,2)$ SCFTs
satisfies a c--theorem as well \cite{Zamolodchikov:1986gt}, and it is 
extremized by the exact  2d R--current \cite{Benini:2012cz}.
It is possible to construct classes of 2d SCFTs  
by the partial topological twist of 4d SCFTs on Riemann surfaces $\Sigma$
with constant curvature $2\kappa$ \cite{Witten:1988xj,Bershadsky:1995vm}.
In order to preserve supersymmetry on the product space $\Sigma \times \mathbb{R}^{1,1}$
background magnetic fields for the global symmetries can be turned on \cite{Festuccia:2011ws,Kutasov:2013ffl}.
These background fluxes must be properly quantized and 
cancel the contributions of the spin connection in the Killing spinor equations.
For generic choices of the fluxes the 2d theory has $\mathcal{N}=(0,2)$ supersymmetry.
The central charge of these 2d SCFTs can been computed from the global anomalies
of the 4d theory \cite{Benini:2015bwz}.

When considering 4d theories with an AdS$_5$ holographic dual  description 
the topological twist can be reproduced at the
gravitational level by turning on properly quantized fluxes for the 
(abelian) gauge symmetries in the bulk \cite{Maldacena:2000mw}.
This triggers a RG flow across dimensions that, when restricting
to the supergravity approximation, connects the original AdS$_5$
description to a warped AdS$_3 \times \Sigma$ geometry.

Instead of constructing this flow one can consider the full 10d 
geometries. Solving the BPS equations in this case should
lead to a warped product AdS$_3 \times \mathcal{M}_7$, where the
general properties of the seven manifold $\mathcal{M}_7$ were originally discussed 
in \cite{Kim:2005ez,Gauntlett:2007ts}.
This approach was taken in \cite{Benini:2015bwz} for the infinite class 
of $Y^{pq}$ toric quiver gauge theories \cite{Benvenuti:2004dy}.
The $Y^{pq}$ theories are examples of 4d $\mathcal{N}=1$ 
SCFTs describing a stack of N D3 branes probing the tip of a toric
Calabi--Yau threefold CY$_3$ over a  5d Sasaki--Einstein (SE) base X$_5$  with $U(1)^3$ isometry
(see \cite{Kennaway:2007tq,Franco:2017jeo} and references therein). 
 
An  interesting aspect of toric gauge theories is the relation between 
the central charge $a$ and the volumes vol(X$_5$). It has been indeed shown that 
the holographic dictionary translates a--maximization into the minimization of
vol(X$_5$) \cite{Martelli:2005tp}. The equivalence between the two formulations has been derived explicitly 
in \cite{Butti:2005vn}, where it was shown that the central charge can be obtained from the geometric
data that describe the probed X$_5$ geometry
and  are related to the $U(1)^3$ isometry of X$_5$.
Such data encode the structure of 
 the moduli space of the 4d SCFT
 in a convex lattice polygon called the toric diagram. 

A similar correspondence between the 2d central charge  $c_r$ and the volumes of  
the seven manifolds is  currently lacking.
A possible obstruction in formulating a volume formula dual to
c--extremization arises from the mixing of the global symmetries
in the 2d exact R--current.

It is indeed possible to compare the structure of the mixing of the exact R--current 
with the abelian symmetries at the IR fixed point in the 4d $\mathcal{N}=1$ 
and in the 2d $\mathcal{N}=(0,2)$ theory obtained from twisted compactification.
As a general result it has been observed that symmetries that trivially
mix with the 4d R--current can mix non-trivially with the 2d one.
This has been explicitly observed in \cite{Benini:2015bwz} for the  $Y^{pq}$ toric quiver gauge theories.
In the 4d case the exact R--symmetry of toric quiver gauge theories is a mixture of the $U(1)^3$ symmetries
of X$_5$. In the field theory language this can be rephrased as saying that the exact R--current 
is a mixing of the flavor symmetries that parameterize the mesonic moduli space and are encoded in the toric diagram.
The other symmetries are of baryonic type. On the geometric side they are associated to the third Betti 
number of X$_5$. On the field theory side they correspond to the non-anomalous combination
of the $U(1) \subset U(N)$ gauge groups, decoupling in the IR.
In the case of X$_5 =Y^{pq}$ there is a single baryonic symmetry. It does not mix with the 4d R--current but it 
can mix in the 2d case \cite{Benini:2015bwz}.
This is expected to be a general behavior and should hold for models with a larger amount
of baryonic symmetries.

This discussion leads to the conclusion that a putative volume formula for $c_r$ should involve symmetries that are not 
necessarily isometries of the seven manifold, so making the generalization of the results in \cite{Martelli:2005tp} to these cases not straightforward.

In this paper, despite the role of the baryonic symmetries in c-extremization, we obtain an alternative formulation 
of $c_r$  in terms of the toric data of the 4d parent theory.
The final formula involves the geometric data, the mixing parameters of the R--current with the other 
global symmetries and the fluxes turned on in the $\Sigma$ directions.
At large $N$ our formulation reproduces the behavior of $c_r$ as a function of the mixing parameters and of the fluxes 
for toric quiver gauge theories topologically twisted on $\Sigma$.

The paper is organized as follows.
In section \ref{review} we review some basic aspects of toric quiver gauge theories and 
of the topological twist, necessary to our analysis. 
In section \ref{secgeom}  
we derive the expression of $c_r$ in terms of the toric data of the 4d theory.
We first study  cases with smooth horizons, correctly reproducing the behavior of 
$c_r$ as a function of the R-charges. We confirm the validity of this formula
by studying many examples of increasing complexity.
In section \ref{perimeter} we consider the case of non--smooth horizons,
describing the prescription for obtaining $c_r$ in terms of the 
toric data of the 4d theory.
In section \ref{FTside} we study the compactification of del Pezzo gauge theories, dP$_2$ and dP$_3$, 
with respectively two and three non-anomalous baryonic symmetries, showing their mixing
in the exact 2d R--current.
Then we study a case with a generic amount of baryonic symmetries, by showing the mechanism in 
necklace quivers, denoted as $L^{pqp}$ theories. 
In section \ref{conclusions} we discuss the interpretation and possible implications of our results. For 2d theories obtained by topologically twisted reduction of dP$_2$ and dP$_3$ toric theories, in appendix \ref{mixing} we report the explicit values for the parameters of $U(1)$ mixing 
for particular choices of the fluxes.
 
%
%
%
%
 \section{Review}
 \label{review}
%
%
%
%
%

We start our discussion by reviewing the main aspects of toric quiver gauge theories and their twisted compactification on constant curvature Riemann surfaces. 

\subsection{Toric quiver gauge theories}
\label{toricrev}
Toric quiver gauge theories \cite{Feng:2000mi} 
describe
the near horizon limit of a stack of N D3 branes probing 
the tip of a CY$_3$ cone over a 5d SE X$_5$, characterized by a $U(1)^3$ action on the metric.  
On the field theory side the dual  ${\cal N}=1$ SCFTs are described by quiver gauge theories whose nodes carry $U(N)$
gauge factors and are connected by oriented arrows, representing bifundamental matter fields. 
\begin{figure}[H]
\begin{center}
  \includegraphics[width=12cm]{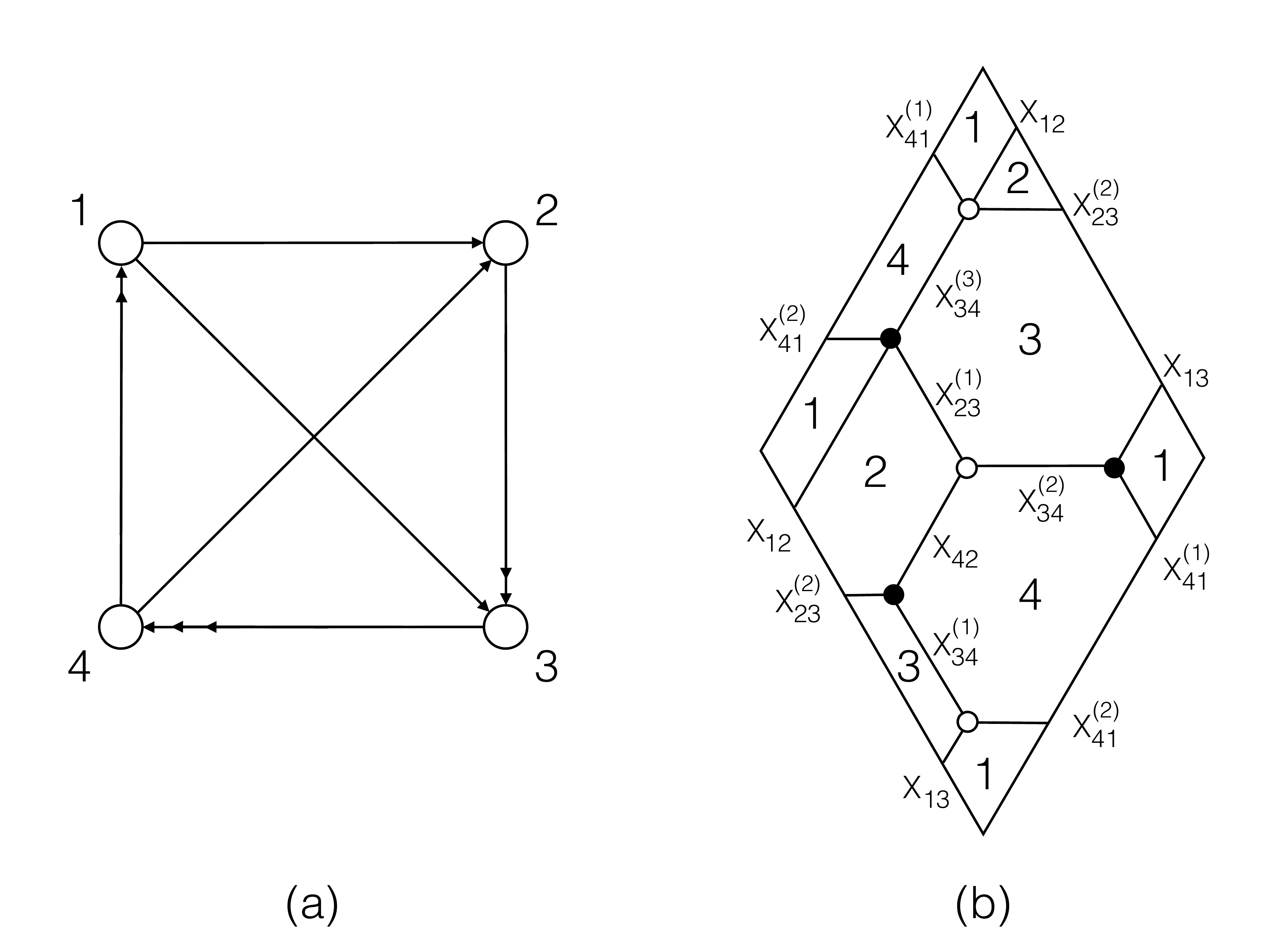}
  \caption{Quiver and dimer of the dP$_1$ model. In quiver (a) the number of arrows on the straight lines indicates the number of fields connecting two nodes.}
  \label{FIGDP1}
\end{center}
\end{figure}
In order to exemplify the discussion we consider the explicit case of a gauge theory living on a stack of N D3 branes probing the first del Pezzo singularity, dP$_1$. 
It has four gauge groups and the corresponding quiver is represented in Figure \ref{FIGDP1}(a). The superpotential 
\begin{equation}\label{superpotential}
W = -\epsilon_{\alpha \beta} X_{12} X_{23}^{(\alpha)} X_{34}^{(3)} X_{41}^{(\beta)}
+\epsilon_{\alpha \beta} X_{23}^{(\alpha)} X_{34}^{(\beta)} X_{42}
+\epsilon_{\alpha \beta} X_{13}  X_{34}^{(\alpha)}  X_{41}^{(\beta)} 
\end{equation}
is subject to the toric condition, which requires that each field appears in exactly two terms having opposite signs \footnote{For an exhaustive review on toric gauge theories we refer to 
\cite{Kennaway:2007tq,Franco:2017jeo}.}. 
This model has a $SU(2) \times U(1)$ flavor symmetry that, together with the $U(1)_R$ R--symmetry, builds up the isometry group of dP$_1$. 
In general, there are also baryonic symmetries associated to the non--trivial second cohomology group of X$_5$.
These symmetries can be obtained from the  $U(1) \subset U(N) $ gauge factors. They are IR free and at low energies 
decouple from the dynamics, becoming global symmetries. 
In quivers with a chiral--like matter content as the ones considered here,  some of these  $U(1)$'s are anomalous.
The non--anomalous abelian factors correspond to the aforementioned baryonic symmetries.
For the specific example of dP$_1$, to begin with there are four $U(1)_i \subset U(N)_i$ global symmetries of baryonic 
type with $T_{i=1, \dots , 4}$ generators. Two combinations are anomalous and one decouples. We are then left with just a single non--anomalous
baryonic symmetry that can be for example identified with the combination $2T_1-T_2+T_3$.

When flowing to the IR fixed point abelian flavor symmetries can mix with the R--current to form 
the exact R--symmetry, whereas the baryonic symmetries do not mix, as discussed in \cite{Bertolini:2004xf,Butti:2005vn}. 
This is a general feature of this family of 4d SCFTs.

For a quiver theory with $n_G$ gauge groups the mixing coefficients of global symmetries into the exact R--symmetries are obtained by maximizing the central charge
\cite{Intriligator:2003jj} 
\begin{equation}
\label{FT1}
a_{FT} = \frac{3}{32} (3\,  {\rm Tr} R^3 -{\rm Tr} R)= \frac{3}{32} \big(2  n_G (N^2-1) + \sum_{i=1}^{n_F} \text{dim}(\rho_i) (3(R_i - 1)^3-(R_i-1)) \big)
\end{equation}
where the first term is the contribution of the gaugini, $n_F$ it the total amount of matter multiplets,
$\text{dim}(\rho_i)$ is the dimension of the corresponding representation and $R_i$ the
R--charge of the scalar component of the $i$--th multiplet. For matter multiplets in the bifundamental and/or adjoint representations, 
at large $N$ the central charge is further simplified by the constraint ${\rm Tr} R=0$ and we read
\begin{equation}
\label{FT}
a_{FT} =  \frac{9}{32} N^2 \left( n_G + \sum_{i=1}^{n_F} (R_i - 1)^3 \right) + \mathcal{O}(1)
\end{equation} 

For toric gauge theories the $R_i$ charges can be determined directly from the geometric data of the singularity \cite{Gubser:1998vd,
Gubser:1998fp,Martelli:2005tp,Martelli:2006yb,Tachikawa:2005tq,Butti:2005vn,Butti:2005ps,Butti:2006nk,Benvenuti:2006xg,
Lee:2006ru,Kato:2006vx,Gulotta:2008ef,Eager:2010yu}, as we now review.

First of all, we recollect how to construct the toric diagram corresponding to a given quiver gauge theory. One embeds the quiver diagram (for the dP$_1$ case see figure \ref{FIGDP1}(a)) in a two dimensional torus. This resulting planar diagram can be dualized by inverting the role of faces and nodes, thus obtaining a bipartite diagram, called dimer, where faces correspond to gauge groups,
edges to fields and nodes to superpotential interactions (for the dP$_1$ model it 
is given in figure \ref{FIGDP1}(b)). The toric condition of the superpotential translates into a bipartite structure of the dimer.
From the dimer one can construct perfect matchings (PM's), that is
collections of edges (fields) characterized by the property that each node is connected to one and only one edge of the set.

 One can introduce a new set of formal variables $\pi_I$ associated to each PM.
These variables are defined by the relations
\begin{equation}
\label{eq:PMbasis}
X_i\equiv\prod_{I} (\pi_I)^{M_{i,I}}
\end{equation}
where the product is taken over all the PMs and
\begin{equation}
\label{eq:matchingmatrix}
M_{i,I}=
\left \{
\begin{array}{ll}
0 & \text{if $X_i$ does not belong to the set of $\pi_I$}\\
1 & \text{if $X_i$ belongs to the set of $\pi_I$}
\end{array}\right.
\end{equation}
The $\pi_I$ provide a convenient set of variables that can be used to parametrize the abelian moduli space of the quiver gauge theory. The advantage of using the PMs variables $\pi_I$ instead of the more natural set of scalar components of the chiral fields $X_i$ comes from the fact that using definition (\ref{eq:PMbasis}) the F-term equations are trivially satisfied. This is a consequence of the fact that, in this basis, each term in the superpotential becomes equal to $\pm\prod_I \pi_I$, with $I$ ranging over all PMs.

To each PM we can associate a signed intersection number, $\pm1$ or $0$, with respect to a basis of 1--cycles of the first homology of the torus.
The signs can be inferred from the bipartite structure of the dimer. For each PM these two intersection numbers are the first two coordinates of 3d vectors $V_I \equiv (\cdot,\cdot,1)$ defining  a convex integral polygon, named toric diagram, embedded into a 2d section of a 3d lattice at height one.

\begin{figure}[H]
\begin{center}
  \includegraphics[width=11cm]{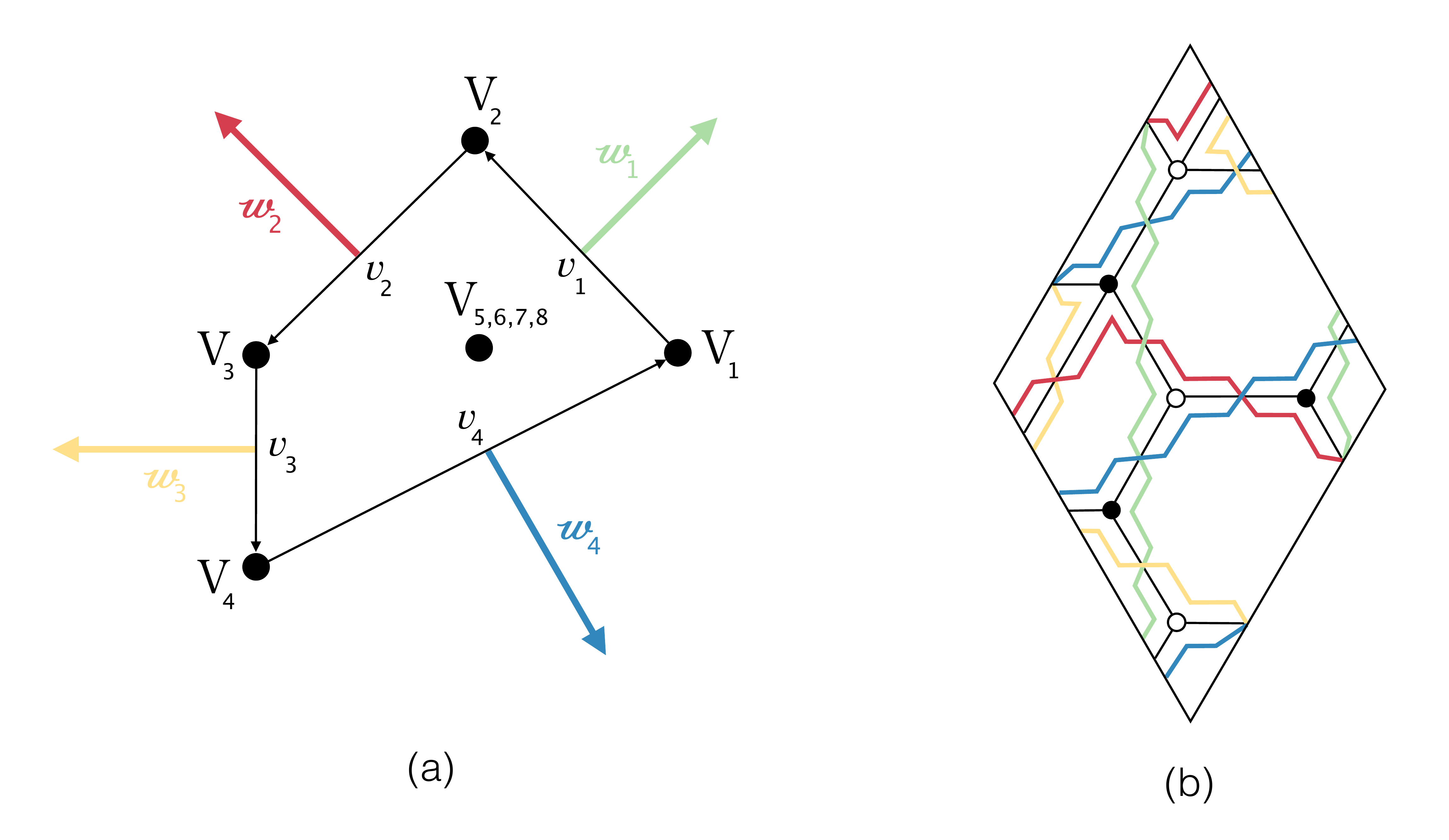}
  \caption{Toric diagram and tiling for the dP$_1$ model. In figure (a) primitive normal vectors $w_I$ of the toric diagram are also indicated and the different colors clarify their relation with the zig-zag paths in the dimer, figure (b).
  This is useful for reading the R-charges of the fields in terms of the 
  charges of the zig-zag paths or of the PM.}
  \label{dP1t}
\end{center}
\end{figure}

In the dP$_1$ case, the  
toric diagram is given by the 3d vectors $V_I$ that are associated to the PM's as follows 
\begin{equation}
\begin{array}{c|c|c|c}
\text{PM} & \text{Primitive vector} &\text{PM} & \text{Primitive vector} \\
\hline
\pi_1     =  \{X_{34}^{(3)}, X_{42}, X_{13}\}              & V_1 = (1,0,1)   &
\pi_5     =  \{X_{12}, X_{13}, X_{42}\}                    & V_5 = (0,0,1)   \\
\pi_2     =  \{X_{23}^{(1)}, X_{34}^{(1)}, X_{41}^{(1)}\}  & V_2 = (0,1,1)   &
\pi_6     =  \{X_{13}, X_{23}^{(1)}, X_{23}^{(2)}\}        & V_6 = (0,0,1)   \\
\pi_3     =  \{X_{12}, X_{34}^{(1)}, X_{34}^{(2)}\}        & V_3 = (-1,0,1)  &
\pi_7     =  \{X_{34}^{(1)}, X_{34}^{(2)},  X_{34}^{(3)}\} & V_7 = (0,0,1)   \\
\pi_4     =  \{X_{23}^{(2)}, X_{34}^{(2)}, X_{41}^{(2)}\}  & V_4 = (-1,-1,1) &
\pi_8     =  \{X_{41}^{(1)}, X_{41}^{(2)}, X_{42}\}        & V_8 = (0,0,1)   \\
\end{array}
\end{equation}
\vskip 3pt
\noindent
and the corresponding toric diagram is drawn in figure \ref{dP1t}(a). 

It is also useful to review the notion of zig--zag paths.
Given the set of primitive vectors $v_I$, one can define primitive normal vectors $w_I$, orthogonal to the edges of the toric diagram,  $v_I \equiv (V_{I+1} - V_{I})$ (see figure \ref{dP1t}(a)). These vectors are in 1--1 correspondence with a set of paths, made out of edges of the dimer, called zig--zag paths and represented in figure \ref{dP1t}(b). 
They are oriented closed loops on the dimer that turn maximally left (right) at the black (white) nodes.
The zig--zag paths correspond to differences of consecutive PM's
lying at the corners (and, if present, on the perimeter) of the toric diagram and are associated to the $U(1)$ global symmetries of the superpotential.

Viceversa, given a particular toric diagram it is possible to identify the main features of the corresponding quiver gauge theory as follows:

\noindent
$\bullet$ The number of $U(N)$ gauge groups describing the quiver is given by twice the area of the toric diagram.

\noindent
$\bullet$ The matter content of the theory (type of bifundamental fields and their degeneracy)
can be inferred from the edges $v_I$ of the toric diagram 
\cite{Feng:2005gw},
up to Seiberg duality, or equivalently toric phases, corresponding to  Yang-Baxter transformations on the zig zag paths 
\cite{Hanany:2005ss}.
In its minimal toric phase a set of bifundamental fields $\Phi_{IJ}$ is assigned to each pair $(I,J)$ with degeneracy $| {\rm det}(v_I, v_J) |$. 

\noindent
$\bullet$ The corresponding spectrum of $R_{IJ}$ charges is determined by assigning a R--charge $\Delta_{\pi_I}$ to each external PM and using the following prescription \cite{Butti:2005ps}
\begin{equation}
\label{PMRij}
\left \{
\begin{array}{lr}
R_{IJ} = \sum_{K=I+1}^J \Delta_{\pi_K}     & \quad I<J 
\vspace{.2cm}\\
R_{IJ} = 2-\sum_{K=J+1}^I \Delta_{\pi_K} & \quad I>J
\end{array}
\right.
\end{equation}
where the charges $\Delta_{\pi_I}$ are subject to the constraint
\begin{equation}
\label{Deltaconstraint}
\sum_{I=1}^d \Delta_{\pi_I} =2 
\end{equation}
to ensure that each superpotential term has R--charge equal to two. A geometric interpretation of (\ref{Deltaconstraint}) can be given in terms of the \emph{isoradial embedding} \cite{Hanany:2005ss}.

The fact that the degeneracy of the fields with a given R--charge $R_{IJ}$ is given by $| {\rm det}(v_I, v_J) |$ has a nice geometric interpretation in terms of zig--zag paths \cite{Butti:2005ps}.

\noindent 
$\bullet$ The number $d$ of external  vertices of the toric diagram also determines the total number of non--anomalous $U(1)$ global symmetries of the gauge theory, which are identified as one R--symmetry,  two flavor symmetries and $(d-3)$ baryonic symmetries.
Analogously to (\ref{Deltaconstraint}), the condition for the superpotential to be neutral with respect to any non--R symmetry translates into
\begin{equation}
\label{nonR}
\sum_{I=1}^d Q_{\pi_I}^{\bf J} = 0    \qquad \quad  {\bf J} = 1, \dots , d-1
\end{equation}
where $Q_{\pi_I}^{\bf J}$ is the charge of the $I$--th PM with respect to the ${\bf J}$--th symmetry.

\noindent
$\bullet$ From the geometric data one can also identify the anomalies of the theory.
The crucial observation is that the areas of the triangles of the toric diagram are 
related to the coefficients of the global anomalies of the field theory as \cite{Benvenuti:2006xg, Lee:2006ru} 
\begin{equation}
\label{toric}
\text{Tr}_{\rm 4d} ({\cal T}_I {\cal T}_J {\cal T}_K) = \frac{N^2}{2} | \det (V_I,V_J,V_K ) |
\end{equation}
where ${\cal T}_I$ are global symmetry generators
and the trace $\text{Tr}_{\rm 4d}$ is taken over the 4d fermions
with the insertion of the 4d chirality operator.
\\
Consequently,  the central charge can be written as 
\cite{Lee:2006ru} 
 \begin{equation}
\label{ag}
a_{geom} \equiv  \frac{9  }{64} N^2 \,  |\det(V_I,V_J,V_K)| \, \Delta_{\pi_I} \,   
\Delta_{\pi_J} \,  \Delta_{\pi_K}
\end{equation}
This expression is equivalent to (\ref{FT}) once we take into account the mapping between the two sets of $R_i$ and $\Delta_{\pi_I}$ charges, eq. (\ref{PMRij}). 

In the case of the dP$_1$ model, using definition (\ref{eq:PMbasis}) we find the following map between fields $\Phi_{IJ}$ obtained from the toric diagram and those given by the quiver description
\begin{equation}
  \begin{array}{c|c|c|c}
    (I, J) & | {\rm det}(v_I, v_J) | & \Phi_{IJ} & R_{IJ}\\
    \hline
    (4, 1) & 3 & \{ X_{13}, X_{34}^{(3)}, X_{42} \} & \Delta_{\pi_1}\\
    (1, 2) & 2 & \{ X_{23}^{(1)}, X_{41}^{(1)} \} & \Delta_{\pi_2}\\
    (2, 3) & 1 & \{ X_{12} \} & \Delta_{\pi_3}\\
    (3, 4) & 2 & \{ X_{23}^{(2)}, X_{41}^{(2)} \} & \Delta_{\pi_4}\\
    (1, 3) & 1 & \{ X_{34}^{(1)} \} & \Delta_{\pi_2} + \Delta_{\pi_3}\\
    (2, 4) & 1 & \{ X_{34}^{(2)} \} & \Delta_{\pi_3} + \Delta_{\pi_4}\\
  \end{array}
\end{equation}
\\
where $\Delta_{\pi_I}$ satisfy (\ref{Deltaconstraint}) and all internal PMs are assigned zero charge under all $U(1)$ symmetries.

\vskip 10pt 
The example we have considered has a smooth horizon where all the external  points 
of the toric diagram  correspond to corners. In this case the prescription for assigning R--charge 
to the bifundamental fields is unambiguously given in (\ref{PMRij}). In the case of singular horizons there are also points
on the perimeter of the toric diagram that do not correspond to corners. 
These points have a degeneracy (given by a binomial coefficient), as they correspond to more than one PM. The assignment of the
R--charges in terms of the external PM's may then become ambiguous. According to the prescription in \cite{Butti:2005ps,Butti:2006nk}, at these points one sets to zero the R--charges of the  
PM's that do not determine any zig-zag path, being then left with an unambiguous assignment of $\Delta_{\pi_I}$ charges. 

\vskip 10pt

The holographic correspondence provides the following relation between the central charge and the X$_5$ volume  \cite{Gubser:1998vd} 
\begin{equation}
\label{MSY}
a_{\text{holo}} \equiv \frac{N^2 \pi^3}{4 \, \text{vol}(\text{X}_5(\mathbf{b}))}
\end{equation}
where the volume is parameterized in terms of the components of the Reeb vector $\mathbf{b}$, a constant norm Killing vector that commutes with the X$_5$  isometries.
It follows that the  $a$-maximization prescription that determines the exact R--current in field theory corresponds to the volume minimization in the gravity dual.
 
When the cone over X$_5$ is toric the central charge can be directly obtained from the toric geometry. 
In fact, the X$_5$ volume can be expressed as \cite{Martelli:2005tp}
\begin{equation}
\text{vol}(\text{X}_5) = \frac{\pi}{6} \sum_{I=1}^{d} \text{vol} (\Sigma_I) 
\end{equation}
where $d$ represents the number of vertices  
and vol$(\Sigma_I)$ corresponds to the volume
of a 3--cycle $\Sigma_I$, on which D3 branes, 
corresponding to dibaryons, are wrapped on \cite{Franco:2005sm}.

Holographic data also determine the $\Delta_{\pi_I}$ charges that can be parameterized in terms of the components of 
the Reeb vector $\mathbf{b}$. Using the explicit parameterization   
\cite{Gubser:1998fp}
\begin{equation}
  \Delta_{\pi_I}(\mathbf{b}) =  \frac{\pi}{3} \frac{ \text{vol} (\Sigma_I (\mathbf{b})) }{\text{vol}(\text{X}_5(\mathbf{b})) }
\end{equation}
it is easy to show the equivalence between $a_{geom}$ in eq. (\ref{ag}) and $a_{\text{holo}}$ in eq. (\ref{MSY}).

\subsection{Twisted compactification}
\label{sectw}
%
%
%
%
In this section we review the main aspects of partial topologically twisted compactifications 
of a 4d $\mathcal{N}=1$ SCFTs on a genus $g$ Riemann surface $\Sigma$, and the computation of the central charge $c_r$
for the corresponding 2d SCFTs, directly from 4d anomaly data.

When placing a 4d ${\cal N}=1$ SCFT on $\Sigma \times \mathbb{R}^{1,1}$,
supersymmetry  is generally broken by the coupling with the $\Sigma$ curvature. In order to (partially) preserve it one performs a
twist \cite{Witten:1988xj} by turning on background gauge fields along $\Sigma$ 
for an abelian 4d R--symmetry $t_R$ that assigns integer charges to the fields.
Choosing its flux to be proportional to the curvature, its contribution to the Killing
spinor equations cancels the contribution from the spin connection and possibly non--trivial solutions for Killing spinors can be found.
More generally, one can also turn on properly quantized background fluxes along the $\Sigma$ directions
for other abelian global symmetries. In this case supersymmetry is preserved if the associated gaugino variations
vanish as well.
Summarizing, the most general twist is performed along the generator 
\begin{equation}
\label{twist}
T = \kappa \,  T_R +\sum_{{\bf I}=1}^{n_A}  b_{\bf I} T_{\bf I} 
\end{equation}
where $\kappa = 0$ for the torus and $\kappa = \pm 1$ for curved Riemann surfaces \footnote{We use conventions of \cite{Hosseini:2016cyf} that differ by a factor 2 from the conventions previously used in \cite{Amariti:2017cyd}.}. Here $n_A$ refers to the number of abelian $T_{\bf I}$ generators of 
non--R global symmetries (both flavor and baryonic ones) and $b_{\bf I}$ are the corresponding background fluxes.
For generic choices of the fluxes $\mathcal{N}=(0, 2)$ supersymmetry is preserved on  $\mathbb{R}^{1,1}$ \cite{Amariti:2017cyd}.

After the  twist the trial 2d R--symmetry generator is a mixture of the abelian generator $T_R$ and the other $T_{\bf I}$  generators  
\begin{equation}
 \label{eq:mixR-symmFlav}
R = T_R + \sum_{{\bf I}=1}^{n_A} \epsilon_{\bf I} T_{\bf I}
\end{equation}
where  $\epsilon_{\bf I}$ are the mixing coefficients and $T_R, T_{\bf I}$ are meant to act on fields reorganized in 2d representations. 

At the IR fixed point the $\epsilon_{\bf I}$ coefficients have to extremize the 2d central charge $c_r= 3 k_{RR} \equiv 3 {\rm Tr}_{\rm 2d}( RR)$ \cite{Benini:2013cda}.
Practically, the relevant anomaly coefficient $k_{RR}$ can be obtained from the anomaly polynomial $I_6$ expressed in terms of the
triangular anomalies of the 4d theory, 
by integrating $I_6$ on $\Sigma$ and matching the resulting expression with the general structure of the anomaly polynomial $I_4$ in two dimensions \cite{Benini:2015bwz}. 
In particular,  one obtains the trial central charge  
\begin{eqnarray}
\label{eq:krrex}
  c_r &=&  3 \, \eta_{\Sigma} \text{Tr}_{\rm 4d} ( T  R^2)   \\
 &=&  3 \, \eta_{\Sigma}  \left[ ( b_{\bf K} k_{{\bf K} \nocomma {\bf I} \nocomma{\bf J}} 
  +
  \kappa k_{R  \nocomma {\bf I} \nocomma{\bf J}} ) \epsilon_{\bf I}
  \epsilon_{\bf J} + 2 ( b_{\bf K} k_{R \nocomma {\bf K} \nocomma {\bf I}} +  \kappa
  k_{R \nocomma R \nocomma {\bf I}} ) \epsilon_{\bf I} + ( b_{\bf K} k_{R \nocomma R
  \nocomma {\bf K}} +   \kappa k_{R \nocomma R \nocomma R} ) \right] \nonumber 
\end{eqnarray}
where $\eta_{\Sigma}=2|g-1|$ for $g \neq 1$ and $\eta_{\Sigma}=1$ for $g=1$, and we have defined $k_{{\bf K} \nocomma {\bf I} \nocomma{\bf J}}  \equiv {\rm Tr}_{\rm 4d} (T_{\bf K} T_{\bf I} T_{\bf J})$, $k_{R \nocomma {\bf I} \nocomma{\bf J}}  \equiv {\rm Tr}_{\rm 4d} (R T_{\bf I} T_{\bf J})$ and similarly for the other trace coefficients. 

The exact central charge is finally obtained by extremizing with respect to the variables $\epsilon_{\bf I}$ \cite{Benini:2012cz}. 
At the fixed point we have
\begin{equation}
\label{mixing2d}
  \epsilon^{\ast}_{\bf I} = - \eta_{\Sigma} k^{- 1}_{{\bf I} \nocomma {\bf J}}  \left( k_{R
  \nocomma {\bf J} \nocomma {\bf K}} b_{\bf K} + \kappa k_{R \nocomma R \nocomma {\bf J}}
  \right)
\end{equation}
where 
\begin{equation}
  k_{{\bf I} \nocomma {\bf J}} 
   = 
   \eta_{\Sigma}  
   \left( 
   \kappa k_{ R \nocomma   {\bf I} \nocomma {\bf J}} 
   + 
   b_{\bf K} k_{{\bf K} \nocomma {\bf I} \nocomma {\bf J}} 
   \right)
\end{equation}

From (\ref{mixing2d}) we observe that coefficients $ \epsilon^{\ast}_{\bf I}$ are generically non--vanishing 
for any choice of the $b_{\bf I}$ fluxes. In particular, this is true for the coefficients associated to baryonic symmetries, which then do mix with the exact R--current in two dimensions, even if they do not
in the original 4d theory. This pattern has been already observed in \cite{Benini:2015bwz} for the $Y^{pq}$ family.
In section \ref{FTside} we will study toric quiver gauge theories with a larger amount of baryonic symmetries, confirming that they generically mix
with the 2d exact R--current after the twisted compactification on $\Sigma$. 

Starting from a 4d toric theory with $n_G$ $U(N)$ gauge groups and $n_F$ massless chiral fermions, to each 2d field surviving the compactification on $\Sigma$ we can associate a $T$--charge $n_i$ and a R--charge $R_i$ according to (see eqs. (\ref{twist}) and (\ref{eq:mixR-symmFlav}))
\begin{equation}
\label{eq:par}
n_i =  \kappa \, r_i + Q_i^{\bf J} b_{\bf J},
\quad
R_i = r_i +  Q_i^{\bf J} \epsilon_{\bf J} \qquad i=1,\dots,n_F
\end{equation}
where $r_i$ is the R--charge respect to the 4d R--current and $Q_i^{\bf J}$ is the charge matrix of the fermions respect to the global $U(1)$ non--R symmetries, inherited from the 4d parent fields.
Therefore, applying prescription (\ref{eq:krrex}) we find that at large $N$ the central charge before extremization is given by
\begin{equation}
\label{anom}
c_r  = 3N^2  \eta_{\Sigma}  \left(\kappa \, n_G + \sum_{i=1}^{n_F} (n_i-\kappa)(R_i-1)^2\right) + {\cal O}(1)
\end{equation}
This formula is general and applies to any 2d SCFT obtained from compactification of a 4d quiver gauge theory on a Riemann surface with curvature $\kappa$. Through $R_i$ it depends parametrically on the mixing coefficients $\epsilon_{\bf I}$ that need to be determined by the 2d extremization procedure.

As reviewed in the previous sub--section, in the case of 4d toric quiver theories we can parametrize the $U(1)$ charges in terms of PM variables $\Delta_{\pi_I}$ and $Q_{\pi_I}^{\bf J}$. When twisting, we can also assign to PMs a further $n_{\pi_I}$ charge  with respect to the twisting $T$ symmetry (\ref{twist}) as (in this case $n_A = d-1$) 
\begin{equation}
 \label{eq:ncharges}
n_{\pi_I} = \kappa \Delta_{\pi_I} + b_{\bf J} Q^{\bf J}_{\pi_I} \quad \qquad {\rm with}  \qquad \sum_{I=1}^d n_{\pi_I} = 2\kappa
\end{equation}
where the constraint on $n_{\pi_I}$ follows from (\ref{Deltaconstraint}) and (\ref{nonR}).

The $r_i$ and $Q_i^{\bf J}$ charge assignments in two dimensions, eq. (\ref{eq:par}), need necessarily to respect the original constraints arising from the condition of superconformal invariance for  the 4d superpotential. In particular, given the superpotential $W = \sum_\alpha W_\alpha$, these constraints imply that for each superpotential term $W_\alpha$ the conditions $\sum_{i \in W_\alpha}r_i = 2$ and $\sum_{i \in W_\alpha} Q_i^{\bf J} =0$ hold. 
Consequently, from (\ref{eq:par}) we read  
\begin{equation}
\label{constrnd}
\sum_{i \in W_\alpha} R_i = 2 \, , \qquad 
\sum_{i \in W_\alpha} n_i = 2 \kappa
\end{equation}
Now, we can think of the dimensional flow from the original 4d theory to the resulting 2d one as being accompanied by the set of toric data $(\Delta_{\pi_I}, Q^{\bf J}_{\pi_I}, n_{\pi_I})$ that parametrize the $U(1)$ charges in 4d and, consequently, that can still be used to parametrize the corresponding charges in two dimensions. Using this parametrization reinterpreted as charge parametrization for 2d fields, constraints (\ref{constrnd}) are traded with 
(\ref{Deltaconstraint}), (\ref{nonR}) and (\ref{eq:ncharges}).

\section{$c_{r}$ from toric geometry}
\label{secgeom}
%
%
%
%
For the class of 2d SCFTs obtained from the topologically twist reduction of toric quiver gauge theories, we now provide a general prescription for determining the central charge $c_r$ directly in terms of the geometry of the toric diagram associated to the original 4d parent theory. This is the main result of the paper, which we are going to check in the successive sub--sections for a number of explicit examples.

%
%
%

\subsection{Reading the 2d central charge from the toric diagram}
\label{central}
%
%
%
%
%
To this end, we consider a toric gauge theory twisted along the abelian generator 
\begin{equation}
\label{twistsym}
T= \sum_{I=1}^{d} a_I {\cal  T}_I \quad \qquad {\rm with} \qquad \sum_{I=1}^d a_I = 2\kappa
\end{equation}
where $I$ runs over the $d$ external points of the toric diagram. To be consistent with the conventions used so far, the abelian ${\cal T}_I$ generators are chosen to assign R--charge one to the superpotential of the 4d theory.
It is always possible to construct such a set of generators by combining the generators of the 4d trial R--current, the  two flavor symmetries and the $(d-3)$ non--anomalous baryonic symmetries that appear in (\ref{twist}).
The new fluxes are subject to the constraint in (\ref{twistsym})
in order to ensure surviving ${\cal N} = (0,2)$ supersymmetry in 2d. They need to be further constrained
in such a way that each flux $b_{\bf I}$ in (\ref{twist}) is properly quantized.

Accordingly, the 2d trial R--symmetry can be written as 
\begin{equation}
\label{2dtrial}
R = \sum_{I=1}^{d} \epsilon_I {\cal T}_I \qquad {\rm with} \qquad \sum_{I=1}^d\epsilon_I=2
\end{equation} 
where the constraint follows from the requirement for $R$ to be a canonical normalized R--current. 

The 2d central charge $c_r$ expressed in terms of the 4d anomaly coefficients $\text{Tr}_{\rm 4d} ({\cal T}_I {\cal T}_J {\cal T}_K)$, the $a_I$ fluxes and the mixing parameter $\epsilon_I$ becomes  (see eq. (\ref{eq:krrex}))
\begin{equation}
\label{c2d}
c_{r} = 3 \, \eta_{\Sigma} \text{Tr}_{\rm 4d} ( T R^2) = 
3 \, \eta_{\Sigma} \text{Tr}_{\rm 4d} ({\cal T}_I {\cal T}_J {\cal T}_K) \,a_I \epsilon_J \epsilon_K 
\end{equation}
In the case of toric theories the anomaly coefficients are given by (\ref{toric}) in terms of the areas of the triangles of the toric diagram. Therefore, the 2d central charge can be rewritten as
\begin{equation}
\label{c2dbis}
c_{r} =   \frac{3 \eta_{\Sigma} N^2}{2}  |\det(V_I,V_J ,V_K)| a_I \epsilon_J \epsilon_K
\end{equation}

In order to complete the map between the 2d field theory and the 4d geometric data we need to find a prescription for parametrizing  
the $a_I$ fluxes and the mixing parameters $\epsilon_I$ in terms of the PM's associated 
to the external vertices of the toric diagram. To this end, we observe that the constraints satisfied by $a_I$ and $\epsilon_I$, eqs. (\ref{twistsym}, \ref{2dtrial}), are the same as the constraints satisfied by $\Delta_{\pi_I}$, eq. (\ref{Deltaconstraint}) and $n_{\pi_I}$, eq. (\ref{eq:ncharges}). Therefore, we are naturally led to identify $\epsilon_I  \equiv \Delta_{\pi_I}$ and $  a_I \equiv n_{\pi_I}$. The central charge $c_r$ for the 2d SCFT obtained from a 4d toric quiver gauge theory topologically twisted on a 2d constant curvature Riemann surface can be then expressed entirely in terms of the toric data by the formula 
\begin{equation}
\label{c2stris}
c_{r} = \frac{3 \eta_{\Sigma} N^2}{2}  |\det(V_I,V_J ,V_K)| n_{\pi_I} \Delta_{\pi_J} \Delta_{\pi_K}
\end{equation}
with $\Delta_{\pi_J} $ and $n_{\pi_I}$ satisfying  constraint (\ref{Deltaconstraint}) and (\ref{eq:ncharges}).
The exact central charge for the 2d SCFT is then obtained by extremizing (\ref{c2stris}) as a function of $\Delta_{\pi_I}$.

Our proposal (\ref{c2stris}) requires some direct check on explicit examples that we report below. However, a holographical confirmation can be already found in
the analysis of the AdS$_5 \to$ AdS$_3$ flow 
engineered in gauged supergravity \cite{Maldacena:2000mw}.  
In this case we need to consider a consistent truncation of 
AdS$_5 \, \times X_5$,
a 5d theory with a gravity multiplet, $n_V$ vector multiplets
and $n_H$ hypermultiplets.
The graviphoton plays the role of the R--symmetry current,  while the
$n_V$ vector    multiplets correspond to the non--R global currents of the holographic dual field theory that remain as massless vector multiplets in a given truncation. In general $n_V \leq n_A$. The hypermultiplets
impose the constraints on the global anomalies.
When flowing to AdS$_3$ and using the Brown-Henneaux formula \cite{Brown:1986nw} in this setup,
it was observed \cite{Karndumri:2013iqa,Benini:2013cda,Amariti:2016mnz}
that $c_{r}$ can be expressed in terms of R--charges ${\hat r}^I$ and fluxes ${\hat a}^J$ as
\begin{equation}
\label{our}
c_{r} =\frac{2 \pi^3 N^2 \eta_{\Sigma}}{3 \text{vol}(\text{X}_5)} C_{IJK} {\hat a}^I {\hat r}^J {\hat r}^K
\end{equation}
where the constraints $\sum {\hat r}^I=2 $ and $\sum {\hat a}^I = 2\kappa$ need to be imposed.
In this formula $C_{IJK}$ are the Chern--Simons coefficients of the dual supergravity, the R--charges ${\hat r}^I$ are obtained from 
the sections of the special geometry corresponding to the (constrained) scalars in the vector multiplets, 
and the prepotentials of $\mathcal{N}=2$ AdS$_5$ gauged  supergravity.
The constants ${\hat a}^I$ are the coefficients of the volume forms in the reduction of the 5d vector multiplets to 3d.

On the other hand, the $C_{IJK}$ coefficients are the holographic duals of the cubic `t Hooft anomaly coefficients, which for toric quiver gauge theories correspond to the  areas of the triangles in the toric diagrams, eq. (\ref{toric}). Therefore 
\begin{equation}
\label{toric2}
C_{IJK} = \frac{N^2}{2} | \det (V_I,V_J,V_K ) |
\end{equation}
If we naturally identify the R--charges $\hat{r}^I$ with the $\Delta_{\pi_I}$  charges assigned to the PM's, and similarly the $\hat{a}_I$ fluxes with the set of $n_I$ fluxes (they satisfy the same constraints
$\sum \Delta_{\pi_I} = 2$ and $\sum n_{\pi_I} = 2\kappa$) we obtain our proposal (\ref{c2stris}).

\subsection{Examples} \label{examples}

In the remaining part of this section we test formula (\ref{c2stris})
on examples of increasing complexity.
As a warm--up we consider the cases of  X$_5=S^5$
corresponding to $\mathcal{N}=4$ SYM and X$_5=T^{1,1}$ corresponding to the conifold.
Then we move to two more complicated cases, namely the second and third del Pezzo surfaces.
We conclude the analysis by considering infinite families of quiver gauge theories
associated to the $Y^{pq}$, $L^{pqr}$ and $X^{pq}$ geometries.

The strategy is the following. For each 4d model we use the general formula (\ref{anom}) to compute the central charge of the corresponding 2d SCFT obtained after twisted compactification. Then, we determine the parametrization of the R--charges and fluxes in terms of the toric data according to our prescription in section \ref{central}. Finally, we check that using this parametrization in (\ref{anom}) we obtain the central charge as given by (\ref{c2stris}). 
%
%
%
\subsubsection*{$\mathcal{N}=4$ SYM}
\begin{figure}[H]
\begin{center}
  \includegraphics[width=9cm]{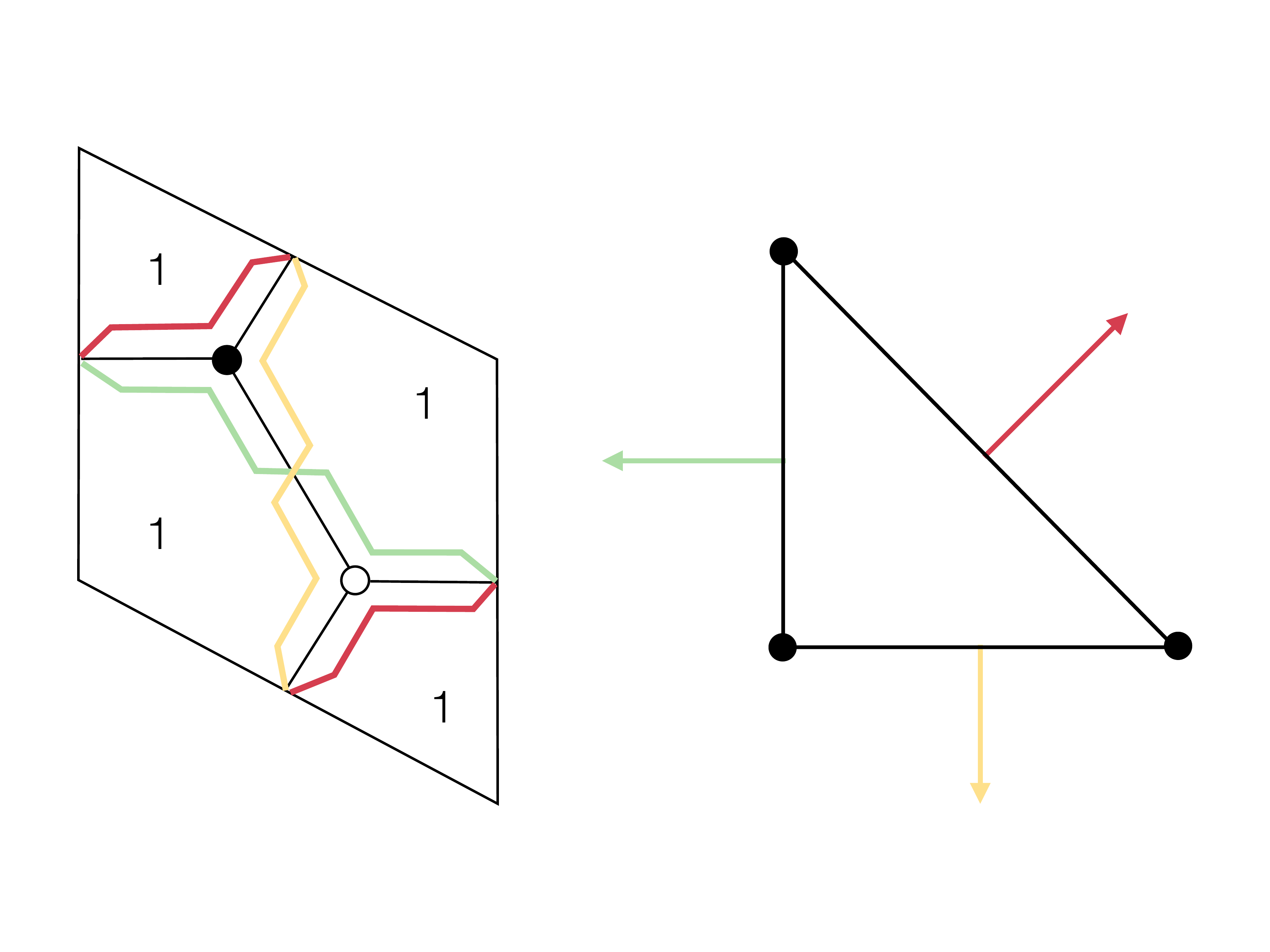}
  \caption{Dimer, zig-zag paths and toric diagram of $S^5$}
  \label{n4fig}
\end{center}
\end{figure}

%
%
%
%
%
The first example that we consider corresponds to the case of X$_5=S^5$.
In this case the dual gauge theory is $\mathcal{N}=4$ SYM and its
twisted compactification on a Riemann surface has been discussed in \cite{Benini:2012cz}.
The 4d field theory can be studied as a toric quiver gauge theory
in $\mathcal{N}=1$ language. In this formulation the global symmetry corresponds to the $U(1)^3$ 
abelian subgroup of  $SO(6)_R$. The quiver has a single node with three adjoint superfields $\Phi_i$ and
superpotential
\begin{equation}
W = \Phi_1 [\Phi_2,\Phi_3]
\end{equation}
The dimer, the zig-zag paths and the toric diagram are
shown in Figure  \ref{n4fig}.

By reducing this theory on $\Sigma$ the topological twist 
is performed along the $U(1)^3$ subgroup of the  $SO(6)_R$.
This corresponds to turning on three fluxes,  one for each 
$U(1)$ factor, constraining their sum to be equal to the curvature $\kappa$. 
From the general expression (\ref{anom}) we can read the 2d central charge at large $N$  
\begin{equation}
\label{cn4}
c_{r} = 3 N^2 \eta_{\Sigma} \, \bigg(\kappa +  \sum_{i=1}^3 (n_{\Phi_i}-\kappa) (R_{\Phi_i}-1 )^2\bigg)
\end{equation}
where $R_{\Phi_i}$ are the R--charges and $n_{\Phi_i}$ the associated fluxes of the three adjoint fields.
These variables are constrained by the relations $R_{\Phi_1} + R_{\Phi_2}+R_{\Phi_3}=2 $
and $n_{\Phi_1}+n_{\Phi_2} +n_{\Phi_3}= 2\kappa$.

Alternatively, we can compute the 2d central charge from (\ref{c2stris}) and find
\begin{equation}
\label{cn4-2}
c_r =
3 N^2 \eta_\Sigma
\big( n_{\pi_1} \Delta_{\pi_2}  \Delta_{\pi_3} +n_{\pi_2} \Delta_{\pi_3}  \Delta_{\pi_1}+n_{\pi_3} \Delta_{\pi_1}  \Delta_{\pi_2}\big)
\end{equation}
In order to check this result against (\ref{cn4}) we
need to express R--charges and fluxes in terms of the ones of the PM's.
This can be done with the prescription discussed in section \ref{toricrev}.
The three zig-zag paths in Figure \ref{n4fig} are the three possible combinations 
of two adjoints, $\Phi_i \Phi_j$. It follows that each adjoint field corresponds to the intersection
of two primitive normal vectors $w_I$ of the toric diagram. 
Furthermore in this case each external PM corresponds to one of the
adjoint fields.
Therefore the charge and the flux assigned to each field correspond to the charge 
and the flux assigned to each external PM
\begin{equation}
\begin{array}{cccc}
R_{\Phi_1} = \Delta_{\pi_1},\quad&
R_{\Phi_2} = \Delta_{\pi_2},\quad&
R_{\Phi_3} = \Delta_{\pi_3}\\
n_{\Phi_1} = n_{\pi_1},\quad&
n_{\Phi_2} = n_{\pi_2},\quad&
n_{\Phi_3} = n_{\pi_3} 
\end{array} 
\end{equation}
By substituting this parameterization in (\ref{cn4}) we can easily prove that in this case
the central charge is equivalent  to (\ref{cn4-2}) if constraints (\ref{Deltaconstraint}) and (\ref{eq:ncharges}) 
are imposed.
%
%
%
%
%
%
\subsubsection*{The conifold}
\begin{figure}[h]
\begin{center}
  \includegraphics[width=13cm]{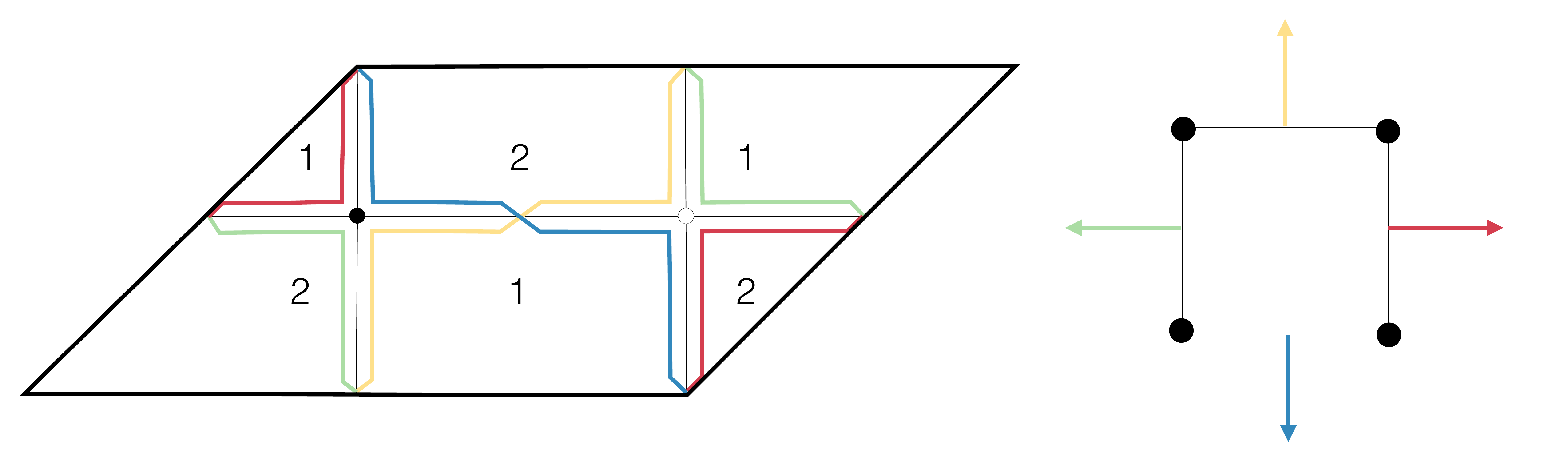}
  \caption{Dimer, zig-zag paths and toric diagram of $T^{1,1}$}
  \label{FIGconifold}
\end{center}
\end{figure}

%
%
%
%
%
As a second example we study the case of the conifold, corresponding to  X$_5 = T^{1,1}$.
The model consists of a $SU(N) \times SU(N)$ gauge theory with two pairs
of bifundamental $a_i$ and anti-bifundamental $b_i$ fields connecting the gauge groups and
interacting through the superpotential 
\begin{equation}
W = \epsilon_{ij} \epsilon_{lk} a_i b_l a_j b_k
\end{equation}
The dimer, the zig-zag paths and the toric diagram are shown in Figure \ref{FIGconifold}.
 In this case the flavor symmetry is $SU(2)^2$ and one baryonic $U(1)$ symmetry is also present.
The R-charges of the four fields, 
$R_{a_i}$ and $R_{b_i}$ are constrained by $R_{a_1}+R_{a_2}+R_{b_1}+R_{b_2}=2$.

When twisting the theory on $\Sigma$ we introduce $T$--fluxes defined in (\ref{eq:par}). In this case they are $n_{a_1}$, $n_{a_2}$, $n_{b_1}$ and $n_{b_2}$,
constrained by $n_{a_1}+n_{a_2}+n_{b_1}+n_{b_2} = 2\kappa$.

The 2d central charge can be written at large $N$, using eq. (\ref{anom})
\begin{equation}
\label{cccon}
c_{r} = 3 N^2 \eta_{\Sigma} \,
\bigg[2 \kappa+ \sum_{i=1}^{2} \bigg( (n_{a_i}-\kappa) (  R_{a_i}-1)^2 +
(n_{b_i}-\kappa) (  R_{b_i}-1)^2 \bigg)\bigg]
\end{equation}
This formula can be reproduced from the geometry of the toric diagram using prescription (\ref{c2stris}). To prove it, we start by ordering the vectors $V_I$ in the toric diagram
as
\begin{equation}
V_1 = (0,0,1),\quad V_2 =(1,0,1),\quad V_3 =(1,1,1) \quad V_4 =(0,1,1)
\end{equation}
The four zig--zag paths in Figure \ref{FIGconifold} are the four possible combinations 
of two bifundamentals, $a_i b_j$. It follows that each bifundamental field corresponds to the intersection
of two consecutive primitive normal vectors of the toric diagram. 
Furthermore in this case each external PM corresponds to one of the
bifundamental fields.
Again the charge and the flux assigned to each bifundamental field correspond to the charge 
and the flux assigned to each external PM
\begin{equation}
\label{parT11}
\begin{array}{cccc}
R_{a_1} = \Delta_{\pi_1},\quad&
R_{b_1} = \Delta_{\pi_2} ,\quad&
R_{a_2} = \Delta_{\pi_3} ,\quad&
R_{b_2} = \Delta_{\pi_4} \\
n_{a_1} = n_{\pi_1} \quad&
n_{b_1} = n_{\pi_2} \quad&
n_{a_2} = n_{\pi_3} \quad&
n_{b_2} = n_{\pi_4} 
\end{array} 
\end{equation}
By substituting parameterization (\ref{parT11}) in (\ref{cccon}) we can check
directly that the central charge $c_{r}$ coincides with the one obtained
from (\ref{c2stris}), under the conditions
\begin{equation}
\label{constraT11}
\sum_{I=1}^{4} \Delta_{\pi_I} = 2, \quad  \quad
\sum_{I=1}^{4} n_{\pi_I} = 2 \kappa
\end{equation}
%
%
%
%
\subsubsection*{dP$_2$}
\begin{figure}[h]
\begin{center}
  \includegraphics[width=5cm]{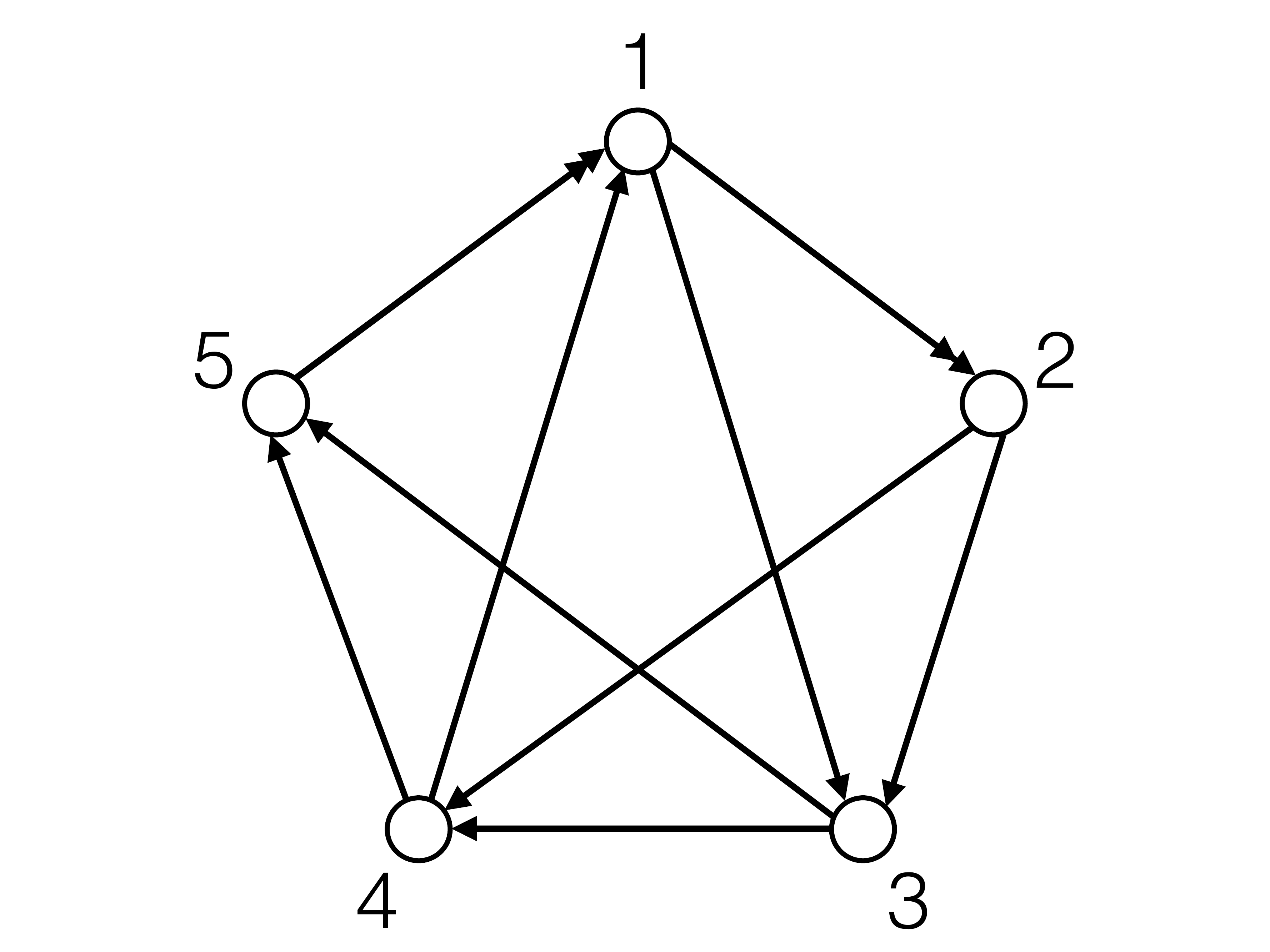}
  \caption{Quiver of the dP$_2^{(I)}$ model}
  \label{FIGDP2}
\end{center}
\end{figure}

%
%
%
%
We now  consider the quiver gauge theory living on a stack of D3 branes probing the
tip of the complex cone over dP$_2$ (see \cite{Benini:2015bwz} for a discussion of the universal twist
of dP$_k$ theories). There are two Seiberg dual realizations of 
such a theory. Here we focus on the case with the minimal number of fields.
This phase is usually referred to as the first phase and denoted as dP$_2^{(I)}$. It is a quiver gauge theory (see figure \ref{FIGDP2}) with five $SU(N)$ gauge groups and 
superpotential
\begin{eqnarray}
  W & = & X_{13} X_{34} X_{41} - Y_{12} X_{24} X_{41} + X_{12} X_{24} X_{45}
  Y_{51} - X_{13} X_{35} Y_{51} \nonumber\\
  &  & + Y_{12} X_{23} X_{35} X_{51} - X_{12} X_{23} X_{34} X_{45} X_{51} . 
\end{eqnarray}
The model has five non anomalous abelian global symmetries. There are a $U(1)_R$ symmetry and two $U(1)$ flavor symmetries
corresponding to the $U(1)^3$ isometry of the SE geometry. There are also five baryonic currents: Two of them are non--anomalous, two are anomalous and one is redundant. 

We perform the $c_r$ calculation from the geometry and we show the
validity of formula (\ref{c2stris}) by matching the geometric result with the one obtained
from the field theory analysis.

\begin{figure}[h]
\begin{center}
  \includegraphics[width=12cm]{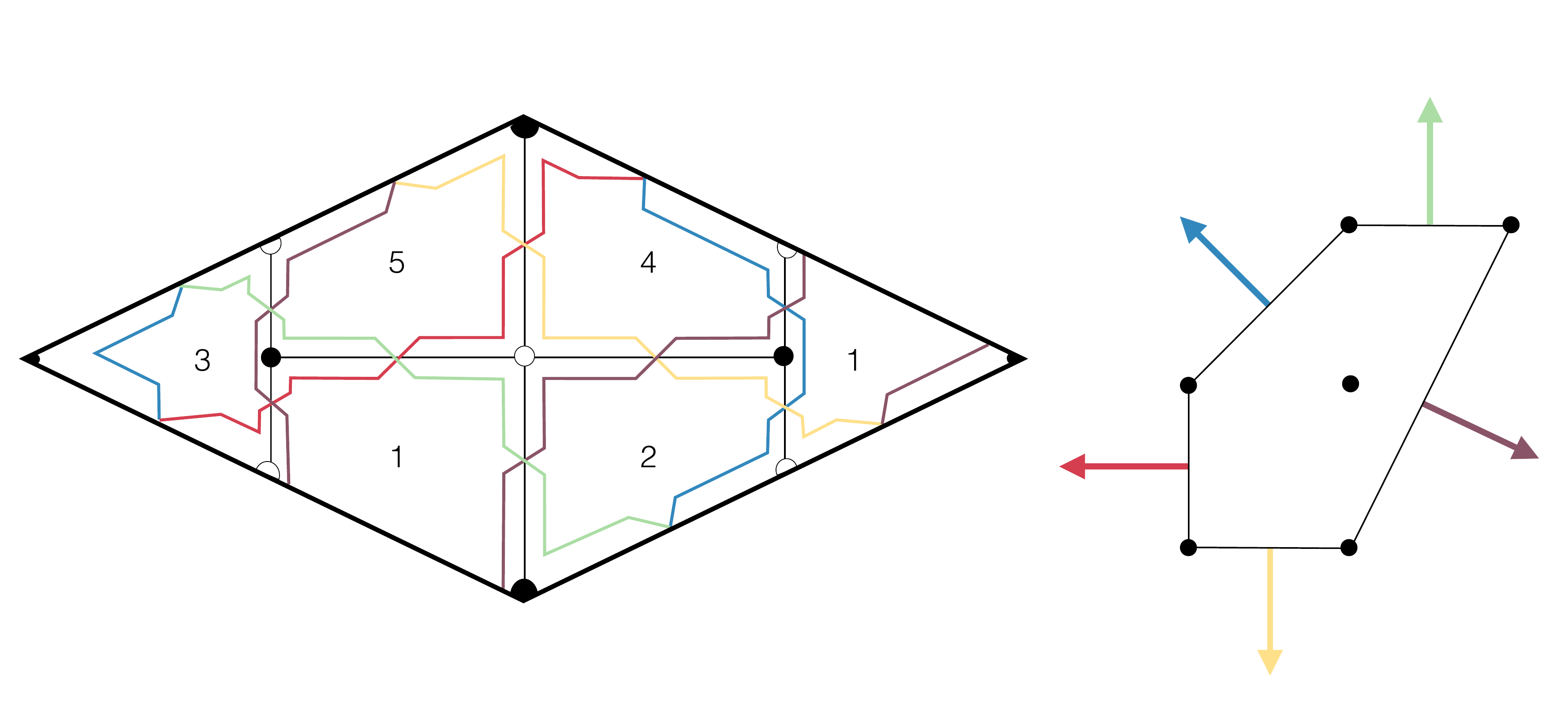}
  \caption{Dimer, zig-zag paths and toric diagram of dP$_2$.}
  \label{FIGdP2zz}
\end{center}
\end{figure}

The dimer, the zig--zag paths and the toric diagram are shown in Figure \ref{FIGdP2zz}.
The toric diagram is identified by the lattice points
\begin{equation}
V_1 = (1,1,1)\quad
V_2 = (0,1,1)\quad
V_3 = (-1,0,1)\quad
V_4 = (-1,-1,1)\quad
V_5 = (0,-1,1)
\end{equation}
The R--charges and the fluxes of the fields can be parameterized 
in terms of the $\Delta_{\pi_I}$ charges and the $n_{\pi_I}$ fluxes as
\begin{equation}
\label{pardP2}
\begin{array}{l|cc}
                    &\hspace{1cm} R_{\phi_i}                    &\hspace{1cm} n_{\phi_i}            \\
\hline
    \phi_1=X_{13}   &\hspace{1cm} \Delta_{\pi_4}+\Delta_{\pi_5} &\hspace{1cm} n_{\pi_4} + n_{\pi_5} \\
    \phi_2=X_{24}   &\hspace{1cm} \Delta_{\pi_5}                &\hspace{1cm} n_{\pi_5}             \\
    \phi_3=X_{51}   &\hspace{1cm} \Delta_{\pi_5}                &\hspace{1cm} n_{\pi_5}             \\
    \phi_4=X_{23}   &\hspace{1cm} \Delta_{\pi_2}                &\hspace{1cm} n_{\pi_2}             \\
    \phi_5=X_{41}   &\hspace{1cm} \Delta_{\pi_1}+\Delta_{\pi_2} &\hspace{1cm} n_{\pi_1} + n_{\pi_2} \\
    \phi_6=Y_{51}   &\hspace{1cm} \Delta_{\pi_2}+\Delta_{\pi_3} &\hspace{1cm} n_{\pi_2} + n_{\pi_3} \\
    \phi_7=Y_{12}   &\hspace{1cm} \Delta_{\pi_3}+\Delta_{\pi_4} &\hspace{1cm} n_{\pi_3} + n_{\pi_4} \\
    \phi_8=X_{45}   &\hspace{1cm} \Delta_{\pi_4}                &\hspace{1cm} n_{\pi_4}             \\
    \phi_9=X_{12}   &\hspace{1cm} \Delta_{\pi_1}                &\hspace{1cm} n_{\pi_1}             \\
    \phi_{10}=X_{35}&\hspace{1cm} \Delta_{\pi_1}                &\hspace{1cm} n_{\pi_1}             \\
    \phi_{11}=X_{34}&\hspace{1cm} \Delta_{\pi_3}                &\hspace{1cm} n_{\pi_3} 
\end{array}
\end{equation}
subject to the constraints $\sum_{I=1}^{5} \Delta_{\pi_I} = 2 $ and $\sum_{I=1}^{5} n_{\pi_I} =2 \kappa$.
This parameterization satisfies the constraints $\sum_{a \in W} R_{\phi_a}=2$
and $\sum_{a \in W} n_{\phi_a}=2 \kappa$.
In this case there are $5$ gauge groups and the central charge is obtained from the formula 
\begin{equation}
\label{cdP2}
c_{r} = 3 N^2 \eta_{\Sigma} \, \bigg(5 \kappa  +  \sum_{i=1}^{11} (n_{\phi_i} -\kappa) (R_{\phi_i}-1)^2 \bigg)
\end{equation}
By substituting parameterization (\ref{pardP2}) in (\ref{cdP2})  we can see show that 
 (\ref{cdP2}) is equivalent to (\ref{c2stris}) once the constraints (\ref{Deltaconstraint}) and (\ref{eq:ncharges}) 
are imposed.
%
%
\subsubsection*{dP$_3$}
%
%

\begin{figure}[H]
\begin{center}
  \includegraphics[width=15cm]{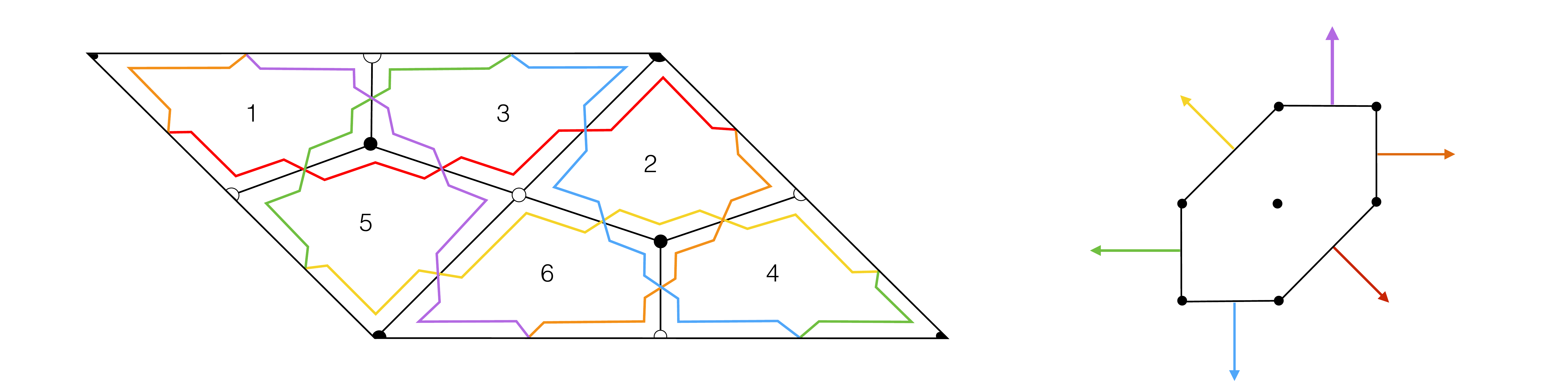}
  \caption{Dimer, zig-zag paths and toric diagram of dP$_3$}
  \label{FIGdP3zz}
\end{center}
\end{figure}

Here we consider the quiver gauge theory living on a stack of D3 branes probing the
tip of the complex cone over dP$_3$. There are four Seiberg dual realizations of 
such a theory, and we focus on the case with the minimal number of fields, usually called the first phase and denoted as dP$_3^{(I)}$.
\begin{figure}[h]
\begin{center}
  \includegraphics[width=5cm]{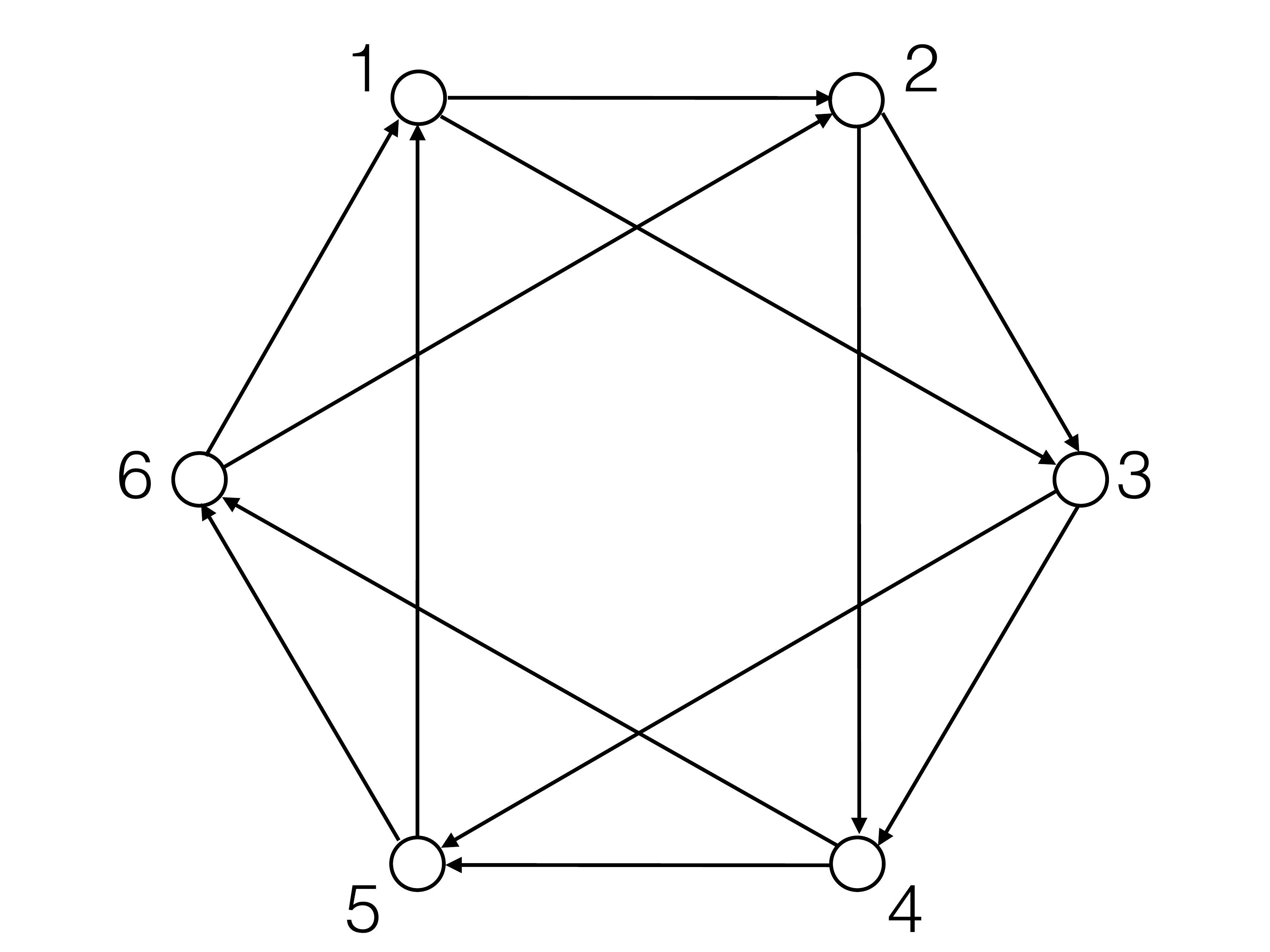}
  \caption{Quiver of the dP$_3^{(I)}$ model}
  \label{FIGDP3}
\end{center}
\end{figure}
The quiver is represented in Figure \ref{FIGDP3}, and it has six gauge groups. The superpotential is
\begin{eqnarray}
  W &=& X_{12} X_{24} X_{45} X_{51} - X_{24} X_{46} X_{62}
  + X_{23} X_{35} X_{56} X_{62} 
  \nonumber\\
  &-& X_{35} X_{51} X_{13} + X_{34} X_{46} X_{61}
  X_{13} - X_{12} X_{23} X_{34} X_{45} X_{56} X_{61} .
\end{eqnarray}
The model possesses six non anomalous abelian global symmetries. There are a $U(1)_R$ symmetry and two $U(1)$ flavor symmetries
corresponding to the $U(1)^3$ isometry of the SE geometry. There are also six baryonic currents: ~Three are non--anomalous, two are anomalous and one is redundant. 
The dimer, the zig-zag paths and the toric diagram are shown in Figure \ref{FIGdP3zz}.
Again we can perform the calculation from the geometry, showing the
validity of formula (\ref{c2stris}).
The toric diagram is identified by the lattice points
\begin{eqnarray}
&&
V_1 =  (1,1,1)
\quad \quad \, \, \, 
V_2 =  (0,1,1)
\quad \,\,\, \, \,
V_3 =  (-1,0,1) \nonumber \\
&&
V_4 =  (-1,-1,1)
\quad
V_5 =  (0,-1,1)
\quad
V_6 = (1,0,1) 
\end{eqnarray}
The R-charges and the fluxes of the fields can be parameterized 
in terms of the $\Delta_{\pi_I}$ charges and of the $n_{\pi_I}$ fluxes as
\begin{equation}
\label{pardP3}
\begin{array}{l|cc}
                    &\hspace{1cm} R_{\phi_i} &\hspace{1cm} n_{\phi_i} \\
  \hline
    \phi_1=X_{12}   &\hspace{1cm} \Delta_{\pi_6}                & \hspace{1cm} n_{\pi_6}            \\
    \phi_2=X_{13}   &\hspace{1cm} \Delta_{\pi_2}+\Delta_{\pi_3} & \hspace{1cm} n_{\pi_2}+n_{\pi_3}  \\
    \phi_3=X_{23}   &\hspace{1cm} \Delta_{\pi_5}                & \hspace{1cm} n_{\pi_5}            \\
    \phi_4=X_{24}   &\hspace{1cm} \Delta_{\pi_1}+\Delta_{\pi_2} & \hspace{1cm} n_{\pi_1}+n_{\pi_2}  \\
    \phi_5=X_{34}   &\hspace{1cm} \Delta_{\pi_4}                & \hspace{1cm} n_{\pi_4}            \\
    \phi_6=X_{35}   &\hspace{1cm} \Delta_{\pi_1}+\Delta_{\pi_6} & \hspace{1cm} n_{\pi_1}+n_{\pi_6}  \\
    \phi_7=X_{45}   &\hspace{1cm} \Delta_{\pi_3}                & \hspace{1cm} n_{\pi_3}            \\
    \phi_8=X_{46}   &\hspace{1cm} \Delta_{\pi_5}+\Delta_{\pi_6} & \hspace{1cm} n_{\pi_5}+n_{\pi_6}  \\
    \phi_9=X_{56}   &\hspace{1cm} \Delta_{\pi_2}                & \hspace{1cm} n_{\pi_2}            \\
    \phi_{10}=X_{51}&\hspace{1cm} \Delta_{\pi_4}+\Delta_{\pi_5} & \hspace{1cm} n_{\pi_4}+n_{\pi_5}  \\
    \phi_{11}=X_{61}&\hspace{1cm} \Delta_{\pi_1}                & \hspace{1cm} n_{\pi_1}            \\
    \phi_{12}=X_{62}&\hspace{1cm} \Delta_{\pi_3}+\Delta_{\pi_4} & \hspace{1cm} n_{\pi_3}+n_{\pi_4}  \\
\end{array}
\end{equation}
with the constraints $\sum_{I=1}^{6} \Delta_{\pi_I} = 2 $ and $\sum_{I=1}^{6} n_{\pi_I} =2 \kappa$.
This parameterization satisfies the constraints $\sum_{a \in W} R_{\phi_a}=2$
and $\sum_{a \in W} n_{\phi_a}=2 \kappa$.
The central charge is obtained from the formula 
\begin{equation}
\label{cdP3}
c_{r} = 3N^2 \eta_{\Sigma} \, \bigg(6  \kappa+ \sum_{i=1}^{12} (n_{\phi_i} -\kappa) (R_{\phi_i}-1)^2 \bigg)
\end{equation}
By substituting parameterization (\ref{pardP3}) in (\ref{cdP3})  we can easily see that 
 (\ref{cdP3}) is equivalent to (\ref{c2stris}) provided the constraints (\ref{Deltaconstraint}) and (\ref{eq:ncharges})
are imposed.
%
%
\subsubsection*{$Y^{pq}$ theories}
%
%
We can prove the validity of (\ref{c2stris}) also for infinite 
families of quiver gauge theories.
The first family that we consider is  X$_5=Y^{pq}$.
These models has been derived in \cite{Benvenuti:2004dy}.  
They are quiver gauge theories with $2p$ gauge groups and 
bifundamental matter. 
For generic values of $p$ and $q$ the models have a $SU(2) \times U(1)$ flavor symmetry and one 
non-anomalous baryonic $U(1)$ symmetry. At the 2d fixed point this baryonic symmetry generically
mixes with the  R-current.

The general prescription to obtain the exact 2d central charge after twisted compactification has been given in \cite{Benini:2015bwz} and detailed explicitly there for some cases of particular interest. 
Knowing the field content of these theories as summarized in Table (\ref{tableY}), at large $N$ we can use the general formula (\ref{anom}) to write
\begin{equation}
\label{centralypq}
c_{r}  = 3N^2 \eta_\Sigma \, \bigg(2 p \kappa+  \sum_{i=1}^6 d_i (n_{\phi_i} -\kappa) (R_{\phi_i}-1)^2\bigg)
\end{equation}
We now show how to reproduce this expression from our geometric formulation (\ref{c2stris}).

For generic values of $p$ and $q$ the toric diagram has four external corners. There are also internal lattice points, associated to the 
anomalous baryonic symmetries, that do not play any role in our analysis.
The corners of the toric diagram are associated to the vectors
\begin{equation}
V_1 = (0,0,1),\quad V_2 =(1,0,1),\quad V_3 =(0,p,1) \quad V_4 =(-1,p-q,1)
\end{equation}
The parameterization of the R--charges and fluxes for the various fields
in terms of the toric data can be read from the following table
\begin{equation}
\label{tableY}
\begin{array}{l|ccc}
            &\hspace{1cm} \text{multiplicity}   &\hspace{1cm} R_{\phi_i}                    &\hspace{1cm} n_{\phi_i}            \\
\hline
 \phi_1=Y   &\hspace{1cm} p+q                   &\hspace{1cm} \Delta_{\pi_1}                &\hspace{1cm} n_{\pi_1}             \\
 \phi_2=U_1 &\hspace{1cm} p                     &\hspace{1cm} \Delta_{\pi_2}                &\hspace{1cm} n_{\pi_2}             \\
 \phi_3=Z   &\hspace{1cm} p-q                   &\hspace{1cm} \Delta_{\pi_3}                &\hspace{1cm} n_{\pi_3}             \\
 \phi_4=U_2 &\hspace{1cm} p                     &\hspace{1cm} \Delta_{\pi_4}                &\hspace{1cm} n_{\pi_4}             \\
 \phi_5=V_1 &\hspace{1cm} q                     &\hspace{1cm} \Delta_{\pi_2}+\Delta_{\pi_3} &\hspace{1cm} n_{\pi_2} + n_{\pi_3} \\
 \phi_6=V_2 &\hspace{1cm} q                     &\hspace{1cm} \Delta_{\pi_3}+\Delta_{\pi_4} &\hspace{1cm} n_{\pi_3} + n_{\pi_4} 
\end{array}
\end{equation}
The charges are subject to constraints (\ref{Deltaconstraint}).    
This parameterization satisfies the constraints $\sum_{a \in W} R_{\phi_a}=2$ and $\sum_{a \in W} n_{\phi_a}= 2\kappa$ at each node of the dimer.
 
It is now easy to check that substituting these expressions for the R--charges and the fluxes in (\ref{centralypq}) and taking into account constraints (\ref{Deltaconstraint}) and (\ref{eq:ncharges}) we reproduce exactly what we would obtain from (\ref{c2stris}).
%
%
\subsubsection*{$L^{pqr}$ theories}
%
%
We now consider a second infinite family, corresponding to X$_5=L^{pqr}$, for $p \neq r$ (the degenerate case $p=r$ will be treated in section \ref{perimeter}).
These models have been derived in 
\cite{Benvenuti:2005ja,Butti:2005sw,Franco:2005sm}. 
They can be described in terms of a necklace quiver, i.e. a set of $p+q$ $SU(N)$ gauge groups
such that each node is connected to its nearest neighbors by a  bifundamental
and an anti-bifundamental fields. In general there may be also additional adjoint chiral multiplets, depending on the value
of $p$ and $q$ and on the Seiberg  dual phase that we are considering.  

 The central charge at large $N$ can be easily obtained from (\ref{anom}) taking into account the field content of these theories in their the minimal phase, as summarized in table (\ref{tableLpqr})
\begin{equation}
\label{centrallpqr}
c_{r} = 3N^2 \eta_\Sigma \, \bigg((p+q) \kappa+ \sum_{i=1}^6 d_i (n_{\phi_i}  -\kappa) (R_{\phi_i}-1)^2\bigg)
\end{equation}
To check the equivalence with the geometric prescrition (\ref{c2stris}) we first assign the external corners of the toric diagrams to the following vectors
\begin{equation}
V_1 = (0,0,1),\quad V_2 =(1,0,1),\quad V_3 =(P,s,1) \quad V_4 =(-k,q,1)
\end{equation}
where $r-Ps - kq=0$ and $p+q=r+s$ and $p\leq r \leq q\leq s$.
R-charges and fluxes parametrized in terms of the PM's are
\begin{equation}
\label{tableLpqr}
\begin{array}{l|ccc}
                &\hspace{1cm} \text{multiplicity} &\hspace{1cm} R_{\phi_i} &\hspace{1cm} n_{\phi_i} \\
\hline
    \phi_1=Y    &\hspace{1cm} q     &\hspace{1cm} \Delta_{\pi_1}                    &\hspace{1cm} n_{\pi_1}             \\
    \phi_2=W_2  &\hspace{1cm} s     &\hspace{1cm} \Delta_{\pi_2}                    &\hspace{1cm} n_{\pi_2}             \\
    \phi_3=Z    &\hspace{1cm} p     &\hspace{1cm} \Delta_{\pi_3}    	            &\hspace{1cm} n_{\pi_3}             \\
    \phi_4=X_1  &\hspace{1cm} r     &\hspace{1cm} \Delta_{\pi_4}                    &\hspace{1cm} n_{\pi_4}             \\
    \phi_5=W_1  &\hspace{1cm} q-s   &\hspace{1cm} \Delta_{\pi_2} + \Delta_{\pi_3}   &\hspace{1cm} n_{\pi_2} + n_{\pi_3} \\
    \phi_6=X_1  &\hspace{1cm} q -r  &\hspace{1cm} \Delta_{\pi_3} + \Delta_{\pi_4}   &\hspace{1cm} n_{\pi_3} + n_{\pi_4} 
\end{array}
\end{equation}
with the constraints $\sum_{I=1}^{4} \Delta_{\pi_I} = 2 $ and $\sum_{I=1}^{4} n_{\pi_I} = 2\kappa$.
This parameterization satisfies the constraints $\sum_{a \in W} R_{\phi_a}=2$
and $\sum_{a \in W} n_{\phi_a}=2 \kappa$.
 
Substituting this parameterization in (\ref{centrallpqr})  we directly obtain an expression equivalent to (\ref{c2stris}),  once constraints (\ref{Deltaconstraint}) and (\ref{eq:ncharges})
are taken into account.
%
%
\subsubsection*{$X^{pq}$ theories}
%
%
Finally we consider the infinite family of models corresponding to X$_5=X^{pq}$. They have been constructed in \cite{Hanany:2005hq}.
In this case there are $2p+1$ gauge groups and taking into account the spectrum of fields and their multiplicities as given in table (\ref{tableX}), the 2d central charge as read from (\ref{anom}) is  
\begin{equation}
\label{centralxpq}
c_{r} = 3N^2 \eta_{\Sigma} \, \bigg( (2p+1) \kappa+  \sum_{i=1}^{10} d_i (n_{\phi_i} -\kappa) (R_{\phi_i} -1)^2 \bigg)
\end{equation}
To check it against the geometric calculation (\ref{anom}), we first label the external corners of the toric diagrams as (we take $p>q$)
\begin{equation}
V_1 = (1,p,1),\quad V_2 =(0,p-q+1,1),\quad V_3 =(0,p-q,1) \quad V_4 =(1,0,1) \quad V_5 =(2,0,1) \nonumber
\end{equation}
The R-charges and fluxes parametrization in terms of the PM's is given by
\begin{equation}
\label{tableX}
\begin{array}{l|ccc}
                &\hspace{1cm} \text{multiplicity} &\hspace{1cm} R_{\phi_i} &\hspace{1cm} n_{\phi_i} \\
\hline
    \phi_1      &\hspace{1cm} p+q-1 &\hspace{1cm} \Delta_{\pi_1}                                    &\hspace{1cm} n_{\pi_1}             \\
    \phi_2      &\hspace{1cm} 1     &\hspace{1cm} \Delta_{\pi_2}                                    &\hspace{1cm} n_{\pi_2}             \\
    \phi_3      &\hspace{1cm} 1     &\hspace{1cm} \Delta_{\pi_3}                                    &\hspace{1cm} n_{\pi_3}             \\
    \phi_4      &\hspace{1cm} p-q   &\hspace{1cm} \Delta_{\pi_4}                                    &\hspace{1cm} n_{\pi_4}             \\
    \phi_5      &\hspace{1cm} p     &\hspace{1cm} \Delta_{\pi_5}                                    &\hspace{1cm} n_{\pi_5}             \\
    \phi_6      &\hspace{1cm} p-1   &\hspace{1cm} \Delta_{\pi_2} + \Delta_{\pi_3}                   &\hspace{1cm} n_{\pi_2} + n_{\pi_3} \\
    \phi_7      &\hspace{1cm} 1     &\hspace{1cm} \Delta_{\pi_3} + \Delta_{\pi_4}                   &\hspace{1cm} n_{\pi_3} + n_{\pi_4} \\
    \phi_8      &\hspace{1cm} q-1   &\hspace{1cm} \Delta_{\pi_2} + \Delta_{\pi_3} + \Delta_{\pi_4}  &\hspace{1cm} n_{\pi_2} + n_{\pi_3} + n_{\pi_4} \\
    \phi_9      &\hspace{1cm} 1     &\hspace{1cm} \Delta_{\pi_1} + \Delta_{\pi_2}                   &\hspace{1cm} n_{\pi_1} + n_{\pi_2} \\
    \phi_{10}   &\hspace{1cm} q     &\hspace{1cm} \Delta_{\pi_4} + \Delta_{\pi_5}                   &\hspace{1cm} n_{\pi_4} + n_{\pi_5} \\
\end{array}
\end{equation}
with the constraints $\sum_{I=1}^{5} \Delta_{\pi_I} = 2 $ and $\sum_{I=1}^{5} n_{\pi_I} =2 \kappa$.
Once again, this parameterization satisfies the constraints $\sum_{a \in W} R_{\phi_a}=2$
and $\sum_{a \in W} n_{\phi_a}=2 \kappa$.

Using this parameterization it is easy to check that result (\ref{centralxpq}) is equivalent to (\ref{c2stris}), once constraints (\ref{Deltaconstraint}) and (\ref{eq:ncharges}) are imposed.

\newpage

%
%
%
%
%
\section{Singular horizons and lattice points lying on the perimeter}
\label{perimeter}
%
%
%
%
In this section we discuss the case of toric diagrams with some
external lattice points that are not corners but lie along the perimeter.
These diagrams are  associated to theories  with non--smooth horizons,
usually arising from the action of an orbifold.

In this case, as discussed in \cite{Butti:2005vn}, the geometric 
procedure to extract the central charge $a$ 
from the toric diagram needs some modification.
The reason is that the lattice points lying on the perimeter are associated to 
a multiple number of PM's. Therefore, this requires a change in the prescription for assigning R--charges to the fields in terms of the charges of the PM's.

The prescription that we propose follows the one described in \cite{Butti:2005ps} and it works as follows.
First divide the PM's in two sets, the ones associated to corners of the toric diagram
and the degenerate ones lying on the perimeter, namely $\pi^c$ and $\pi^p$ respectively.
Then we  associate a R--charge $\Delta_{\pi^c_I}$ to the PM's at the corners,  as done before.
For the PM's on the perimeter, instead, we proceed as follows.
Observing that at each point on the perimeter only one of the degenerate PM's 
enters the definition of the zig-zag paths, we assign a non zero charge $\Delta_{\pi^p_I}$ to this PM and set the
charge of all the other PM's associated to the same $I$-th lattice point to zero.
With this modification of charge assignments we can then parameterize
the R--charges $R_i$ and the fluxes $n_i$ unambiguously as described in section \ref{secgeom}.

We have checked in a large set of examples that by applying this prescription 
the 2d central charge computed from the field theory 
analysis, eq. (\ref{anom}), matches with the one computed using formula (\ref{c2stris}).
In the following we report the explicit check for a couple of examples in the $L^{pqp}$ 
class.
%
%
%
\subsection{$L^{222}$}
\begin{figure}[h]
\begin{center}
  \includegraphics[width=15cm]{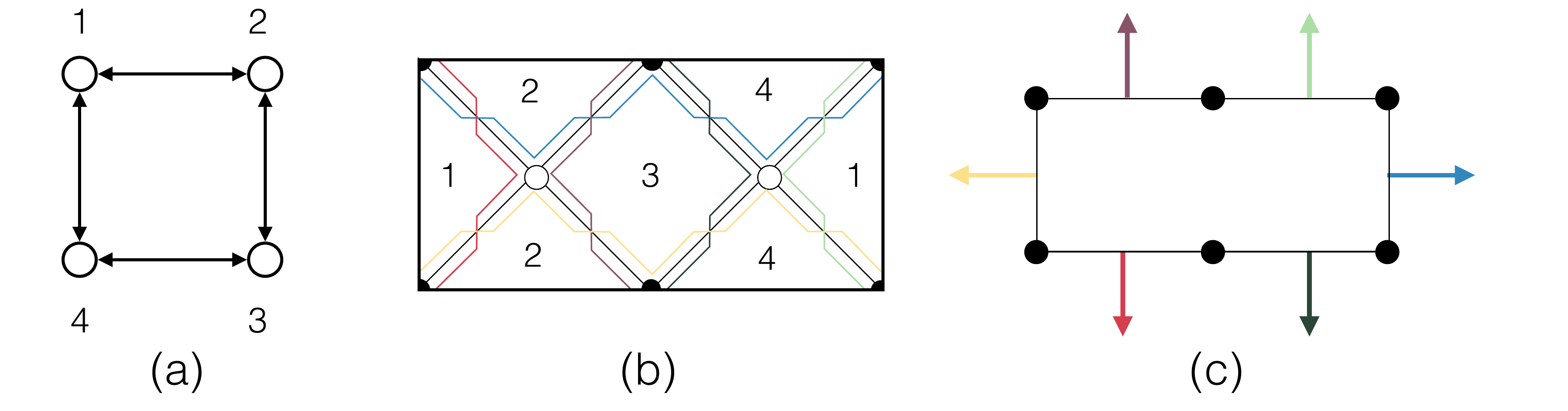}
  \caption{Dimer, zig zag paths and toric diagram of L$^{222}$}
  \label{L222qdt}
\end{center}
\end{figure}

%
%
%
For this particular representative of the $L^{pqr}$ family the quiver diagram, the dimer with the zig--zag paths and the toric diagram are depicted
in Figure \ref{L222qdt}. The superpotential of this model is
\begin{equation}
W = X_{12}X_{23}X_{32}X_{21}
-X_{23}X_{34}X_{43}X_{32}
+X_{34}X_{41}X_{14}X_{43}
-X_{41}X_{12}X_{21}X_{14}
\end{equation}
The central charge can be obtained from formula (\ref{anom}) once we take into account the specific field content of the theory that can be read from the quiver diagram or in table (\ref{L222table}). We obtain
\begin{equation}
\label{central222}
c_{r} = 3N^2 \eta_{\Sigma} \, \bigg( 4 \kappa+  \sum_{i=1}^{8}  (n_{\phi_i} -\kappa) (R_{\phi_i} -1)^2 \bigg)
\end{equation}

In order to match this expression with (\ref{c2stris}) we first observe that the PM's are related to the lattice points as follows
\begin{equation}
\begin{array}{c|c|c|c}
\text{PM} & \text{Lattice point} &\text{PM} & \text{Lattice point} \\
\hline
\pi_1     =  \{ X_{12}, X_{34}\}   & V_1 = (0,0,1) &
\pi_2     =  \{ X_{21}, X_{34} \}  & V_2 = (1,0,1) \\
\pi_3     =  \{ X_{12}, X_{43} \}  & V_2 = (1,0,1) &
\pi_4     =  \{ X_{21}, X_{43} \}  & V_3 = (2,0,1)\\ 
\pi_5     =  \{ X_{32}, X_{14} \}  & V_4 = (2,1,1) &
\pi_6     =  \{ X_{32}, X_{41} \}  & V_5 = (1,1,1) \\
\pi_7     =  \{ X_{23}, X_{14} \}  & V_5 = (1,1,1) &
\pi_8     =  \{ X_{23}, X_{41} \} & V_6 = (0,1,1) \\
\end{array}
\end{equation}
The two points on the perimeter, identified as $V_2$ and $V_5$, are degenerate since they correspond to two different PM's.
As proposed in sub--section \ref{central} we set the charges and the fluxes of one of the two PM's of each degenerate point to zero.
According to our prescription we set $\Delta_{\pi_3} =\Delta_{\pi_7}= 0$ and  $n_{\pi_3} = n_{\pi_7} = 0$. The other non--vanishing charges and fluxes are constrained by the relations 
\begin{equation}
\label{constr222}
\begin{array}{ccccccccccccc}
\Delta_{\pi_1}&+&\Delta_{\pi_2}&+&\Delta_{\pi_4}&+&\Delta_{\pi_5}&+&\Delta_{\pi_6}&+&\Delta_{\pi_8} &= &2
\\
n_{\pi_1}&+&n_{\pi_2}&+&n_{\pi_4}&+&n_{\pi_5}&+&n_{\pi_6}&+&n_{\pi_8} &=& 2 \kappa
\end{array}
\end{equation}
From here we can read the charges and the fluxes of every single field
\begin{equation}
\label{L222table}
\begin{array}{c|cc}
&\hspace{1cm}R_{\phi_i}&\hspace{1cm}n_{\phi_i}\\
\hline
\phi_1 = X_{12} &\hspace{1cm} \Delta_{\pi_{1}} & \hspace{1cm}n_{\pi_{1}}  \\
\phi_2 = X_{21} &\hspace{1cm}\Delta_{\pi_{4}}  + \Delta_{\pi_{2}}& \hspace{1cm}n_{\pi_{4}}+n_{\pi_{2}} \\
\phi_3 = X_{23} &\hspace{1cm}\Delta_{\pi_{8}} & \hspace{1cm}n_{\pi_{8}} \\
\phi_4 = X_{32} &\hspace{1cm}\Delta_{\pi_{5}}+\Delta_{\pi_{6}} &\hspace{1cm} n_{\pi_{5}} +n_{\pi_{6}} \\
\phi_5 = X_{34} &\hspace{1cm}\Delta_{\pi_{1}}+\Delta_{\pi_{2}}  & \hspace{1cm}n_{\pi_{1}}+n_{\pi_{2}} \\
\phi_6 = X_{43} &\hspace{1cm}\Delta_{\pi_{4}}& \hspace{1cm}n_{\pi_{4}}  \\
\phi_7 = X_{41} &\hspace{1cm}\Delta_{\pi_{6}} +\Delta_{\pi_{8}}& \hspace{1cm}n_{\pi_{6}} +n_{\pi_{8}} \\
\phi_8 = X_{14} &\hspace{1cm}\Delta_{\pi_{5}} & \hspace{1cm}n_{\pi_{5}} 
\end{array}
\end{equation}
This parameterization satisfies the constraints $\sum_{a \in W} R_{\phi_a}=2$
and $\sum_{a \in W} n_{\phi_a}=2 \kappa$.
 
By substituting this parametrization in (\ref{central222}) we can easily prove that it is equivalent to (\ref{c2stris}) once constraints (\ref{constr222})
are imposed.
%
%
%
\subsection{$L^{131}$}
\begin{figure}[H]
\begin{center}
  \includegraphics[width=15cm]{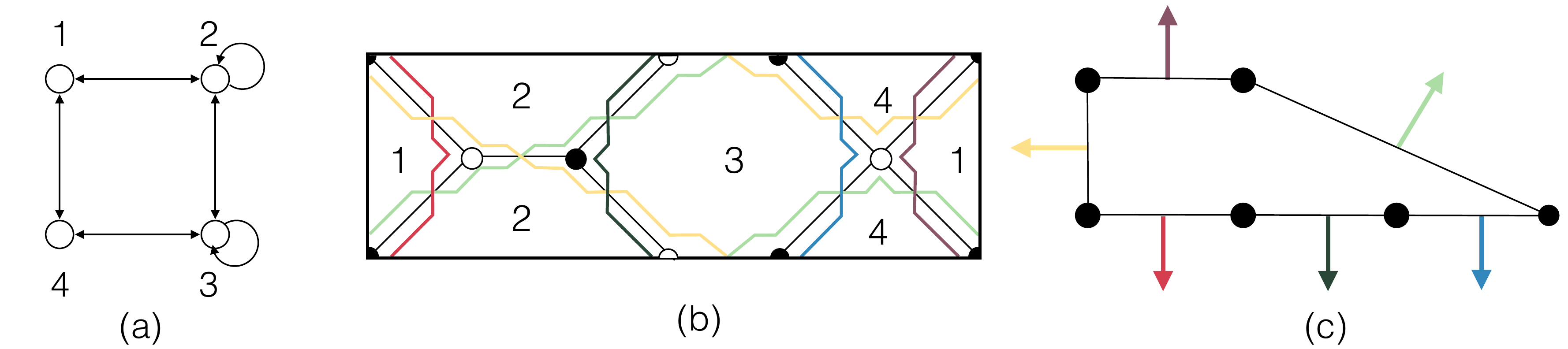}
  \caption{Dimer, zig-zag paths and toric diagram of L$^{131}$}
  \label{L131}
\end{center}
\end{figure}

%
%
%
As a second example, we consider the $L^{131}$ model associated to the quiver, dimer and toric diagram drawn
in Figure \ref{L131}. In this case the superpotential reads
\begin{eqnarray}
W &=&  X_{12} X_{21} X_{14} X_{41} - X_{12} X_{22} X_{21} + X_{32} X_{22}  X_{23}
\nonumber \\
&-&X_{23} X_{33} X_{32} + X_{43} X_{33} X_{34} - X_{14} X_{43} X_{34}
  X_{41}
\end{eqnarray}

Given the particular field content, the central charge computed from (\ref{anom}) reads  
\begin{equation}
\label{central131}
c_{r} = 3N^2 \eta_{\Sigma} \, \bigg( 4 \kappa+  \sum_{i=1}^{10}  (n_{\phi_i} -\kappa) (R_{\phi_i} -1)^2 \bigg)
\end{equation}
In this case the PM's are related to the lattice points as follows
\begin{equation}
\begin{array}{c|c|c|c}
\text{PM} & \text{Lattice point} &\text{PM} & \text{Lattice point} \\
\hline
\pi_1     =  \{X_{12}, X_{23}, X_{34}  \}& V_1 = (0,0,1) &
\pi_2     =  \{ X_{21}, X_{23}, X_{34}\}  & V_2 = (1,0,1) \\
\pi_3     =  \{ X_{12} ,X_{32}, X_{34}\}  & V_2 = (1,0,1) &
\pi_4     =  \{ X_{12}, X_{23}, X_{43}\}  & V_2 = (1,0,1)\\ 
\pi_5     = \{  X_{21} ,X_{32}, X_{34}\}  & V_3 = (2,0,1) &
\pi_6     =  \{ X_{21} ,X_{23}, X_{43}\} & V_3 = (2,0,1) \\
\pi_7     =   \{ X_{12} ,X_{32}, X_{43}\} & V_3 = (2,0,1) &
\pi_8     =   \{ X_{21}, X_{32}, X_{43}\} & V_4 = (3,0,1) \\
\pi_9     =   \{ X_{14}, X_{22} ,X_{33}\} & V_5 = (1,1,1) &
\pi_{10} =  \{  X_{41}, X_{22} ,X_{33}\} & V_6 = (0,1,1)
\end{array}
\end{equation}
There are still two perimeter points, this time with degeneracy three. According to our prescription described in sub--section \ref{central}  we set $\Delta_{\pi_2} =\Delta_{\pi_3}=\Delta_{\pi_5} =\Delta_{\pi_6}= 0$ and 
correspondingly 
$n_{\pi_2} = n_{\pi_3} = n_{\pi_5} =n_{\pi_6} = 0$. The remaining  charges and
fluxes satisfy
\begin{equation}
\label{constr131}
\begin{array}{ccccccccccccc}
\Delta_{\pi_1}&+&\Delta_{\pi_4}&+&\Delta_{\pi_7}&+&\Delta_{\pi_8}&+&\Delta_{\pi_9}&+&\Delta_{\pi_{10}} &=& 2\\
n_{\pi_1}&+&n_{\pi_4}&+&n_{\pi_7}&+&n_{\pi_6}&+&n_{\pi_9}&+&n_{\pi_{10}} &=& 2 \kappa
\end{array}
\end{equation}
The R-charges $R_{\phi_i}$ and the fluxes $n_{\phi_i}$ of the fields can be expressed in terms 
of the charges $\Delta_{\pi_I}$ and the fluxes $n_{\pi_I}$ of the PM's as
\begin{equation}
  \begin{array}{c|cccc}
    & \qquad & R_{\phi_i} & \qquad & n_{\phi_i}\\
    \hline
    \phi_1 = X_{12} &  & \Delta_{\pi_1} + \Delta_{\pi_4} + \Delta_{\pi_7} &  &
    n_{\pi_1} + n_{\pi_4} + n_{\pi_7}\\
    \phi_2 = X_{21} &  & \Delta_{\pi_8} &  & n_{\pi_8}\\
    \phi_3 = X_{22} &  & \Delta_{\pi_9} + \Delta_{\pi_{10}} &  & n_{\pi_9} +
    n_{\pi_{10}}\\
    \phi_4 = X_{23} &  & \Delta_{\pi_1} + \Delta_{\pi_4} &  & n_{\pi_1} +
    n_{\pi_4}\\
    \phi_5 = X_{32} &  & \Delta_{\pi_7} + \Delta_{\pi_8} &  & n_{\pi_7} +
    n_{\pi_8}\\
    \phi_6 = X_{33} &  & \Delta_{\pi_9} + \Delta_{\pi_{10}} &  & n_{\pi_9} +
    n_{\pi_{10}}\\
    \phi_7 = X_{34} &  & \Delta_{\pi_1} &  & n_{\pi_1}\\
    \phi_8 = X_{43} &  & \Delta_{\pi_4} + \Delta_{\pi_7} + \Delta_{\pi_8} &  &
    n_{\pi_4} + n_{\pi_7} + n_{\pi_8}\\
    \phi_9 = X_{41} &  & \Delta_{\pi_{10}} &  & n_{\pi_{10}}\\
    \phi_{10} = X_{14} &  & \Delta_{\pi_9} &  & n_{\pi_9}
  \end{array}
\end{equation}

This parameterization satisfies the constraints $\sum_{a \in W} R_{\phi_a}=2$
and $\sum_{a \in W} n_{\phi_a}=2 \kappa$.
 
 It is now easy to substitute this parameterization in (\ref{central131}) and check that the resulting expression is equivalent to (\ref{c2stris}) once we take into account constraints (\ref{constr131}).
%
%
%
%
%
%
 \section{Mixing of the baryonic symmetries}
\label{FTside}
%
%
%
%
%
In this section, by studying the twisted compactification of some of the 4d $\mathcal{N}=1$ toric quiver gauge theories
discussed above, we provide further evidence that both flavor and baryonic symmetries mix with the R--current 
at the 2d fixed point. We compute the central charge with the formalism reviewed 
in section \ref{sectw} showing its positivity for many choices of the curvature and the fluxes.
%
%
%
%
%
\subsection*{dP$_2$}
%
%
%
We begin by identifying the global currents of the dP$_2$ model.
There are a UV R-current $R_0$, two flavor currents $F_{1,2}$ and two non--anomalous baryonic currents $B_{1,2}$. Having the model five gauge groups, to begin with we  
have five classically conserved baryonic currents, associated to the decoupling of the gauge abelian factors 
$U(1)_i \subset U(N)_i$. 
As usual one of such currents is redundant. Among the other global baryonic $U(1)$'s some of the combinations
can be anomalous at quantum level. 
After the identification of the two non--anomalous baryonic currents the charges of the fields respect to all the global currents are
\begin{equation}
 \begin{array}{c|ccccccccccc}
    &Y_{51} & X_{51} & X_{23} & X_{35} & X_{41} & X_{34} & X_{13} & X_{24} &
    X_{45} & X_{12} & Y_{12} 
    \\
\hline

R_0& 2 & 0 & 2 & 0 & 2 & 0 & 0 & 0 & 0 & 0 & 0 \\
 F_1&-2 & 1 & -3 & 1 & -2 & 1 & 1 & 1 & 0 & 1 & 1 \\
 F_2&1 & 1 & -1 & 1 & 0 & 2 & -2 & 1 & -3 & 1 & -1 \\
 B_1&-1 & -1 & -1 & 1 & 0 & 0 & 0 & -1 & 1 & 1 & 1 \\
 B_2&1 & 1 & -1 & 0 & -1 & 2 & -1 & 1 & -2 & 0 & 0 \\
 \end{array}
\end{equation}
Any linear combination 
\begin{eqnarray}
  R_{\tmop{trial}} & = & R_0 + \epsilon_1 T_{F_1}+ \epsilon_2 T_{F_2} + \eta_1 T_{B_1}+ \eta_2 T_{B_2} 
\end{eqnarray}
is still an  R--current. Such an ambiguity is fixed by maximizing the central charge  
with respect to the mixing parameters $ \epsilon_i$ and $\eta_i$ \cite{Intriligator:2003jj}
\begin{eqnarray}
  \frac{\partial a}{\partial \epsilon_i} & = & \frac{3}{32} [9 \tmop{Tr}
  (R_{\tmop{trial}}^2 F_i) - \tmop{Tr} (F_i)] = 0  \label{eq:extrF}\\
  \frac{\partial a}{\partial \eta_i} & = & \frac{3}{32} [9 \tmop{Tr}
  (R_{\tmop{trial}}^2 B_i) - \tmop{Tr} (B_i)] = 0.  \label{eq:extrB}
\end{eqnarray}
By using the relations $\tmop{Tr} (B_i B_j B_k) = 0 = \tmop{Tr} (B_i)$ 
equations (\ref{eq:extrB}) reduce to a linear system in the $\eta_i$ variables. 
Substituting the solution back into (\ref{eq:extrF}) we are left with two free mixing parameters.
Therefore,  one can always  linearly combine the global symmetries in
such a way that at the fixed point $\eta_1= \eta_2 =0$. This signals the fact that the baryonic symmetries do 
not mix with the 4d exact R--current.

Solving the rest of equations we obtain
\begin{equation}
\epsilon_1 = \frac{1}{8} \left(\sqrt{33}-1\right), 
\quad 
\epsilon_2 = \frac{1}{16} \left(3 \sqrt{33}-19\right)
\end{equation}

We can proceed by twisting the theory on $\Sigma$.
The partial topological twist is performed along the generator
\begin{equation}
T = \kappa T_R + b_1 T_{F_1} + b_2 T_{F_2} + b_3 T_{B_1} + b_4 T_{B_2} 
\end{equation}
The central charge $c_r $  of the 2d theory can be obtained from (\ref{eq:krrex}).
The final formulae are quite involved and we report some non trivial cases in appendix \ref{mixing}.
In this case we observe that the mixing parameters  are non-vanishing for generic choices of the 
constant curvature  and
of the fluxes $b_{\bf I}$. This signals the fact that the baryonic symmetries mix with the 2d exact R--current.

\begin{figure}[H]
\begin{center}
\begin{tabular}{ccc}
  \includegraphics[width=3.6cm]{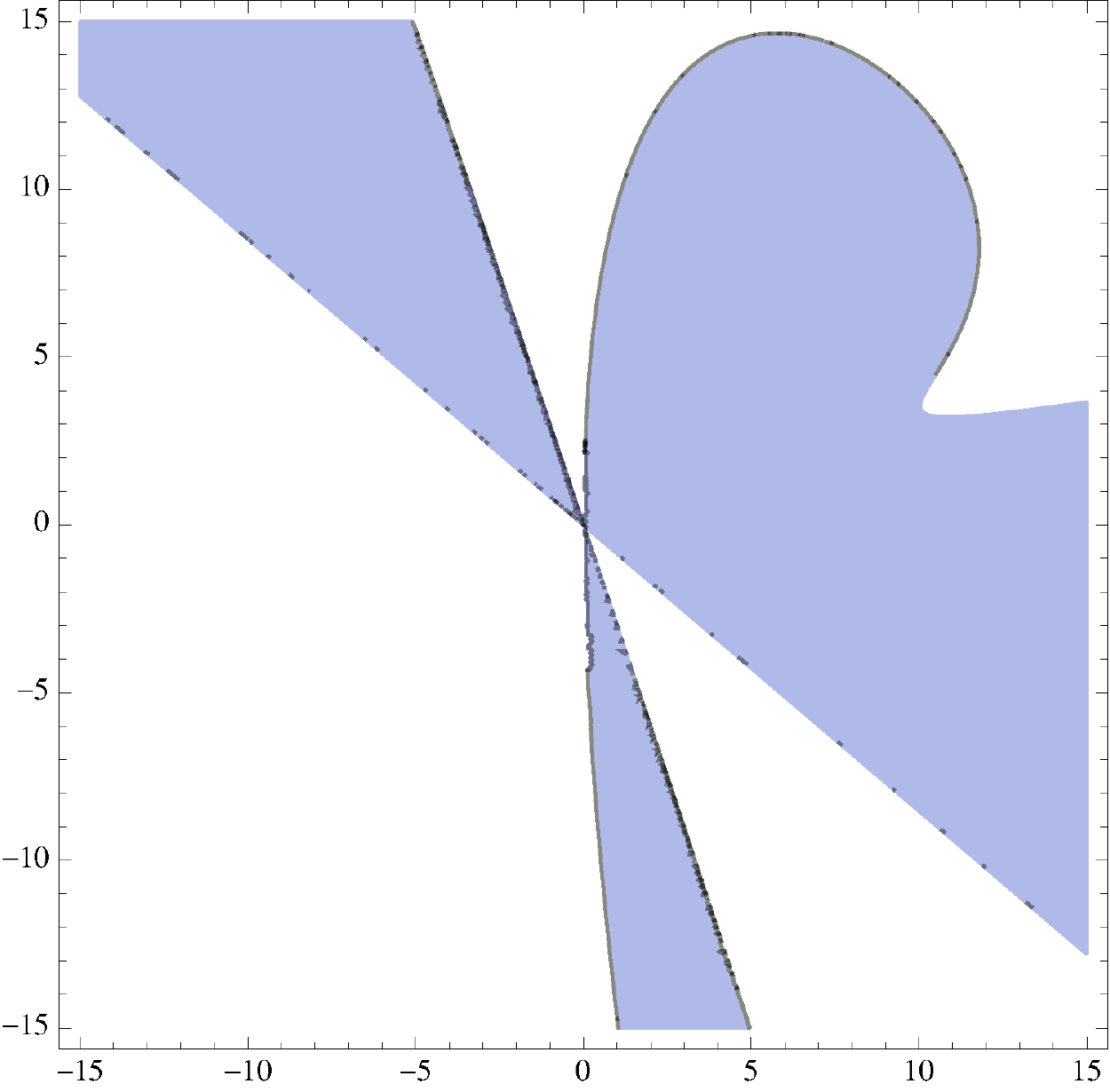}
&
  \includegraphics[width=3.6cm]{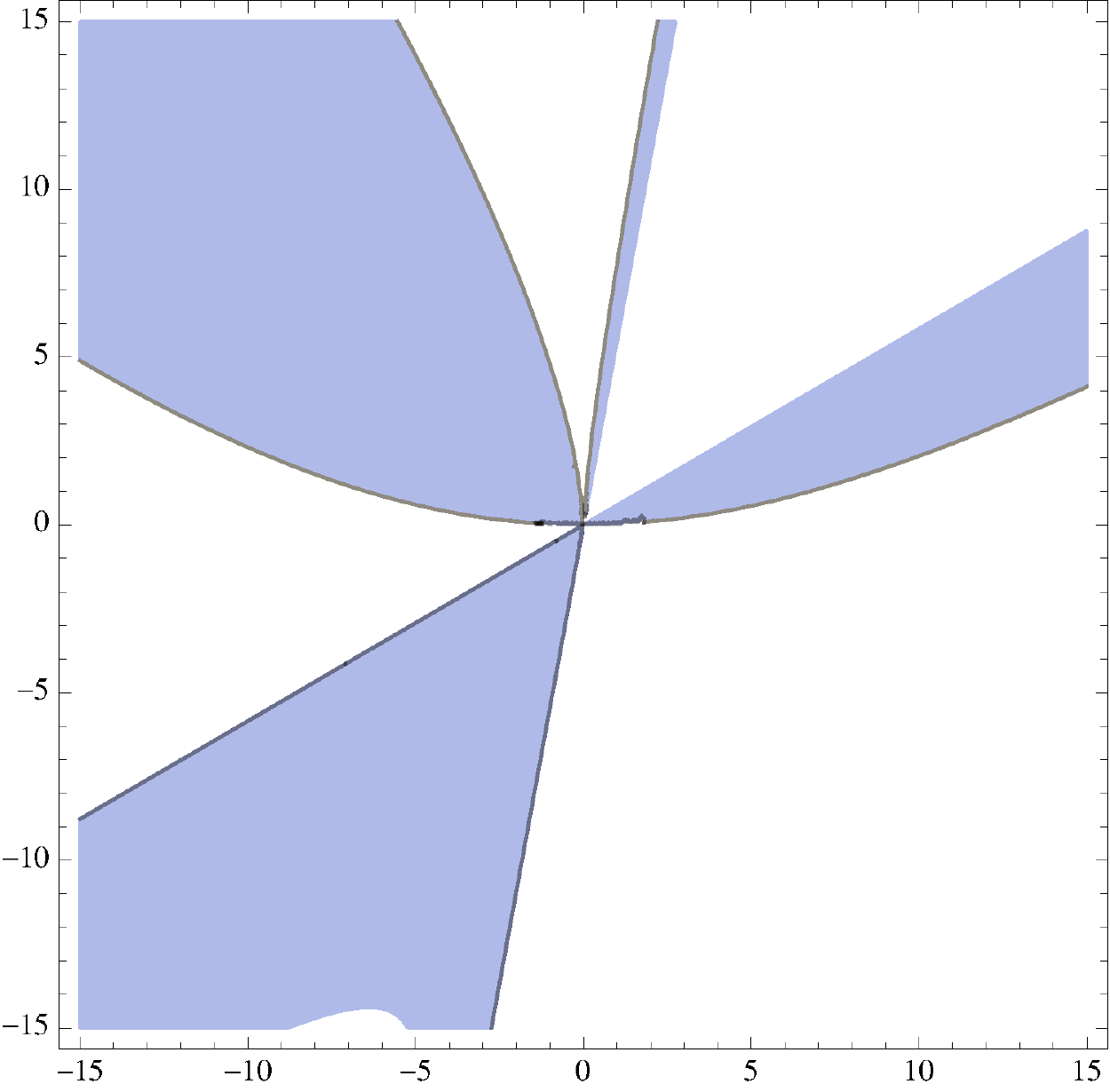}
&
  \includegraphics[width=3.6cm]{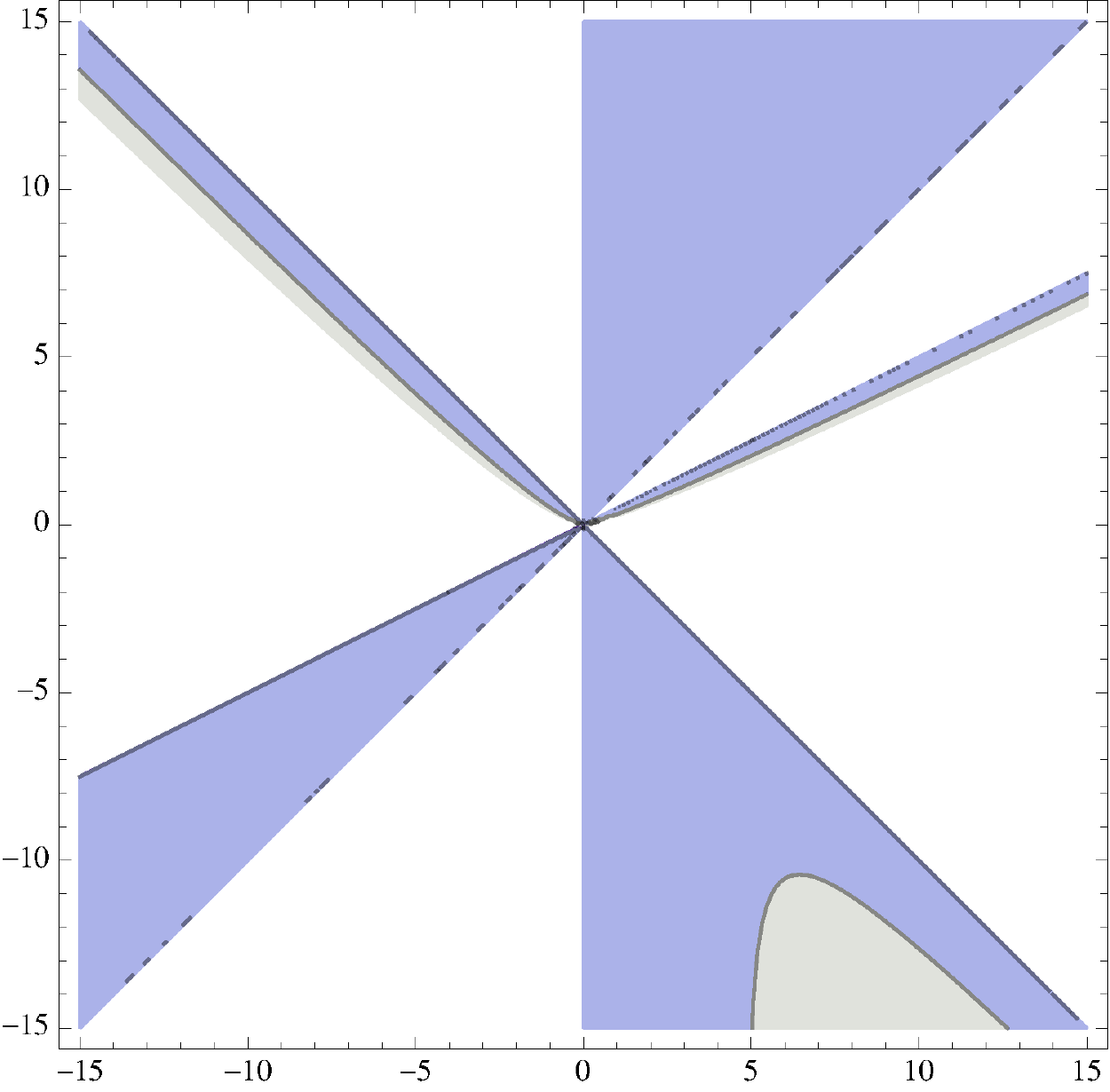}\\
\end{tabular}
  \caption{Central charge of dP$_2$ on $\Sigma=\mathbb{T}^2$  for different values of the integer fluxes.
  We plot the regions of fluxes $b_i$ in which the central charge assumes a positive value.
 In the first case we have fixed $b_1=x$, $b_2=y$ and $b_3=b_4=0$.
  In the second case we have fixed $b_3=x$ and $b_4=y$ and $b_1=b_2=0$.
   In the third case we have fixed $b_2=x$, $b_3=y$ and $b_1=b_4=0$.}
  \label{dP2T}
\end{center}
\end{figure}

\begin{figure}[H]
\begin{center}
\begin{tabular}{ccc}
  \includegraphics[width=3.6cm]{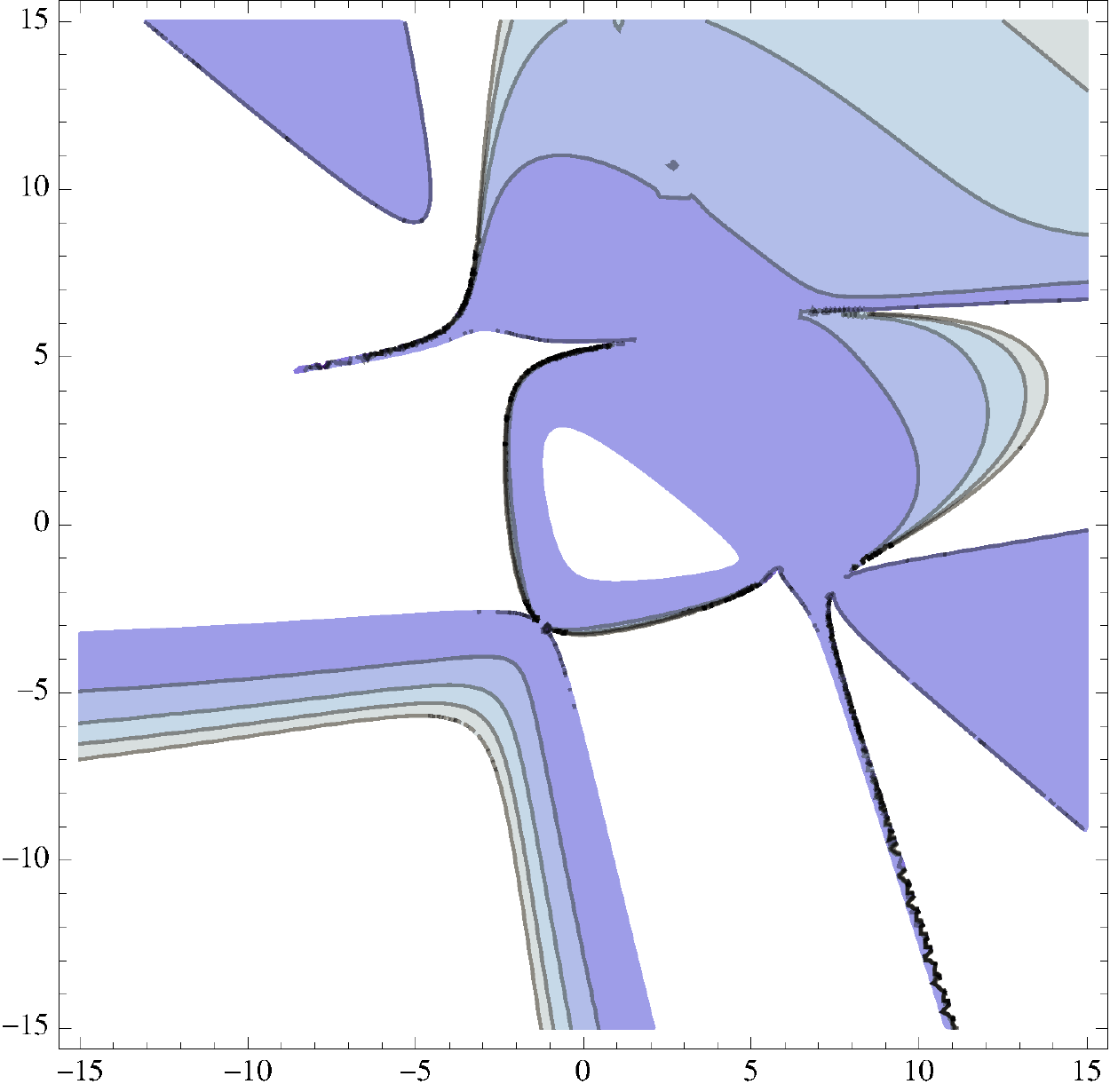}
&
  \includegraphics[width=3.6cm]{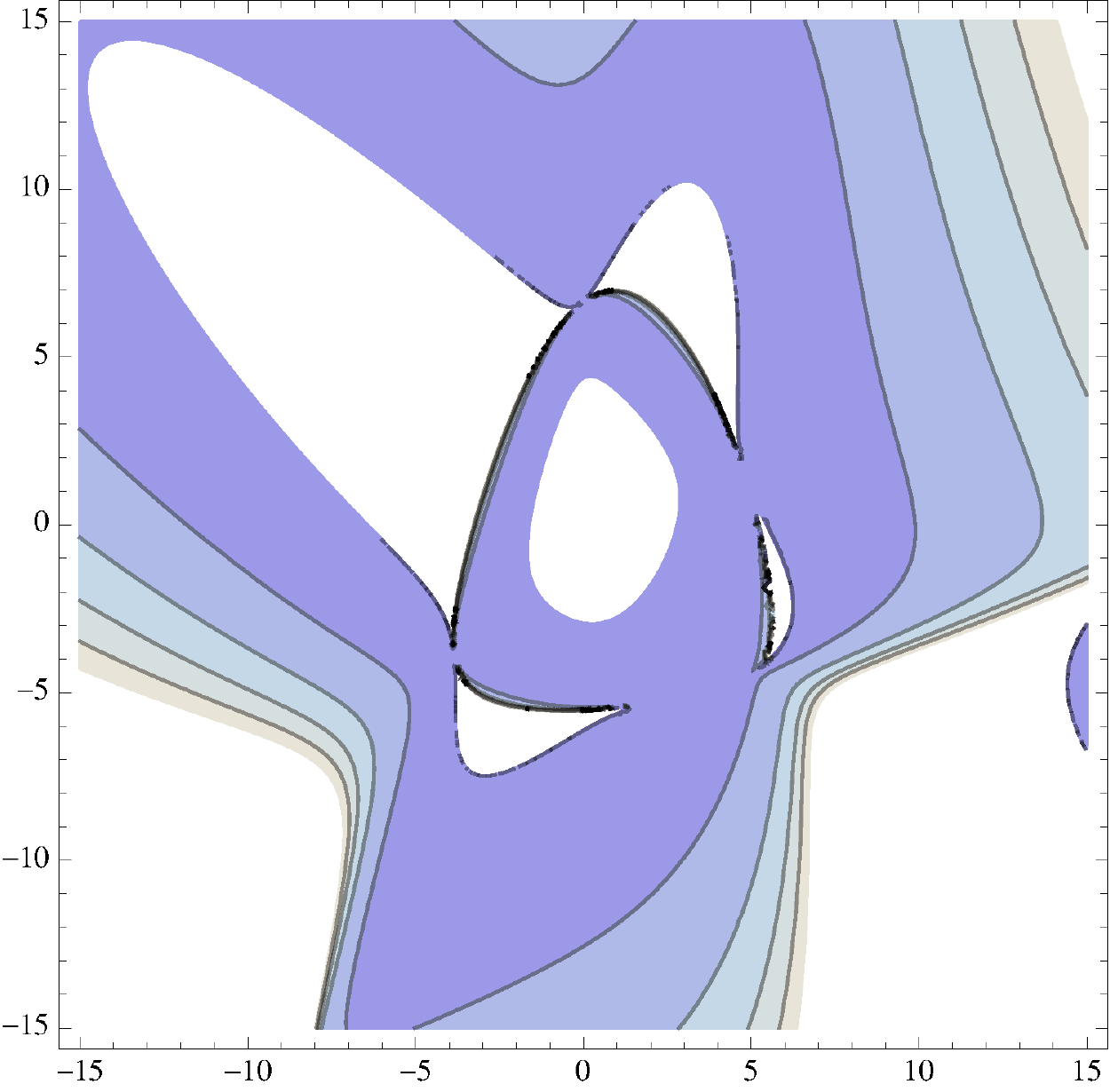}
&
  \includegraphics[width=3.6cm]{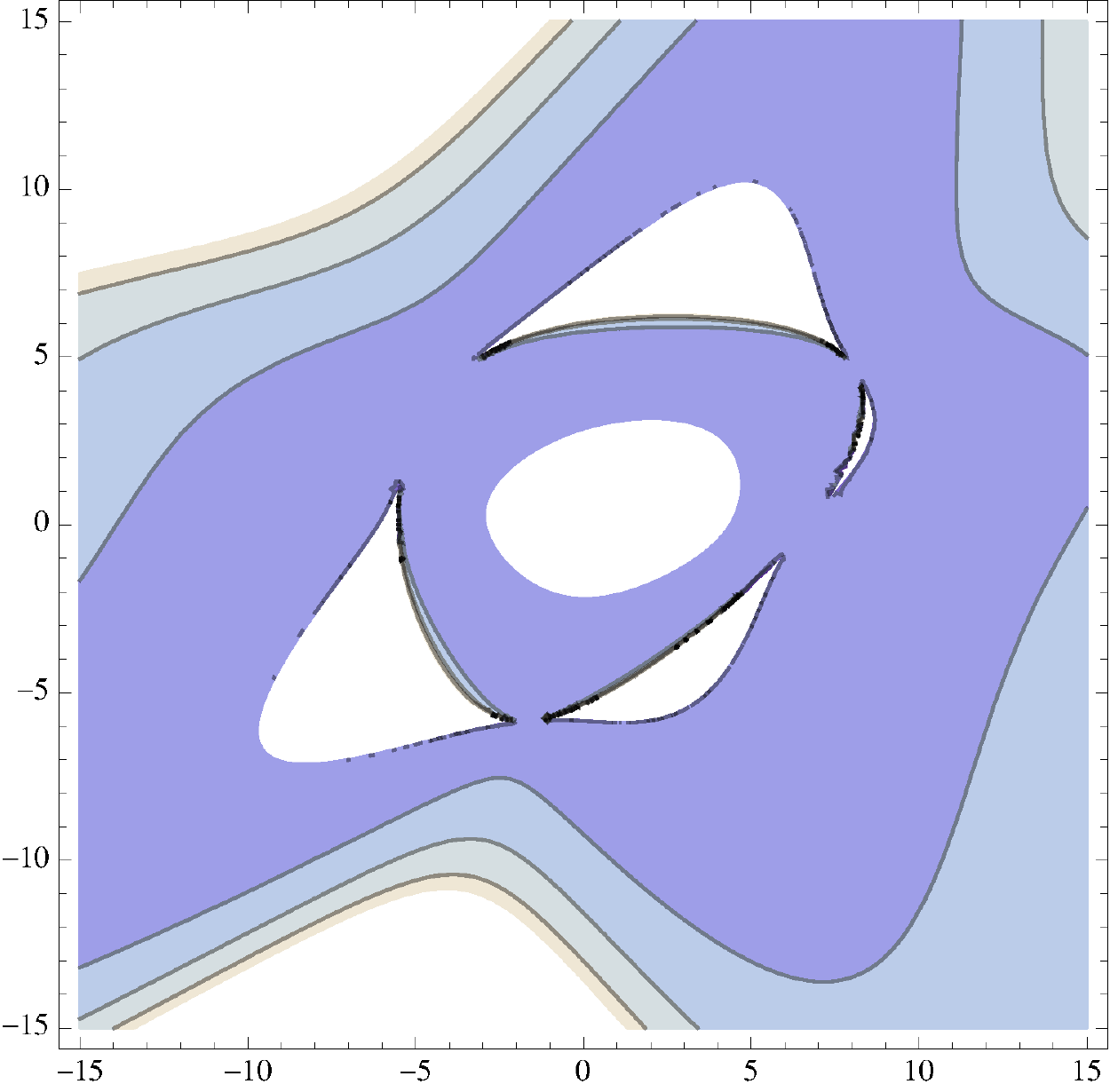}\\
\end{tabular}
  \caption{Central charge of dP$_2$ on $\Sigma=\mathbb{S}^2$  for different values of the integer fluxes.
  We plot the regions of fluxes $b_i$ in which the central charge assumes a positive value.
 In the first case we have fixed $b_1=x$, $b_2=y$ and $b_3=b_4=0$.
  In the second case we have fixed $b_3=x$ and $b_4=y$ and $b_1=b_2=0$.
   In the third case we have fixed $b_2=x$, $b_3=y$ and $b_1=b_4=0$.}
  \label{dP2S}
\end{center}
\end{figure}

\begin{figure}[H]
\begin{center}
\begin{tabular}{ccc}
  \includegraphics[width=3.6cm]{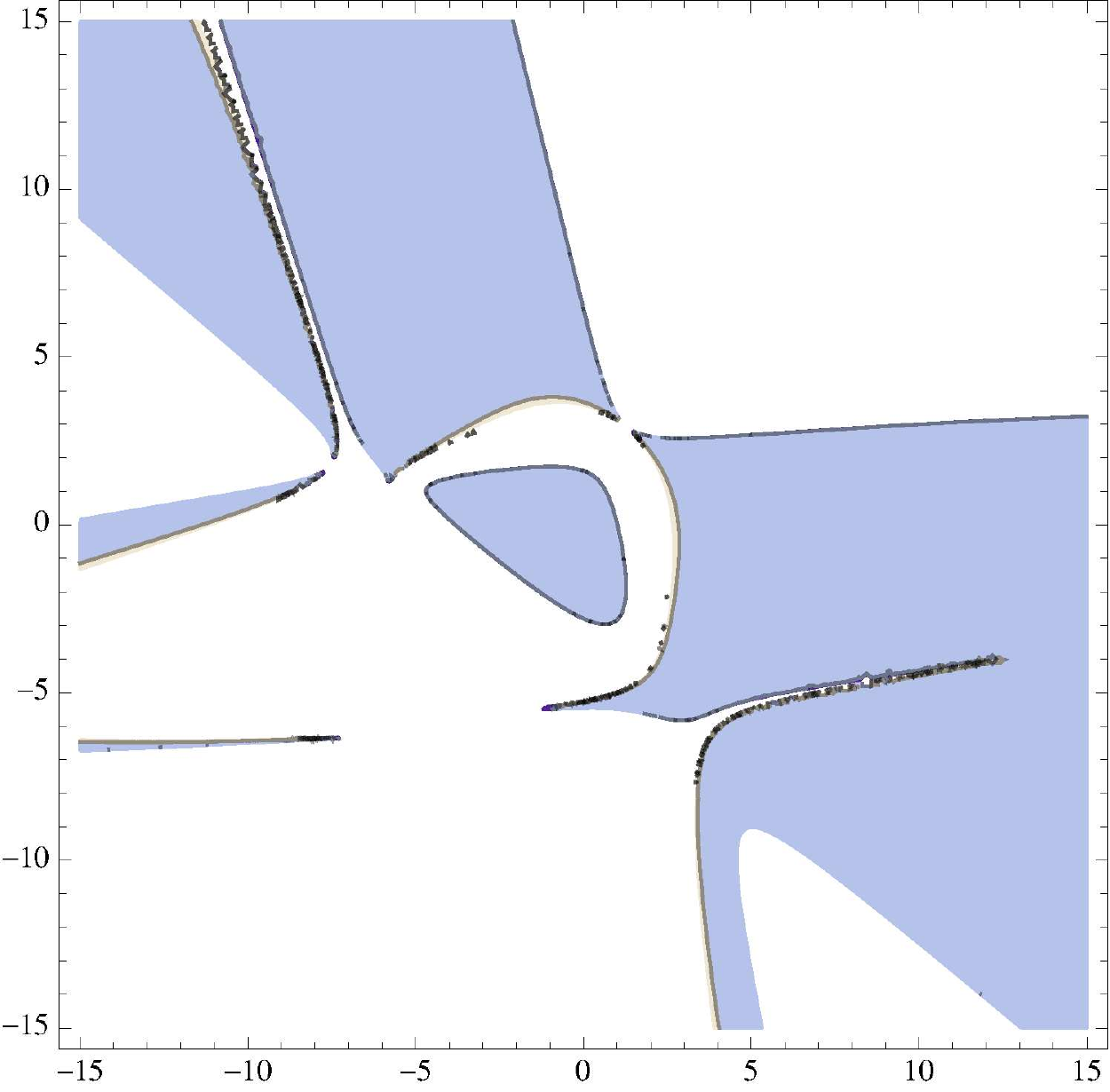}
&
  \includegraphics[width=3.6cm]{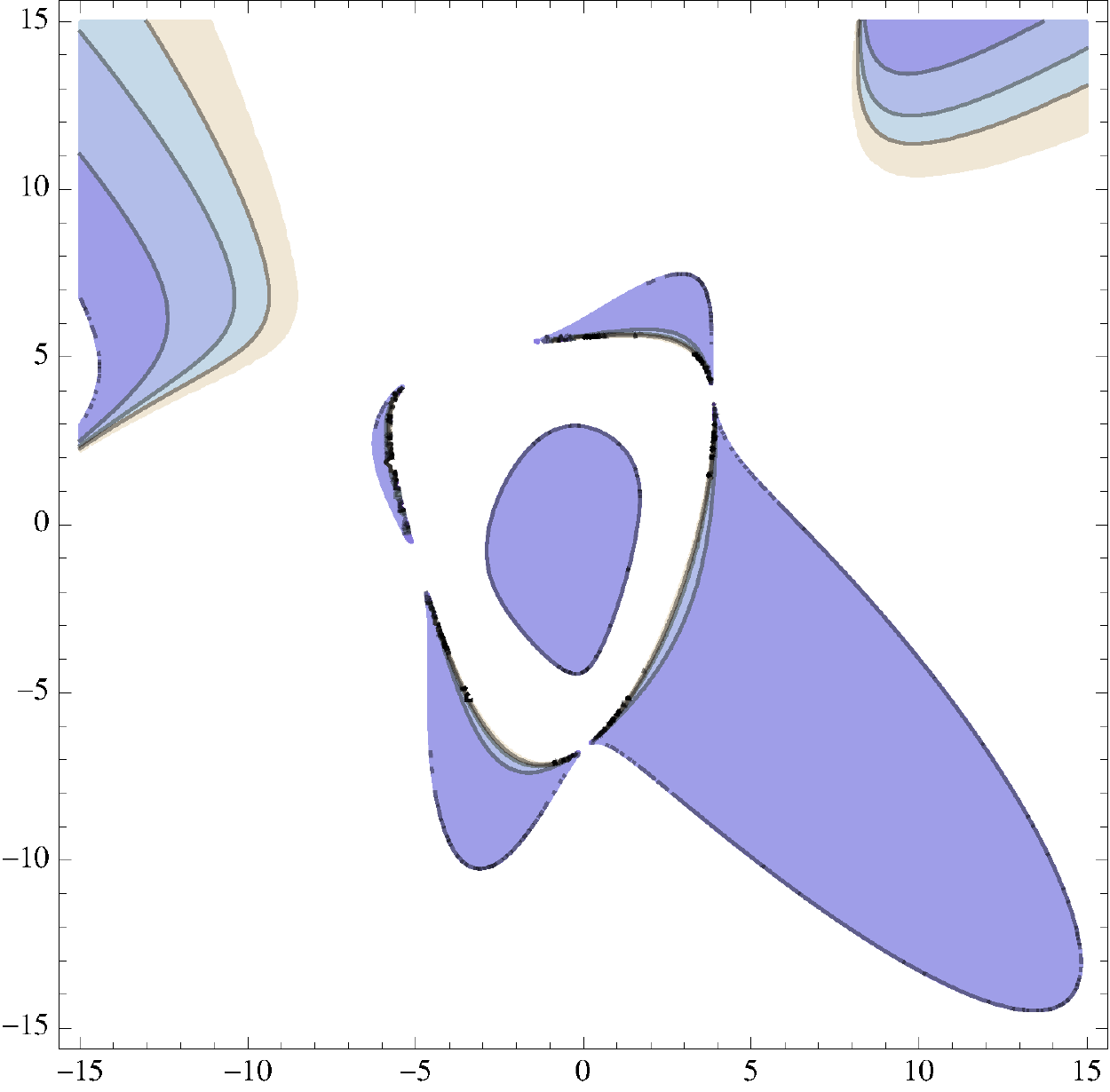}
&
  \includegraphics[width=3.6cm]{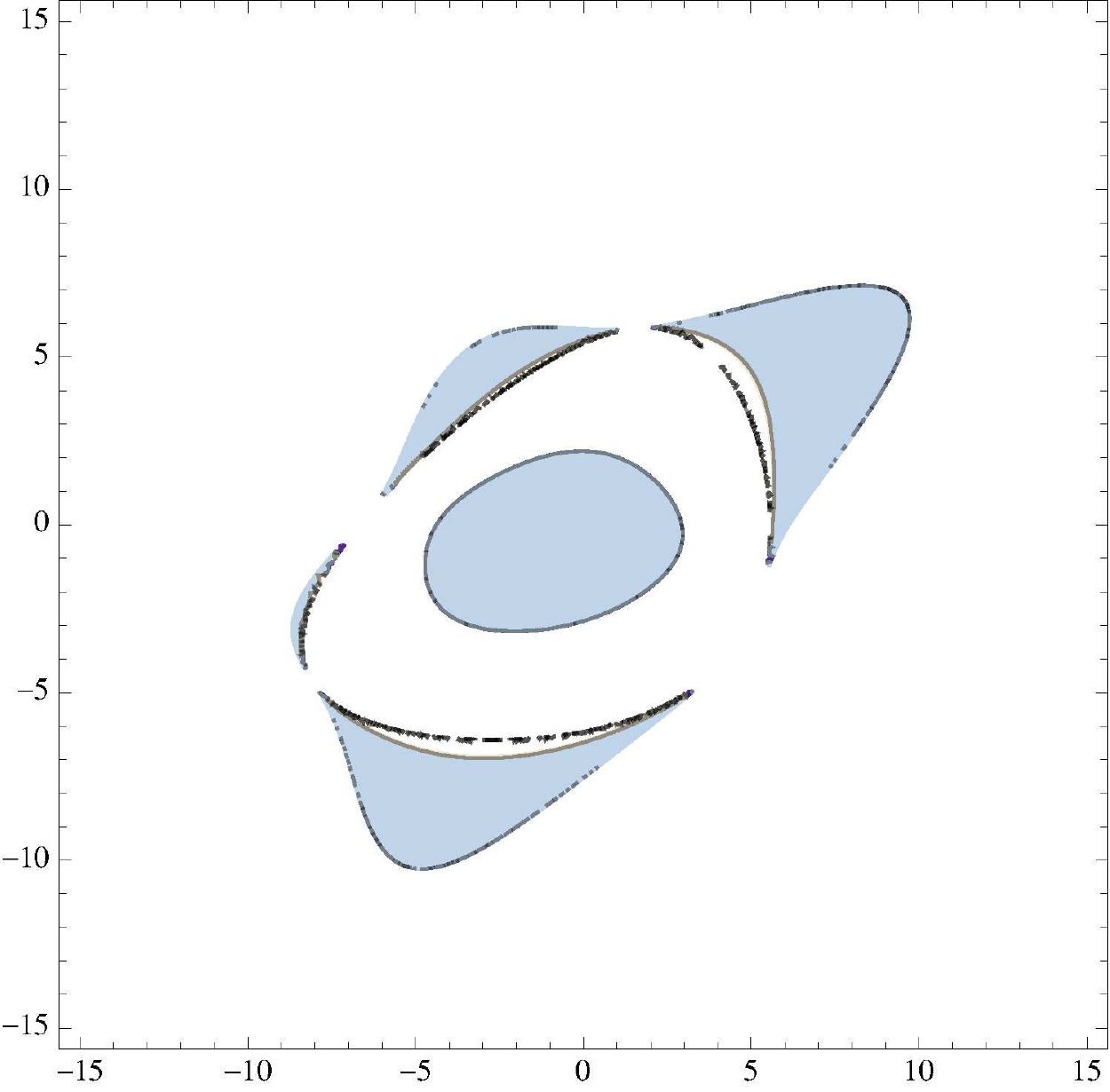}\\
\end{tabular}
  \caption{Central charge of dP$_2$ on $\Sigma=\mathbb{H}^2$  for different values of the integer fluxes.
  We plot the regions of fluxes $b_i$ in which the central charge assumes a positive value.
 In the first case we have fixed $b_1=x$, $b_2=y$ and $b_3=b_4=0$.
  In the second case we have fixed $b_3=x$ and $b_4=y$ and $b_1=b_2=0$.
   In the third case we have fixed $b_2=x$, $b_3=y$ and $b_1=b_4=0$.}
  \label{dP2H}
\end{center}
\end{figure}

We conclude by showing in 
Figure \ref{dP2T}, \ref{dP2S} and   \ref{dP2H}  the central charge for different values of the discrete fluxes
for dP$_2$ compactified on $\Sigma=\mathbb{T}^2$, $\Sigma=\mathbb{S}^2$ and $\Sigma=\mathbb{H}^2$, respectively.
%
%
%
\subsection*{dP$_{3}$}
%
%
%
We now consider the quiver gauge theory living on a stack of D3 branes probing the
tip of the complex cone over dP$_3$. The global currents of the model
are a UV R-current $R_0$, two flavor currents $F_{1,2}$ and three non-anomalous baryonic currents $B_{1,2,3}$.

We can identify the baryonic currents as follows. 
The model has six gauge groups and classically there are six conserved baryonic currents. Two of them are anomalous and one
is redundant. One is left with three non--anomalous baryonic symmetries.
We can choose the charges of the fields with respect of the global symmetries as
\begin{equation}
   \begin{array}{c|cccccccccccc}
    & X_{12} & X_{13} & X_{23} & X_{24} & X_{34} & X_{35} & X_{45} & X_{46} &
    X_{56} & X_{51} & X_{61} & X_{62}\\
    \hline
 R_0 & 2 & 0 & 0 & 0 & 0 & 2 & 0 & 2 & 0 & 0 & 0 & 0 \\
 F_1 & -1 & 1 & 0 & -1 & 1 & -2 & 1 & -1 & 0 & 1 & -1 & 2 \\
 F_2 & 3 & 0 & -1 & -4 & 1 & 0 & 1 & 2 & -1 & 0 & -3 & 2 \\
 B_1 & 2 & 1 & -1 & -1 & 0 & 0 & 0 & 1 & 1 & -1 & -2 & 0 \\
 B_2 & 1 & 0 & -1 & -1 & 0 & 1 & 1 & 0 & -1 & -1 & 0 & 1 \\
 B_3 & -1 & -1 & 0 & 1 & 1 & 0 & -1 & -1 & 0 & 1 & 1 & 0 \\
   \end{array},
   \nonumber
\end{equation}
Any linear combination 
\begin{eqnarray}
\label{trialdP3}
  R_{\tmop{trial}} & = & R_0 + \epsilon_1 T_{F_1}+ \epsilon_2 T_{F_2} + \eta_1 T_{B_1}+ \eta_2 T_{B_2} + \eta_3 T_{B_3} 
\end{eqnarray}
is still an  R-current.
The mixing coefficients in (\ref{trialdP3}) are fixed by maximizing the central charge, and in this case we find
\begin{equation}
\epsilon_1= \frac{2}{3}, \quad \epsilon_2= -\frac{1}{3},\quad \eta_i=0
\end{equation}
Again we chose a parameterization such that at the superconformal fixed point
the contribution of the baryonic symmetries vanishes.

The partial topological twist on $\Sigma$ is performed along the generator
\begin{equation}
T = \kappa T_R + b_1 T_{F_1} + b_2 T_{F_2} + b_3 T_{B_1} + b_4 T_{B_2} + b_5 T_{B_3} 
\end{equation}
The central charge $c_r $ is obtained from (\ref{eq:krrex}).
The final expressions are too complicated and we do not learn much in writing them explicitly. The main point is that,
as in the case of dP$_2$,  the mixing parameters are non--vanishing (see appendix \ref{mixing} for some
examples)
for generic choices of the curvature and the $b_{\bf I}$ fluxes.
 This signals the fact that the baryonic symmetries mix with the 2d exact R--current.

\begin{figure}[H]
\begin{center}
\begin{tabular}{ccc}
  \includegraphics[width=3.6cm]{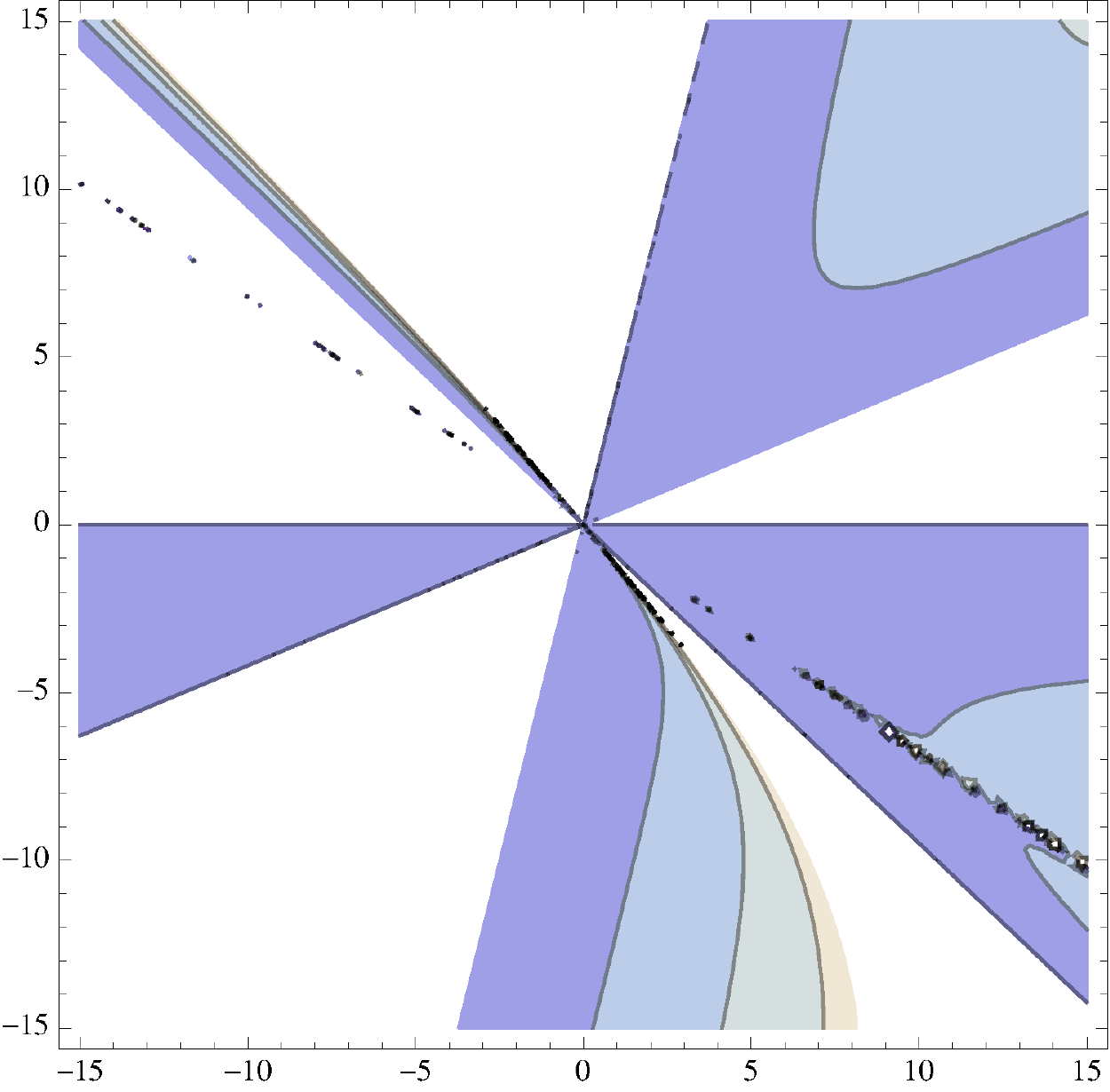}
&
  \includegraphics[width=3.6cm]{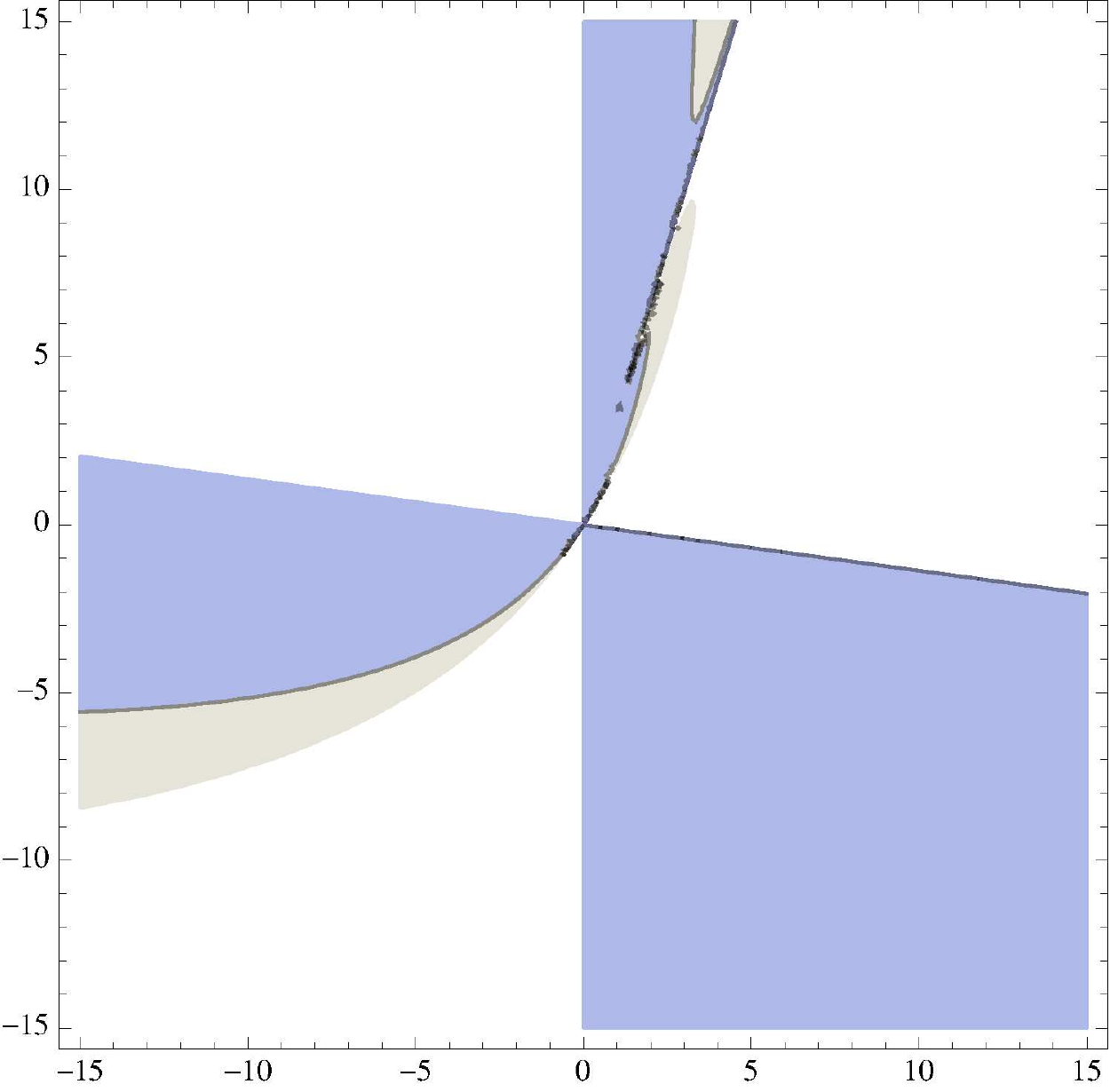}
&
  \includegraphics[width=3.6cm]{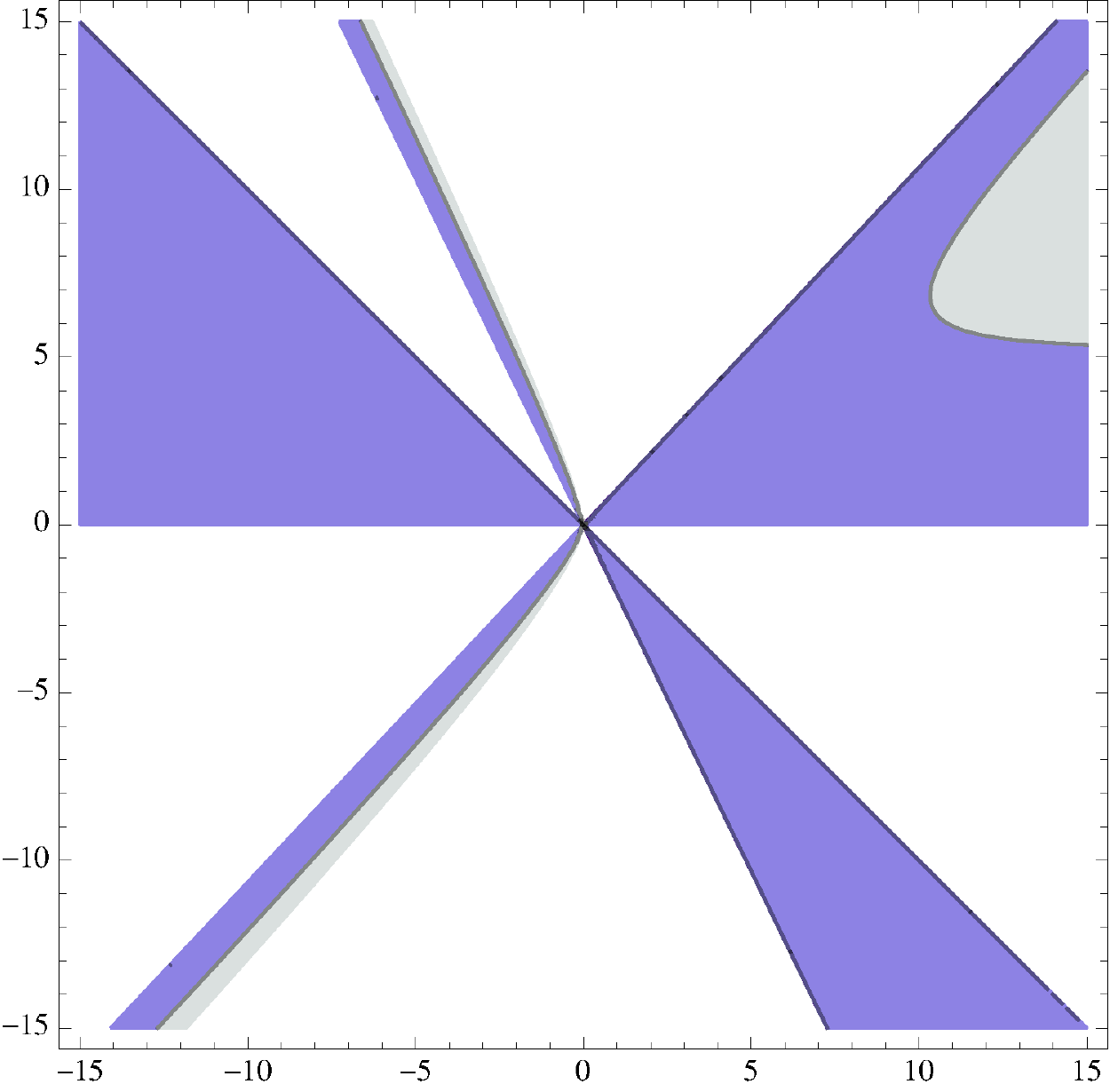}\\
\begin{tabular}{c}
$b_1=x$, $b_2=y$ \\
$b_3=b_4=b_5=0$
\end{tabular}
&
\begin{tabular}{c}
$b_3=x$, $b_4=y$ \\ 
$b_1=b_2=b_5=0$ 
\end{tabular}
&
\begin{tabular}{c}
$b_2=x$, $b_3=y$ \\
 $b_1=b_4=b_5=0$
\end{tabular}
\\
    \end{tabular}
  \caption{Central charge of dP$_3$ on $\Sigma=\mathbb{T}^2$  for different values of the integer fluxes.
  We plot the regions of fluxes $b_i$ in which the central charge assumes a positive value.
  The variable $x$ is plotted on the horizontal axis while $y$ on the vertical one.}
  \label{dP3T}
\end{center}
\end{figure}

\begin{figure}[H]
\begin{center}
\begin{tabular}{ccc}
  \includegraphics[width=3.6cm]{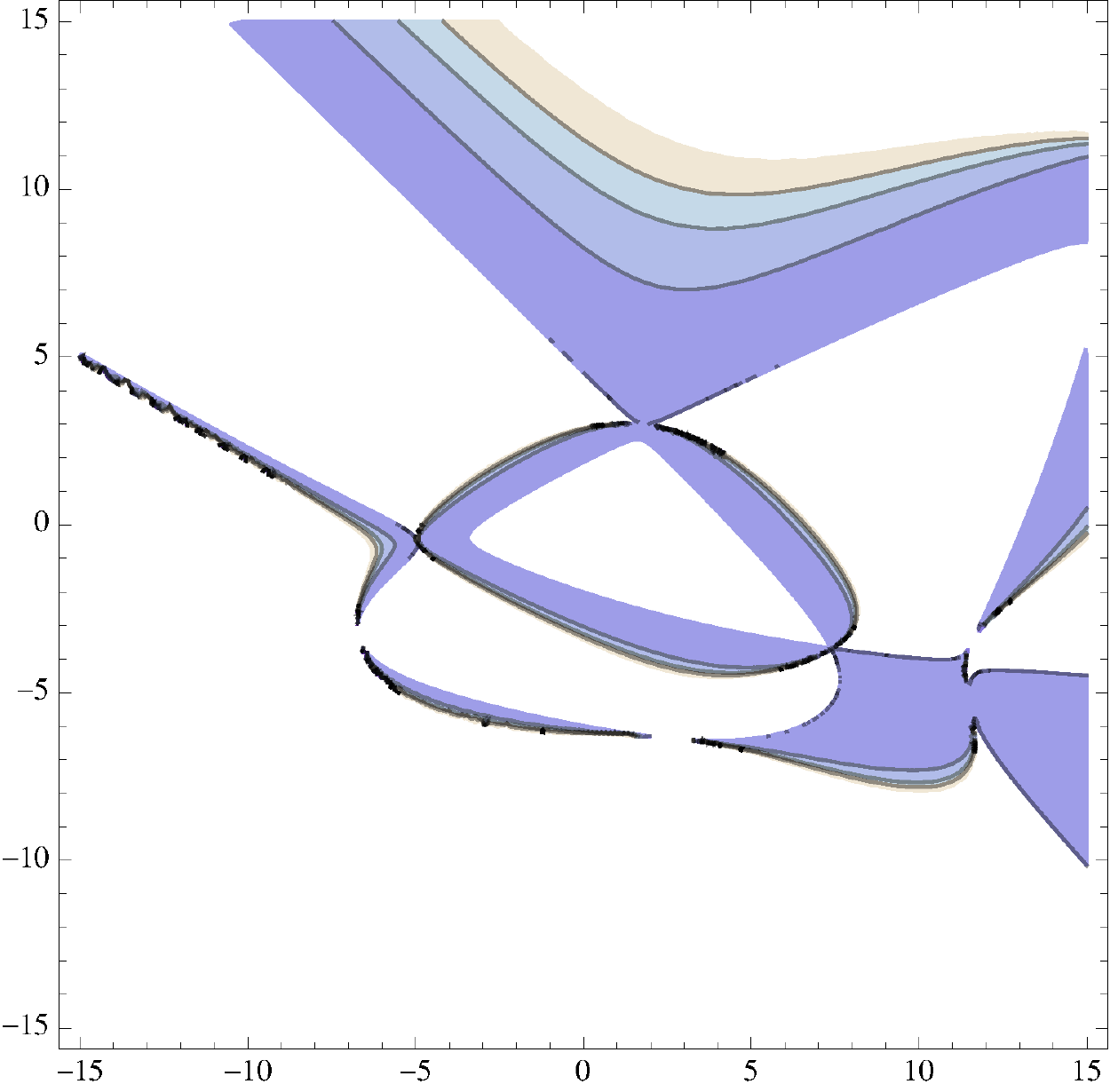}
&
  \includegraphics[width=3.6cm]{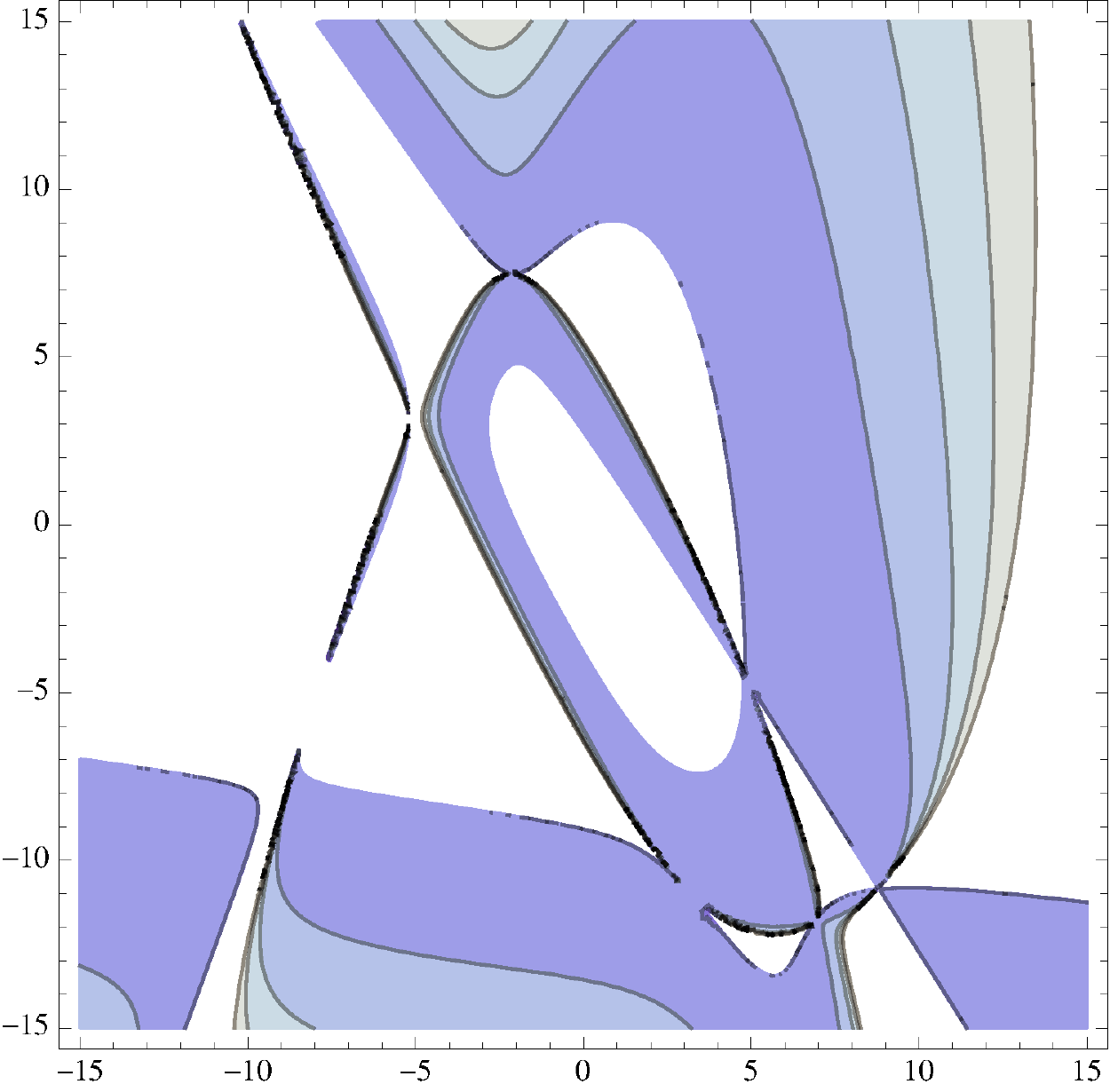}
&
  \includegraphics[width=3.6cm]{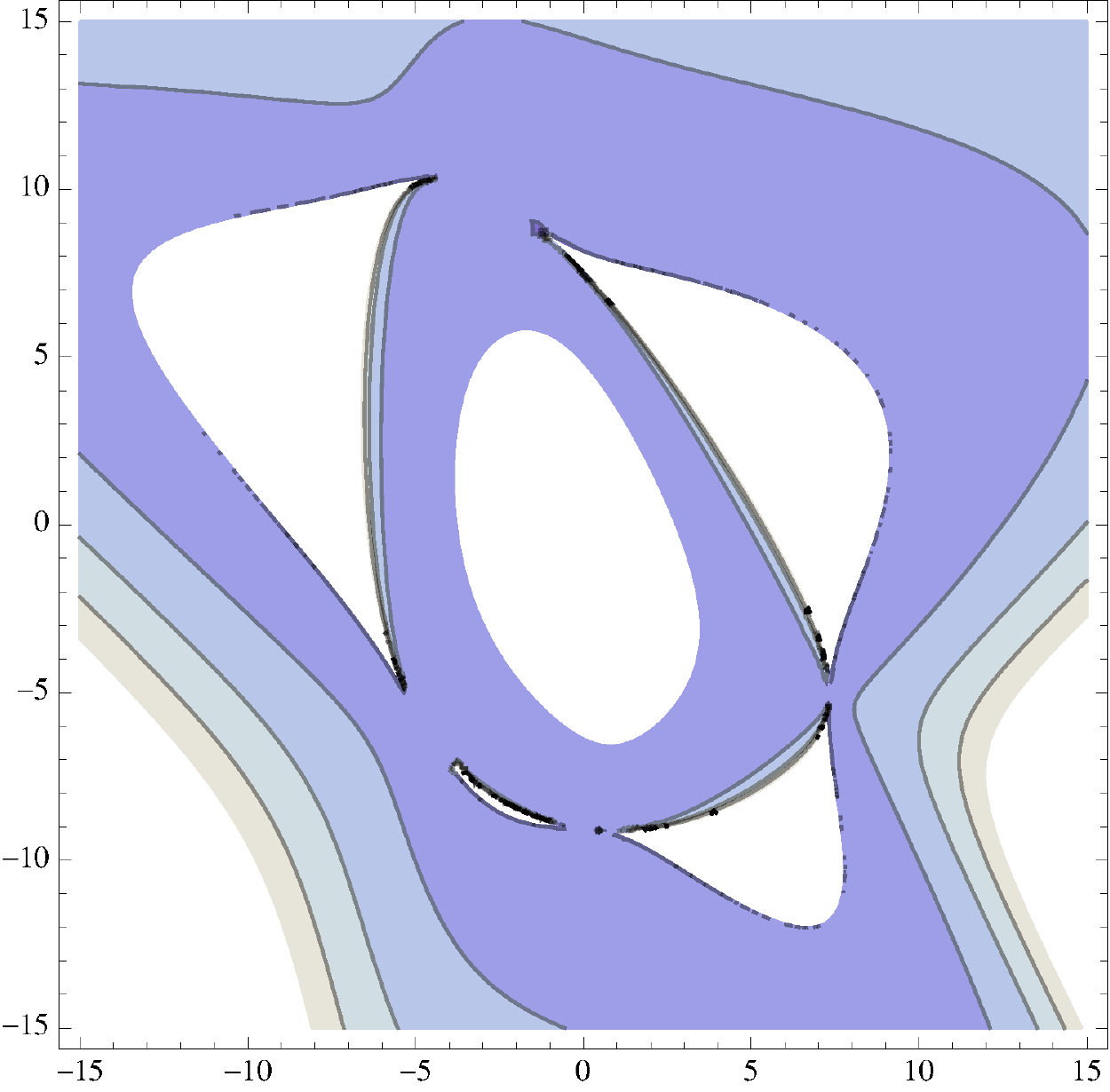}\\
\end{tabular}
  \caption{Central charge of dP$_3$ on $\Sigma=\mathbb{S}^2$  for different values of the integer fluxes.
  We plot the regions of fluxes $b_i$ in which the central charge assumes a positive value.
 In the first case we have fixed $b_1=x$, $b_2=y$ and $b_3=b_4=b_50$.
  In the second case we have fixed $b_3=x$ and $b_4=y$ and $b_1=b_2=b_50$.
   In the third case we have fixed $b_2=x$, $b_3=y$ and $b_1=b_4=b_5=0$.
   }
  \label{dP3S}
\end{center}
\end{figure}

\begin{figure}[H]
\begin{center}
\begin{tabular}{ccc}
  \includegraphics[width=3.6cm]{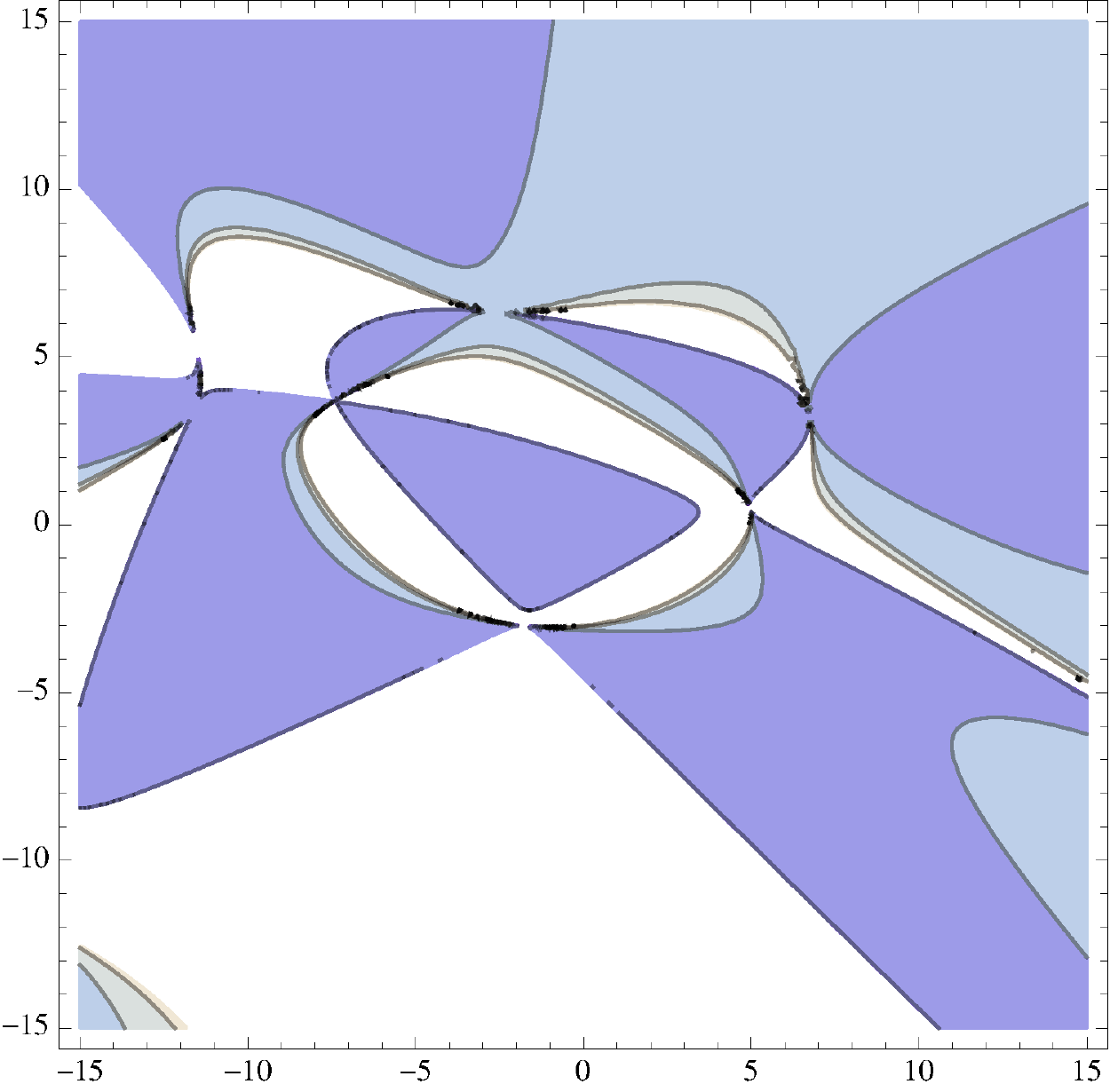}
&
  \includegraphics[width=3.6cm]{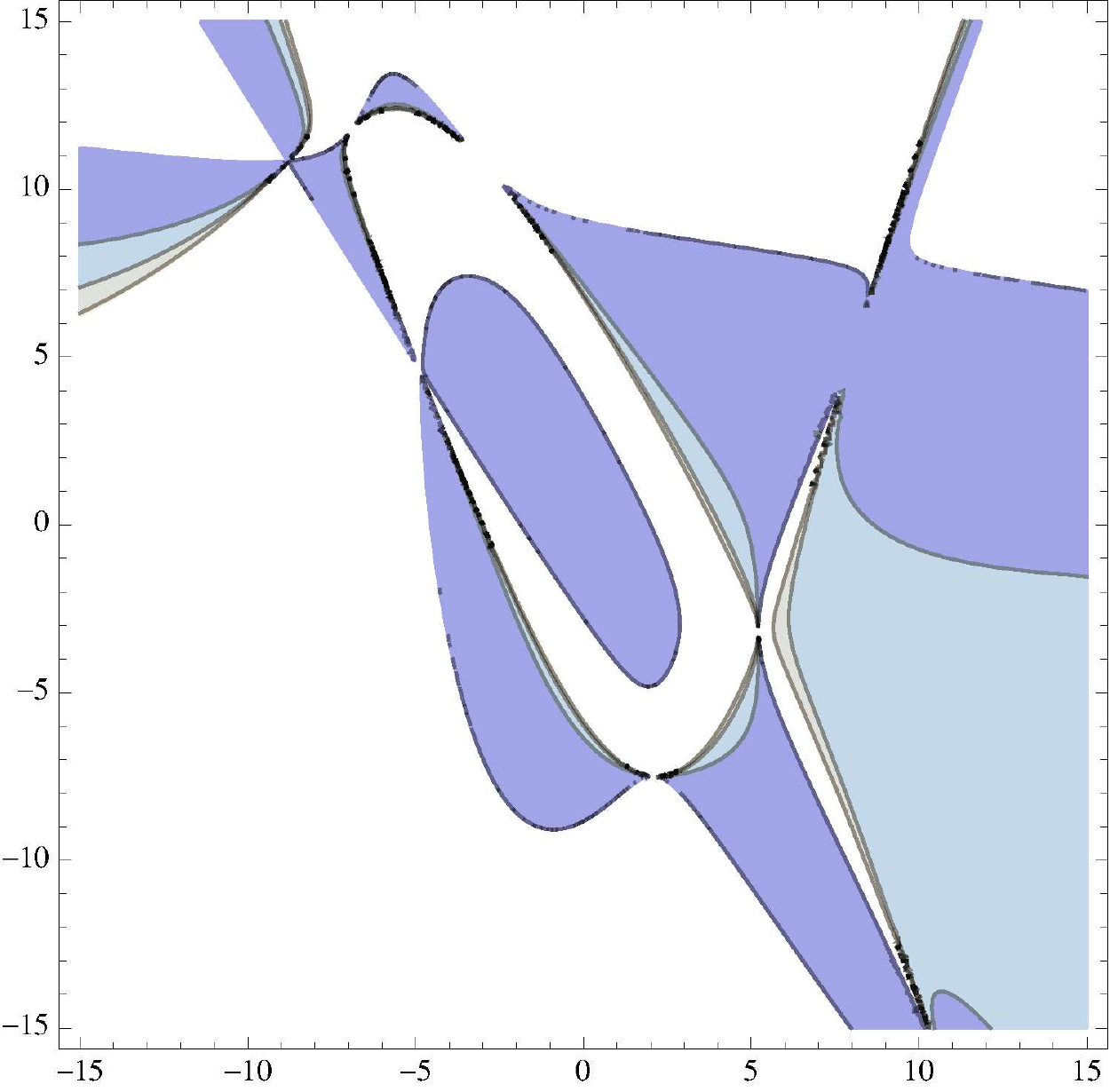}
&
  \includegraphics[width=3.6cm]{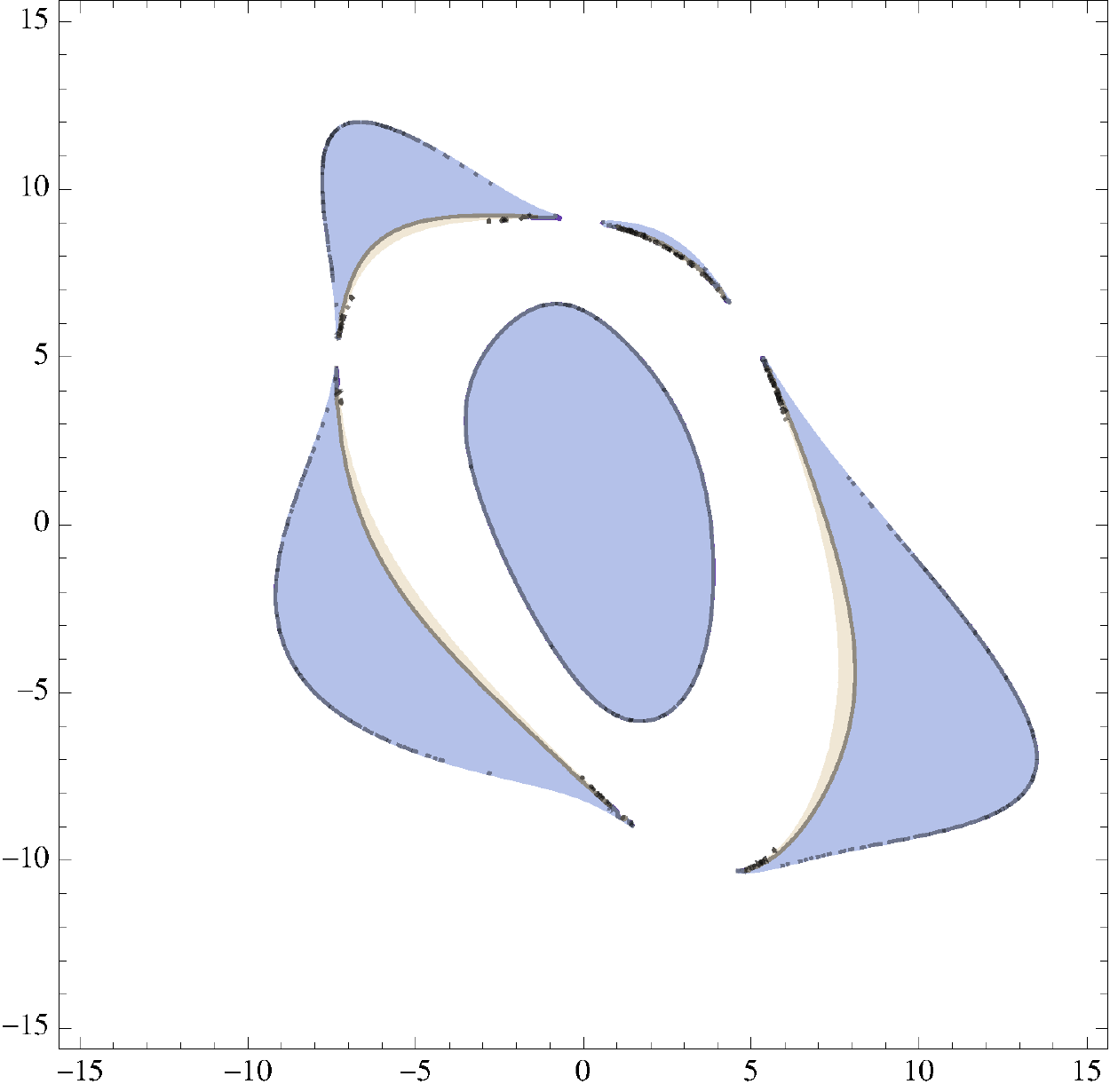}\\
\end{tabular}
  \caption{Central charge of dP$_3$ on $\Sigma=\mathbb{H}^2$   for different values of the integer fluxes.
  We plot the regions of fluxes $b_i$ in which the central charge assumes a positive value.
 In the first case we have fixed $b_1=x$, $b_2=y$ and $b_3=b_4=b_50$.
  In the second case we have fixed $b_3=x$ and $b_4=y$ and $b_1=b_2=b_50$.
   In the third case we have fixed $b_2=x$, $b_3=y$ and $b_1=b_4=b_5=0$.
   }
 \label{dP3H}
\end{center}
\end{figure}

We conclude by showing in 
Figure \ref{dP3T}, \ref{dP3S} and   \ref{dP3H} the central charge for different values of the discrete fluxes
for dP$_2$ compactified on $\Sigma=\mathbb{T}^2$, $\Sigma=\mathbb{S}^2$ and $\Sigma=\mathbb{H}^2$, respectively.
%
%
%
%
%
\subsection*{$L^{pqp}$ theories} 
%
%
%
%
%
%
As a last example we consider models with a higher number of baryonic symmetries, i.e.  
$L^{pqp}$ models \cite{Benvenuti:2005ja,Butti:2005sw,Franco:2005sm}. 
In order to have a comprehensive discussion we pick up a particular (Seiberg dual) phase, 
that can be easily visualized by the description of the system in terms of D4 and NS branes in type IIA string theory
(the other phases are obtained by exchanging the NS branes).
We consider a stack of N D4 branes extended along $x_{0123}$ and wrapping the compact direction  $x_6$.
Then we consider two sets of $p$ NS and $q$ NS' branes. The NS branes are extended along 
$x_{012345}$ and the NS' along $x_{012389}$. We order the NS branes and then the NS' branes clockwise along $x_6$.

\begin{figure}[h]
\begin{center}
  \includegraphics[width=10cm]{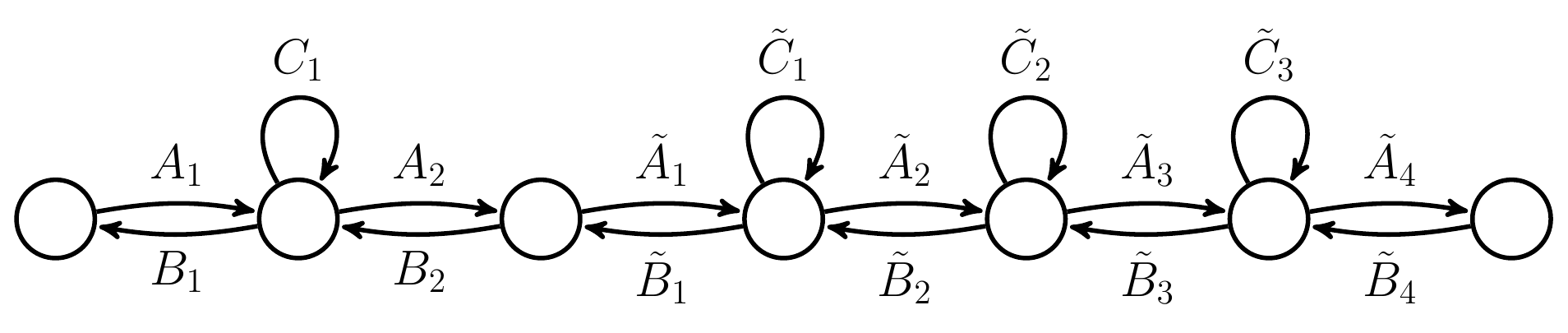}
  \caption{Quiver of the $L^{242}$ model. The first and the last node are
  identified.}
  \label{FIGL242}
\end{center}
\end{figure}

Each gauge group is associated to a segment of N D4 branes on $x_6$,
suspended between two consecutive NS branes. The resulting field theory is a $U(N)$ necklace quiver gauge theory with different types of nodes.
By counting clockwise on $x_6$ we have
\begin{itemize}
\item a set of $p-1$ nodes with an adjoint of type $C$;
\item a node without any adjoint ;
\item  a set of $q-1$ nodes with an adjoint of type $\tilde C$;
\item a node without any adjoint.
\end{itemize}
The $2p$ bifundamental matter fields crossing a NS brane are of type $A$ or $B$ depending on their orientation and
the second set of $2q$ bifundamental fields, crossing a NS' brane, are of type $\tilde A$ and $\tilde B$.
We can visualize the situation in the quiver of Figure \ref{FIGL242} for the case or $p=2$ and $q=4$.

These theories are characterized by one R--current, two flavor currents 
and $p+q$ baryonic currents 
\cite{Benvenuti:2005ja,Butti:2005sw,Franco:2005sm}. One baryonic current is redundant, being the quiver necklace. 
The vector--like nature of the field content ensures that 
the other $p+q-1$ currents are all conserved at quantum level.
The charge assignment of flavor and R-currents is
summarized in the following table\footnote{We refer to $R_0$ as a trial R-charge obtained after the 
maximization on the baryonic charges, as described after (\ref{eq:extrB}).}
\begin{equation}
  \begin{array}{ccccc}
    \tmop{field} & \tmop{mult} & F_1 & F_2 & R_0 \\
    \hline
    A & p & 0 & 1 & 0\\
    B & p & - 1 & 0 &0 \\
    \tilde{A} & q & 1 & 0 &1 \\
    \tilde{B} & q & 0 & - 1 & 1 \\
    C & p - 1 & 1 & - 1 & 2 \\
    \tilde{C} & q - 1 & - 1 & 1 & 0 
     \end{array}
\end{equation}
The baryonic currents are associated to the $U(1)_i \subset U(N)_i$ gauge factors
and the charges are read from the representation of each field under the gauge groups.
Fundamental fields of $SU(N)_i$ have charge $+1$ and anti-fundamental fields of $SU(N)_i$  have charge $-1$
under the baryonic $U(1)_i$. 

The 4d R-charge mixes with the global symmetries through the combination
\begin{equation}
  R = R_0 + \epsilon_1 F_1 + \epsilon_2  T_{F_2} + \sum_{i=1}^{p+q-1} \eta_i T_{B_i}
\end{equation}
with mixing parameters
\begin{equation}
  \epsilon_1 = \frac{\sqrt{p^2-p q+q^2}-2 p+q}{3 (p-q)}, \quad 
  \epsilon_2 = \frac{p}{\sqrt{p^2-p q+q^2}+2 p-q}, \quad 
  \eta_i = 0
  \end{equation}
 determined by the $a$--maximization. 
 
When the theory is partially topologically twisted on $\Sigma$ along the generator
\begin{equation}
T = \kappa T_R + b_1 T_{F_1} + b_2 T_{F_2} + \sum_{i=1}^{p+q-1} b_{i+2} B_i  
\end{equation}
the central charge $c_r $ is obtained from (\ref{eq:krrex}) and extremized respect to the $\epsilon_I$ parameters.
The final formulas are too involved and we do not report them here.
The mixing parameters are non-vanishing for generic choices of the curvature and
of the fluxes $b_{\bf I}$. This signals the fact that the baryonic symmetries mix with the 2d exact R-current.

\begin{figure}[H]
\begin{center}
\begin{tabular}{ccc}
  \includegraphics[width=3.5cm]{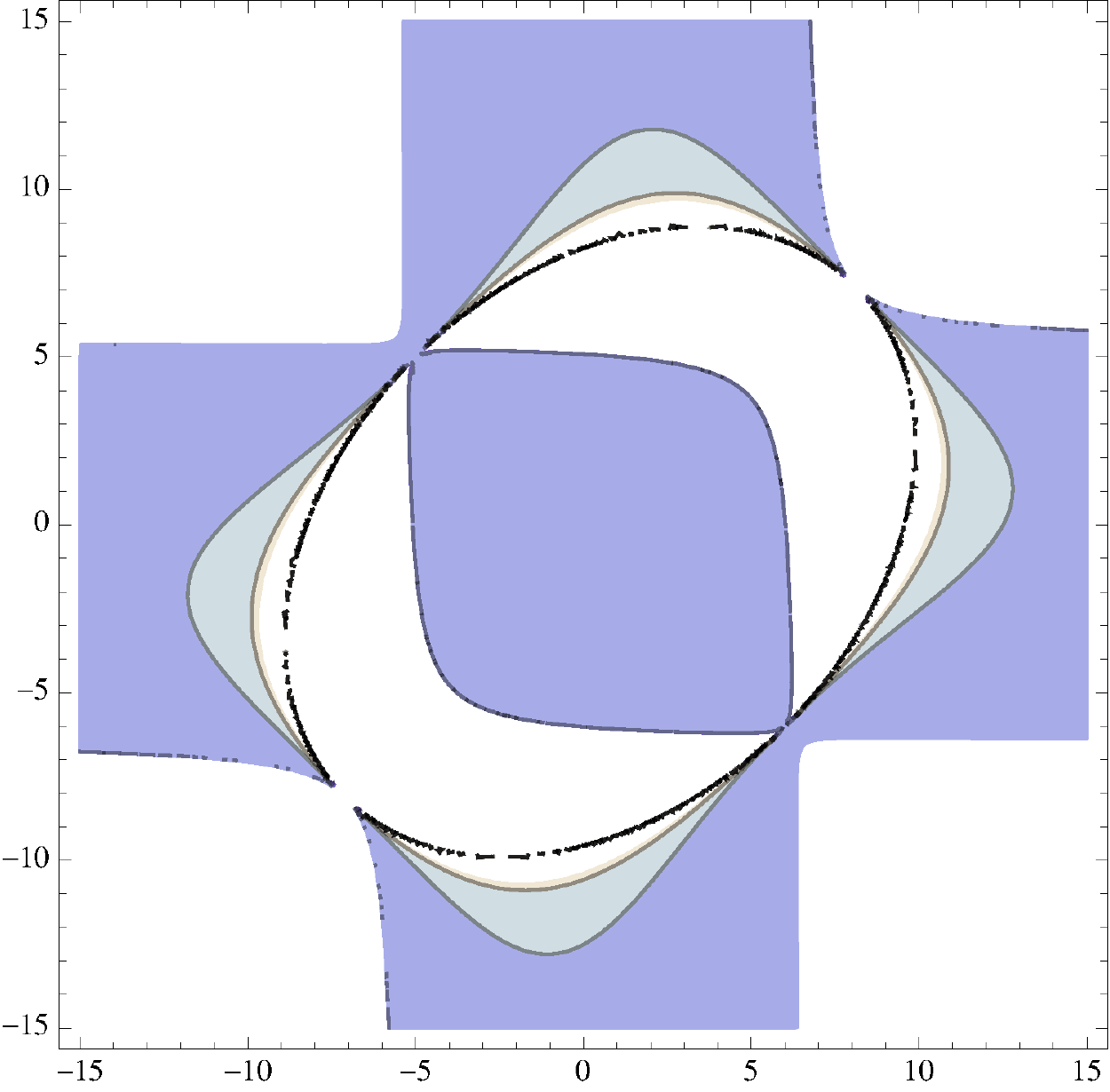}
&
  \includegraphics[width=3.5cm]{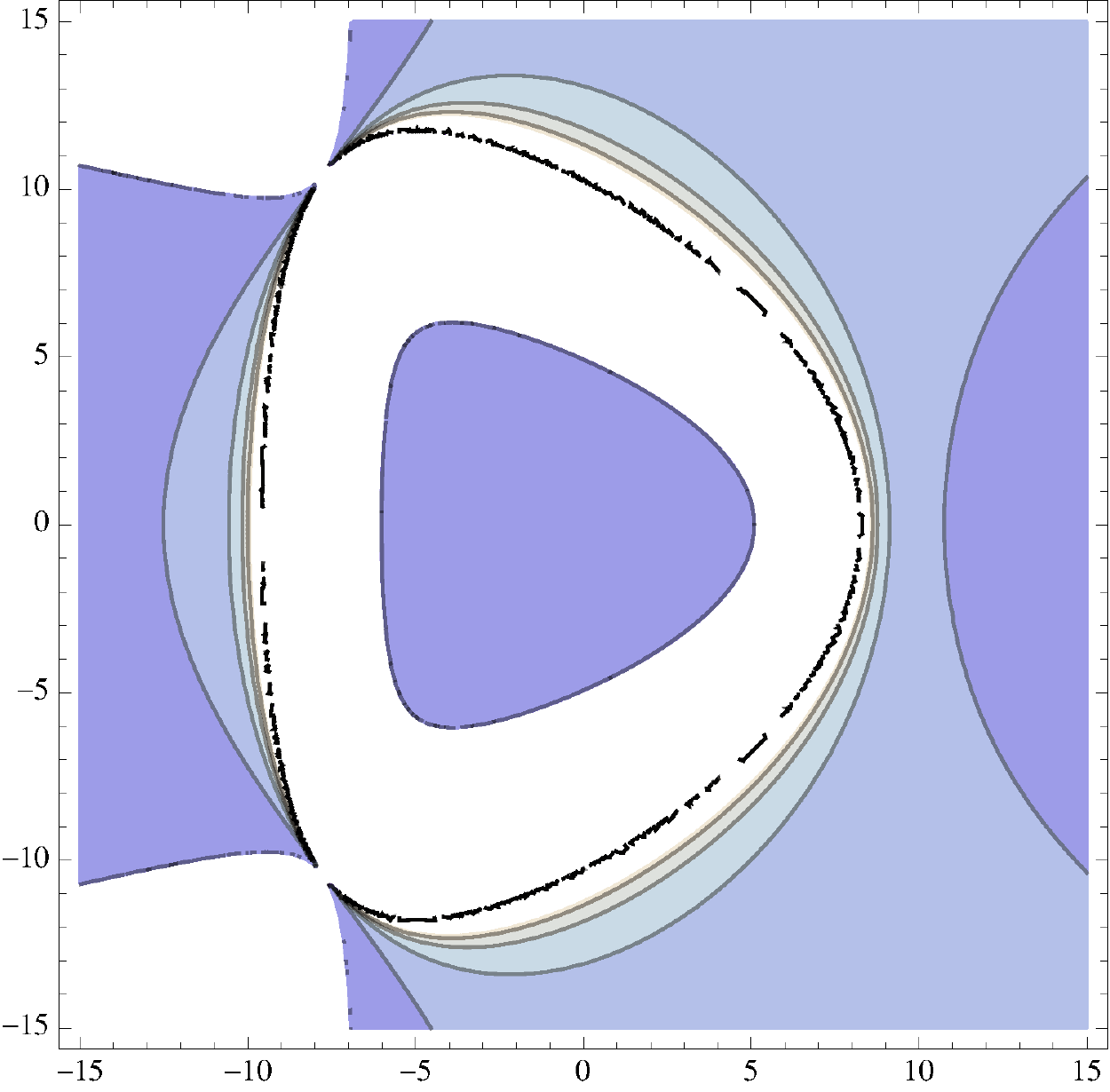}
&
  \includegraphics[width=3.5cm]{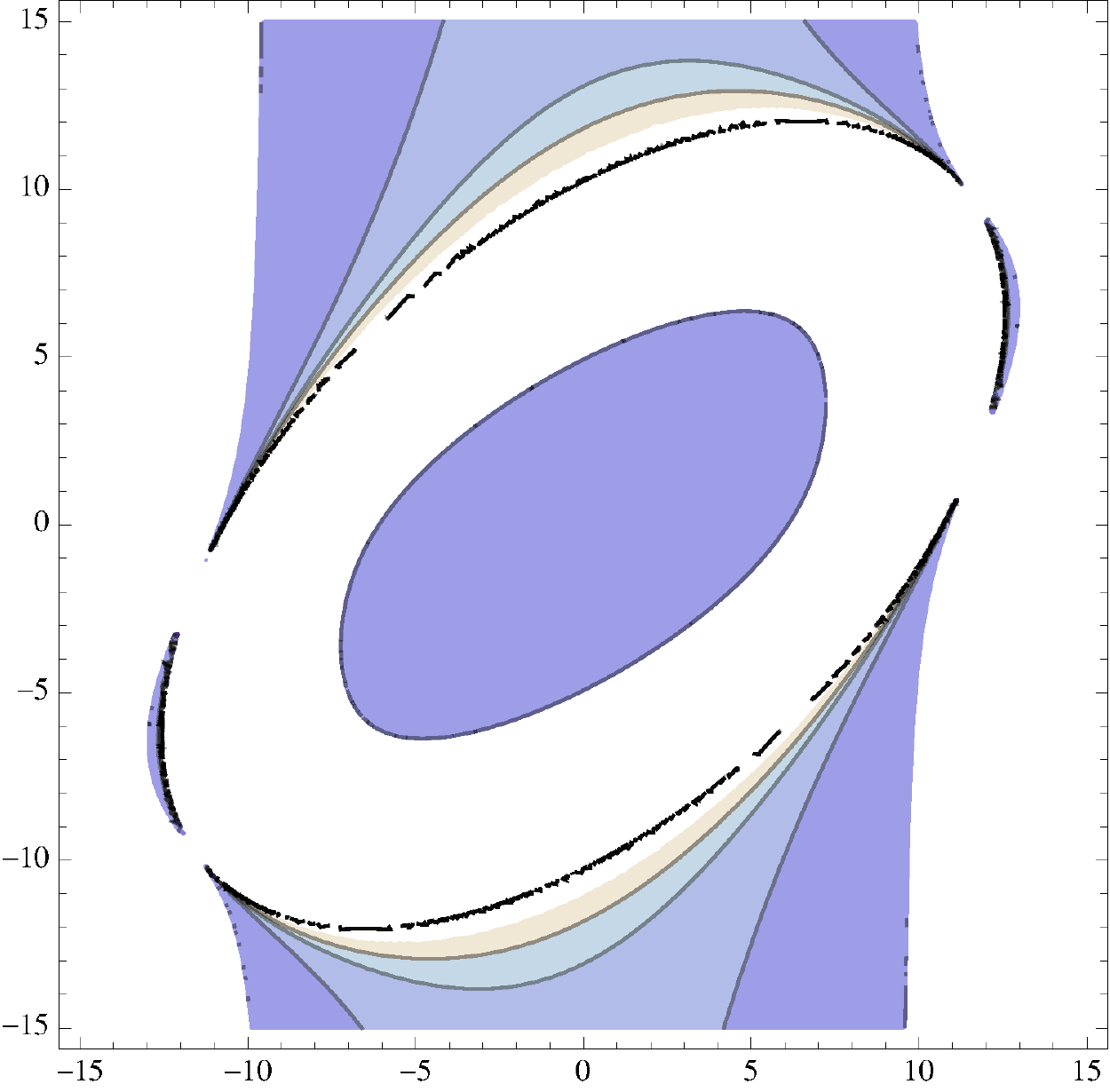}\\
\end{tabular}
  \caption{Central charge of L$^{222}$ on $\Sigma=\mathbb{H}^2$  for different values of the integer fluxes.
  We plot the regions of fluxes $b_i$ in which the central charge assumes a positive value.
 In the first case we have fixed $b_1=x$, $b_2=y$ and $b_3=b_4=b_5=0$.
  In the second case we have fixed $b_3=x$ and $b_4=y$ and $b_1=b_2=b_5=0$.
   In the third case we have fixed $b_2=x$, $b_4=y$ and $b_1=b_3=b_5=0$.}
  \label{L222ccH}
\end{center}
\end{figure}

\begin{figure}[H]
\begin{center}
\begin{tabular}{ccc}
  \includegraphics[width=3.5cm]{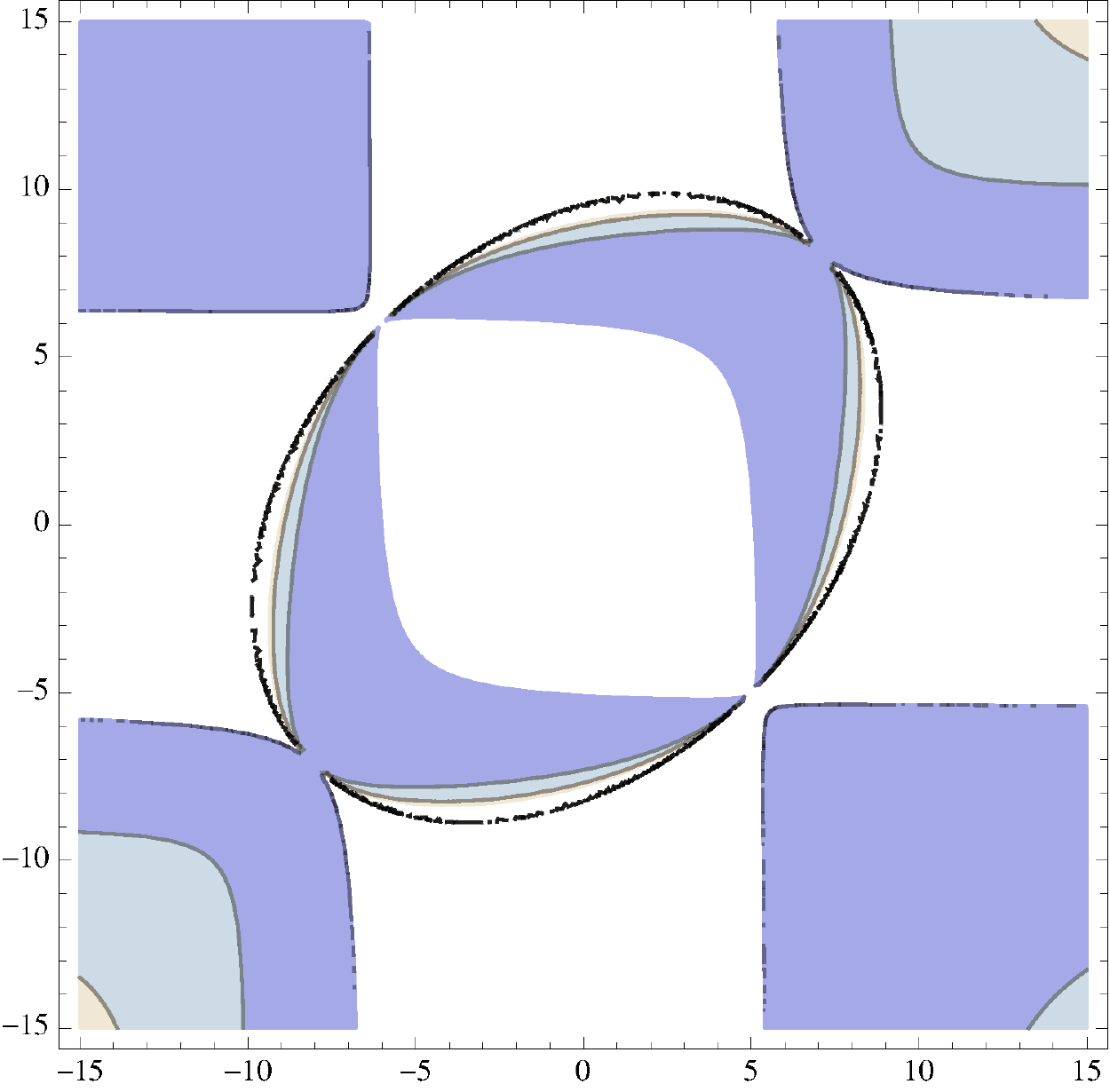}
&
  \includegraphics[width=3.5cm]{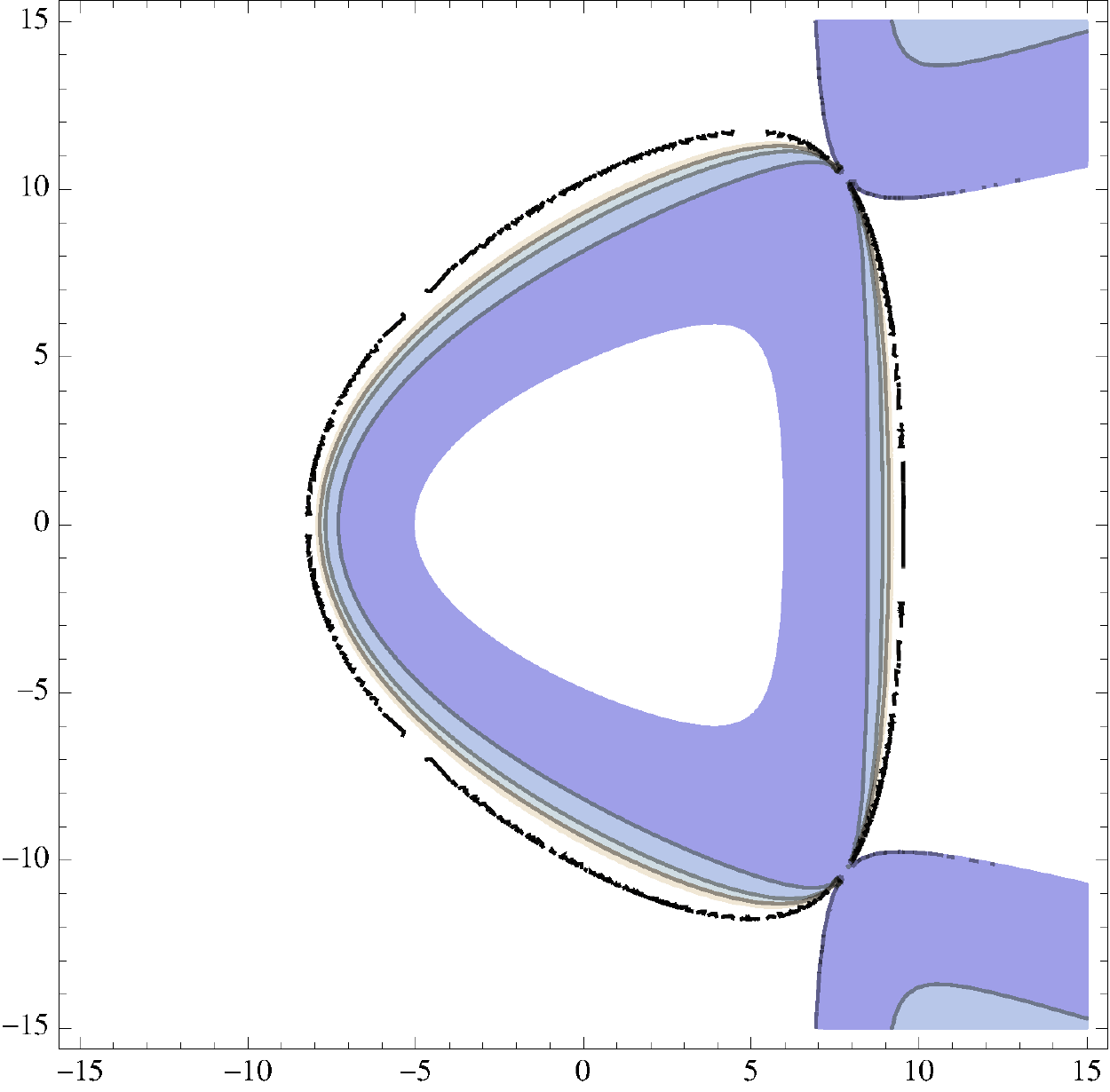}
&
  \includegraphics[width=3.5cm]{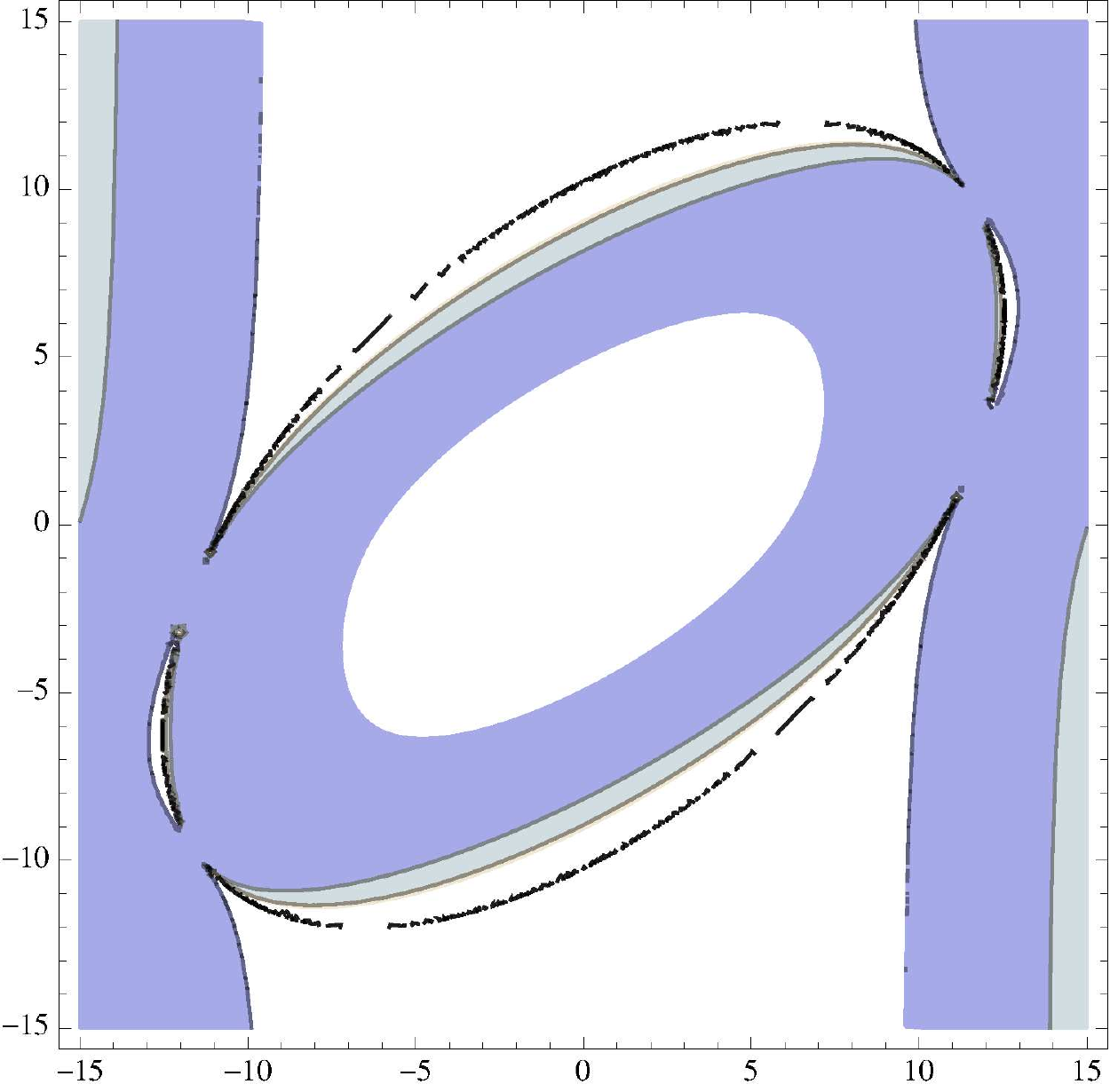}\\
\end{tabular}
  \caption{Central charge of L$^{222}$ on $\Sigma=\mathbb{S}^2$ for different values of the integer fluxes.
  We plot the regions of fluxes $b_i$ in which the central charge assumes a positive value.
   In the first case we have fixed $b_1=x$, $b_2=y$ and $b_3=b_4=b_5=0$.
  In the second case we have fixed $b_3=x$ and $b_4=y$ and $b_1=b_2=b_5=0$.
   In the third case we have fixed $b_2=x$, $b_4=y$ and $b_1=b_3=b_5=0$.}
  \label{L222ccS}
\end{center}
\end{figure}

In the following we show some numerical result for the 2d central charge for the $L^{222}$
and the $L^{131}$ gauge theories. In both cases there are four gauge groups and 
three non--anomalous baryonic symmetries.
In both case we observe that the baryonic symmetries mix for generic values of the
with the  R--current at the 2d fixed point.
\newpage
In Figure \ref{L222ccH} and \ref{L222ccS} we represent the central charge for different values of the discrete fluxes
for $L^{222}$ compactified on $\Sigma=\mathbb{H}^2$ and $\Sigma=\mathbb{S}^2$, respectively.

In the case of the torus reduction ($\kappa =0$)  the formulae are simpler and we can provide the
analytical expression for $c_r$ extremized with respect to the mixing parameters, in terms of the $b_i$ fluxes
\begin{equation}
\begin{array}{rcl}
\tiny 
c_{r}^{L^{222},T^2} = {\frac{6 (b_3^2+b_1 b_3+b_2 b_3-2 b_4 b_3+2 b_4^2+b_5^2+2 b_1 b_2-(b_1+b_2+2 b_4) b_5) (b_3^2+b_5^2+b_2 (b_5-b_3)+b_1 (2 b_2-b_3+b_5))}{(b_1-b_2)(b_4^2-b_3 b_4+b_1 b_3+b_2 b_3-(b_1+b_2+b_4) b_5)}
} \end{array}
\end{equation}
\normalsize
The parameters $\epsilon_i^*$ are generically non vanishing for both the flavor and 
the baryonic global symmetries.

\begin{figure}[H]
\begin{center}
\begin{tabular}{ccc}
  \includegraphics[width=3.5cm]{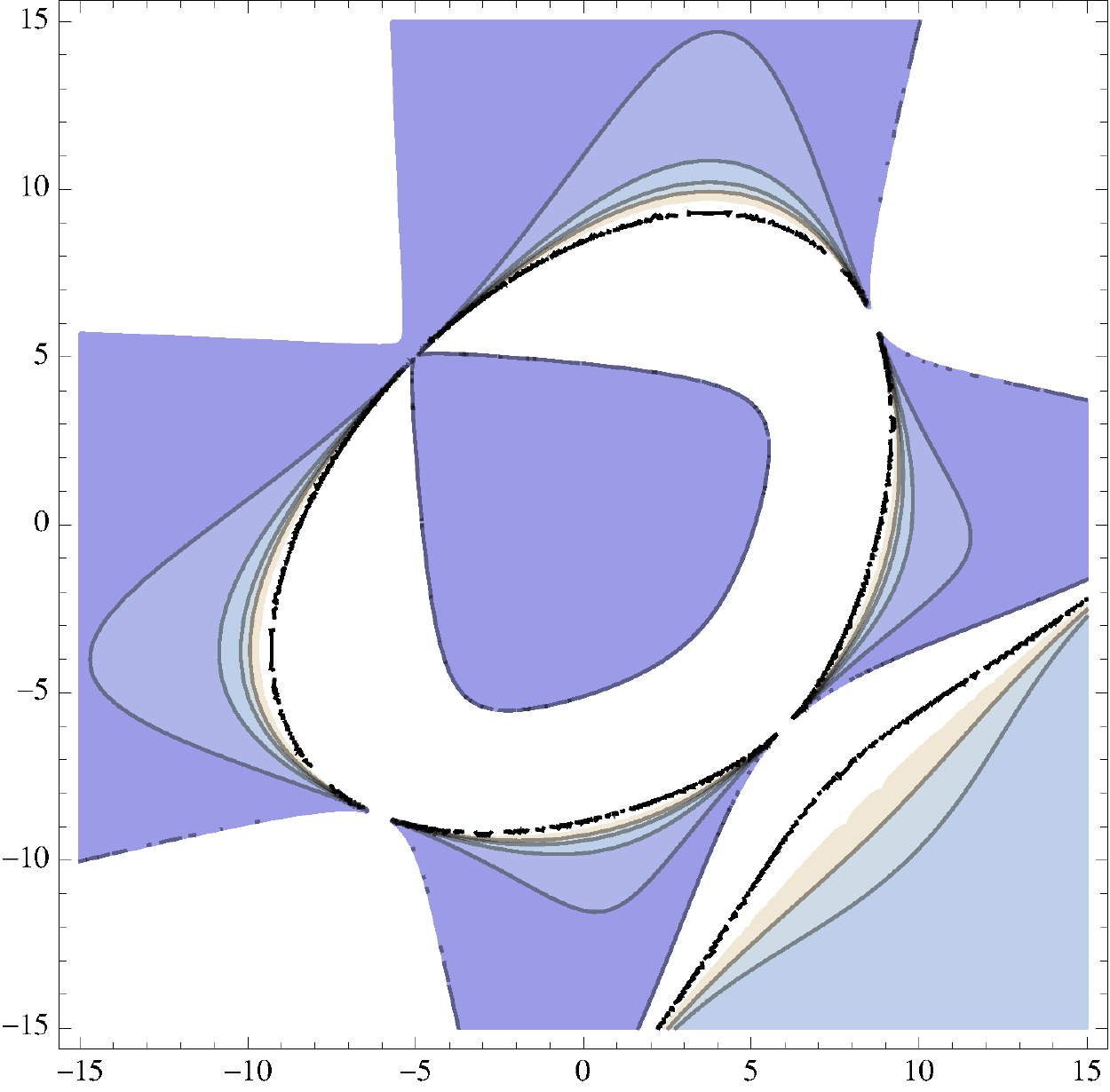}
&
  \includegraphics[width=3.5cm]{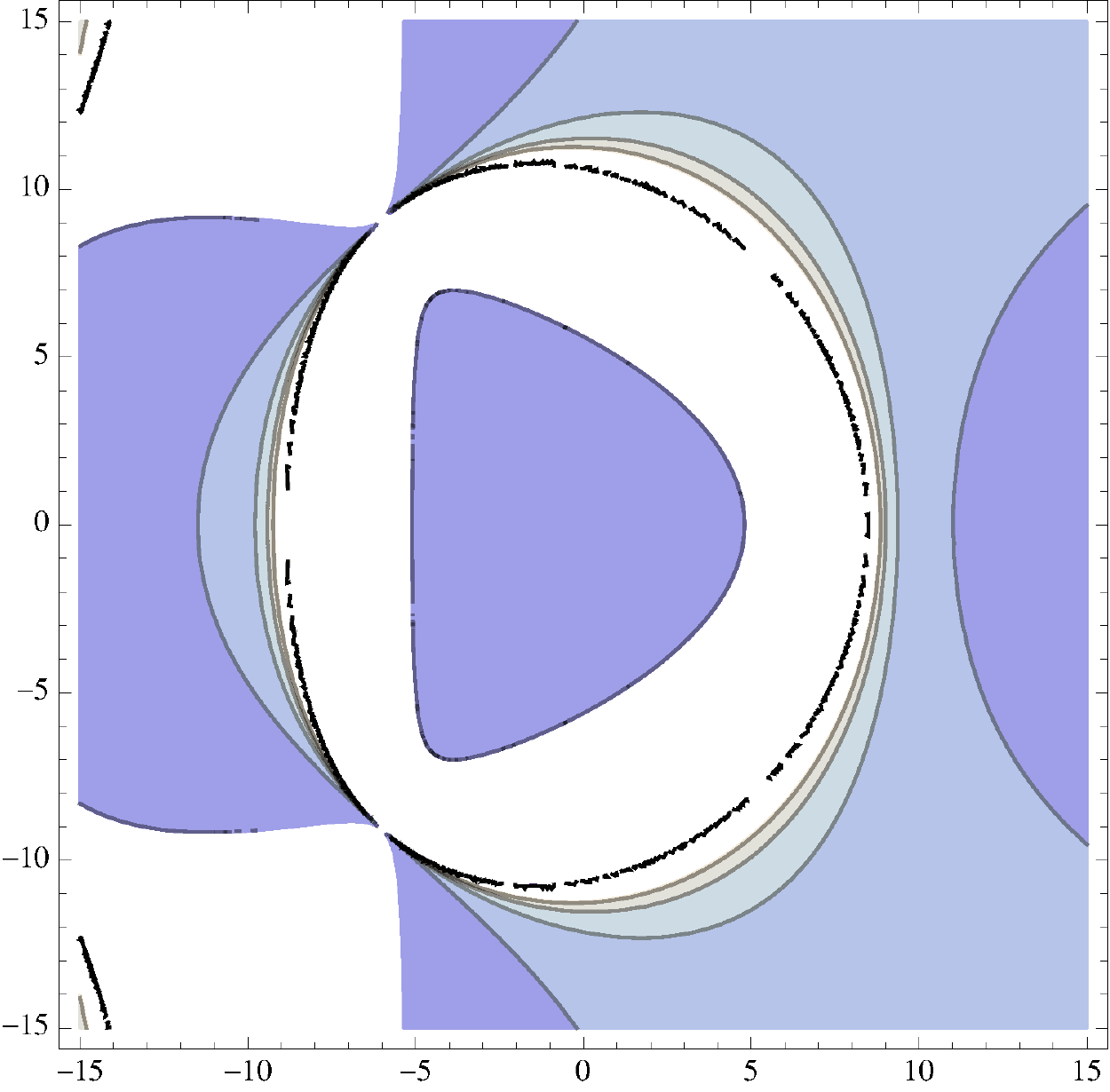}
&
  \includegraphics[width=3.5cm]{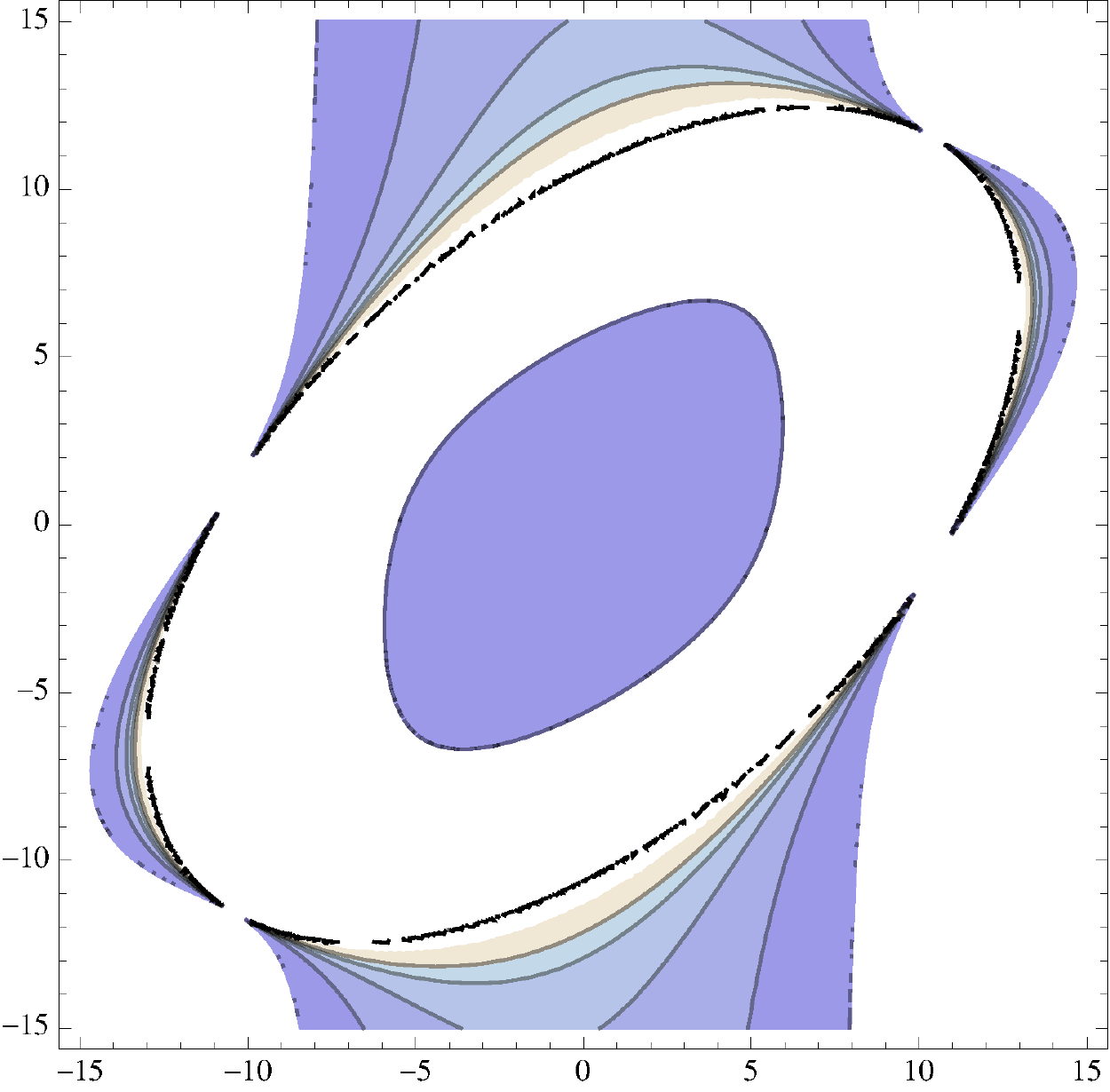}\\
\end{tabular}
  \caption{Central charge of L$^{131}$ on $\Sigma=\mathbb{H}^2$ for different values of the integer fluxes.
  We plot the regions of fluxes $b_i$ in which the central charge assumes a positive value.
 In the first case we have fixed $b_1=x$, $b_2=y$ and $b_3=b_4=b_5=0$.
  In the second case we have fixed $b_3=x$ and $b_4=y$ and $b_1=b_2=b_5=0$.
   In the third case we have fixed $b_2=x$, $b_4=y$ and $b_1=b_3=b_5=0$.}
  \label{L131ccH}
\end{center}
\end{figure}

\begin{figure}[H]
\begin{center}
\begin{tabular}{ccc}
  \includegraphics[width=3.5cm]{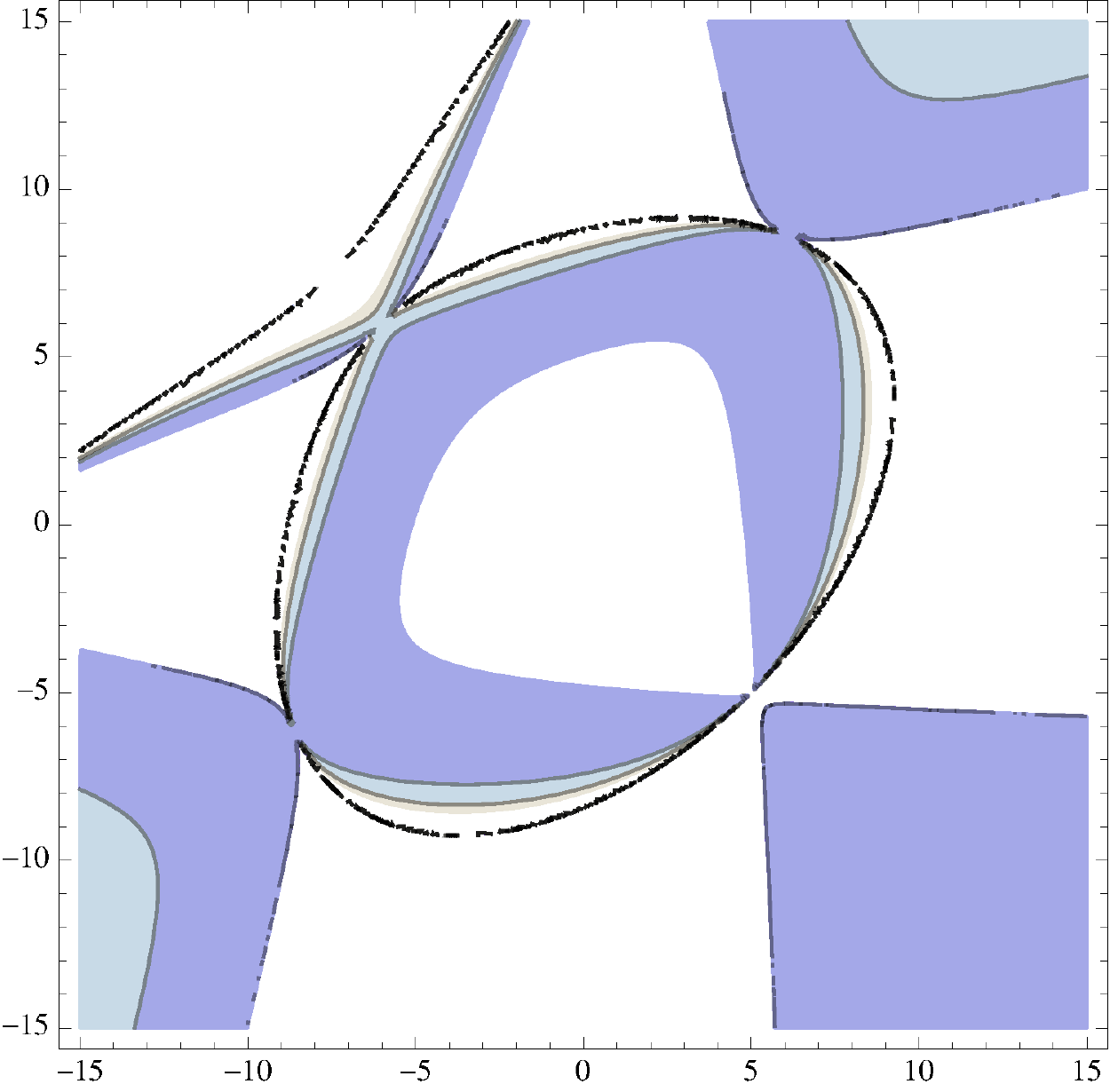}
&
  \includegraphics[width=3.5cm]{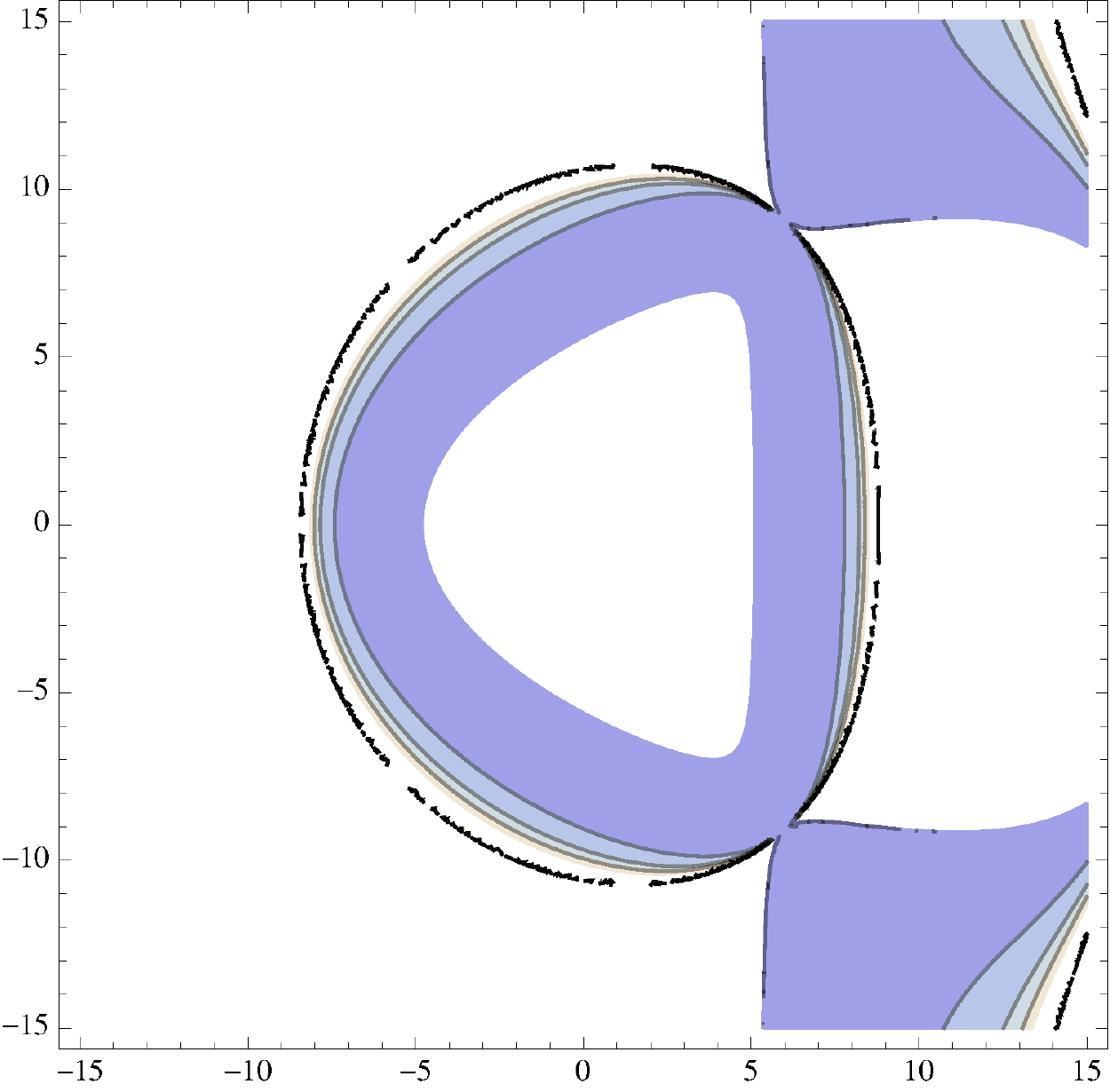}
&
  \includegraphics[width=3.5cm]{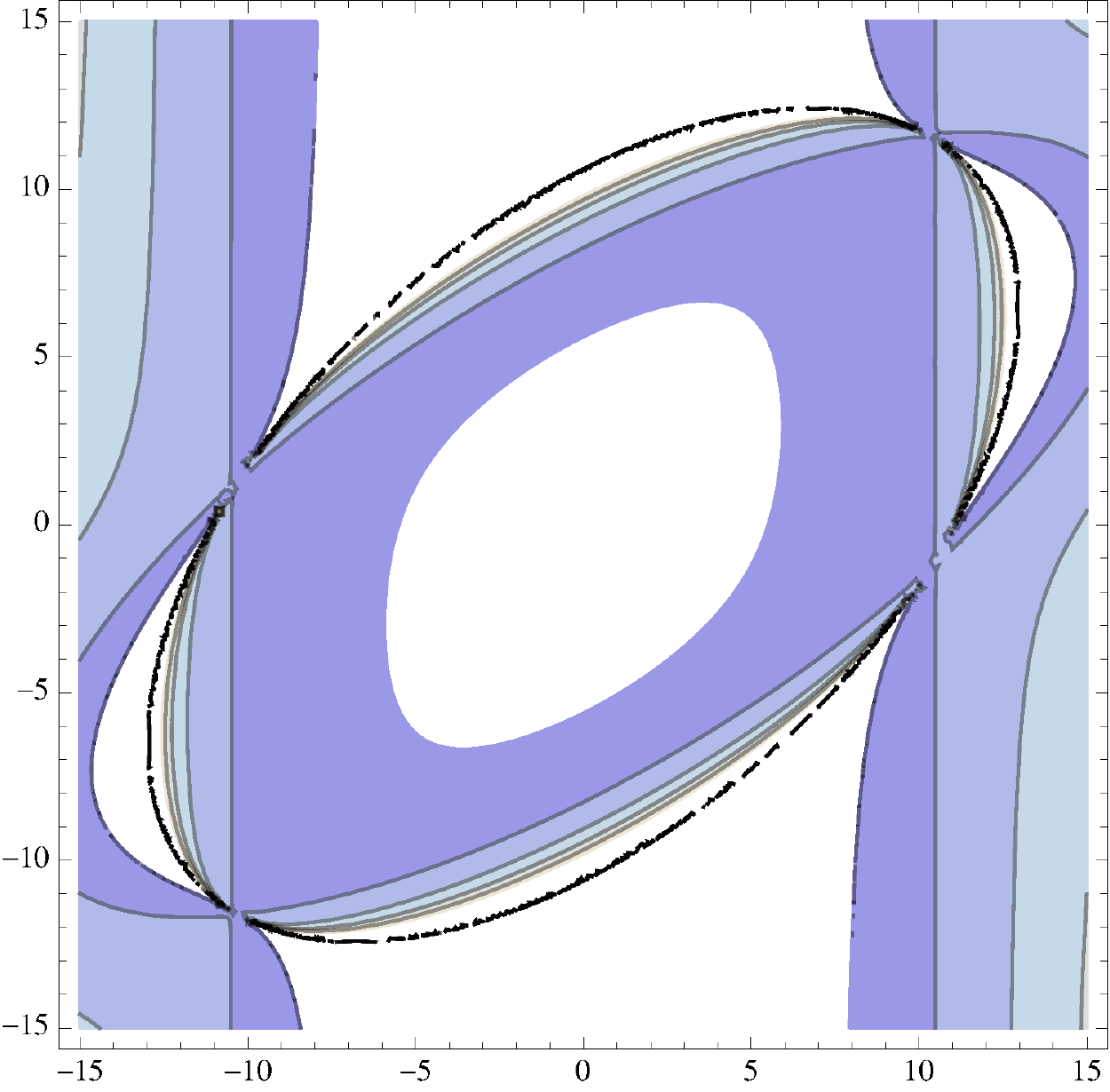}\\
\end{tabular}
  \caption{Central charge of L$^{131}$ on $\Sigma=\mathbb{S}^2$  for different values of the integer fluxes.
  We plot the regions of fluxes $b_i$ in which the central charge assumes a positive value.
  In the first case we have fixed $b_1=x$, $b_2=y$ and $b_3=b_4=b_5=0$.
  In the second case we have fixed $b_3=x$ and $b_4=y$ and $b_1=b_2=b_5=0$.
   In the third case we have fixed $b_2=x$, $b_4=y$ and $b_1=b_3=b_5=0$.}
  \label{L131ccS}
\end{center}
\end{figure}

\newpage
In Figure \ref{L131ccH} and \ref{L131ccS} we represent the central charge for different values of the discrete fluxes
for $L^{131}$ compactified on $\Sigma=\mathbb{H}^2$ and $\Sigma=\mathbb{S}^2$ respectively.

In the case of the torus reduction ($\kappa =0$)  the formulae are simpler and we can provide the
analytical expression for $c_r$ extremized with respect to the mixing parameters, in terms of the $b_i$ fluxes
\begin{equation}
\begin{array}{rcl}
\tiny 
c_{r}^{L^{131},T^2} &=&
6 \left(\frac{(b_1^2-3 b_2 b_1+b_2^2-b_3^2+(b_1+b_2) b_3){}^2}{(b_1-b_2)((b_1^2+(b_3-b_2) b_1+b_2^2+b_4^2+b_5^2+b_2 b_3-b_3 b_4-b_4 b_5)}
-\frac{b_1^2+4 b_2 b_1-2 b_3 b_1-b_2^2+2 b_3^2-2 b_2 b_3}{b_1-b_2}
\right)
\end{array}
\end{equation}
\normalsize
Also in this case we observe that the parameters $\epsilon_i^*$ are generically non vanishing for both the flavor and 
the baryonic global symmetries.
%
%
%
%
%
\section{Further directions}
\label{conclusions}
%
%
%
%
%
In this paper we have studied c-extremization for 2d $\mathcal{N}=(0,2)$ 
SCFTs arising from the twisted compactification of 4d SCFTs on compact,
constant curvature Riemann surfaces.
The SCFTs under investigation consist of infinite families of quiver gauge theories
holographically dual to D3 branes probing the tip
of CY$_3$ cones over X$_5$ basis admitting a $U(1)^3$ toric action.
In such cases we have been able to develop a simple geometric formulation for 
the 2d central charge $c_r$ in terms of its mixing with the global currents.

This formulation borrows many ideas and constructions developed in the 4d parent 
theory, in which it has been demonstrated that the conformal anomaly $a$ 
is proportional to the inverse of the X$_5$ volume.
The geometric analogy with the 4d formulation that we have discussed may be helpful in understanding the 
possible relation between $c_r$ and the volume of the seven 
manifold \cite{Kim:2005ez,Gauntlett:2007ts} in the conjectured AdS$_3 \times \mathcal{M}_7$ correspondence (or eight manifolds in M--theory). It should be interesting to investigate this direction further.

In our analysis we have shown that the 2d central charge,
expressed in terms of the mixing parameters, can be 
reformulated in the language of the toric geometry underlining the moduli space of the 4d theory.
Nevertheless we did not give a general discussion on the extremization of this function.
This point certainly deserves a separate and deep analysis.
Indeed, the existence of an extremum is not guaranteed, as discussed in \cite{Benini:2012cz}.
The main obstructions are due to the absence of a normalizable vacuum of
the 2d CFT and to the presence of accidental symmetries at the IR fixed point.
The study of this problem would be simplified by the knowledge of the spectrum and the interactions of the 2d models.
Progresses in such directions have been made in 
\cite{Almuhairi:2011ws,Kutasov:2013ffl,Gadde:2015wta}.
On the geometric side it would be interesting to see if 
some of the tools developed in 4d (e.g. the zonotope discussed 
in \cite{Kato:2006vx})  can be useful for  
the analysis of the extremization properties of the 2d central charge.

As a last comment we wish to mention that 
recently infinite families of 2d SCFTs have been obtained by
exploiting the role of the toric geometry \cite{Franco:2015tna,Franco:2015tya}.
These theories, denoted as brane brick models, are 
expected to describe the worldvolume theory 
of stacks of D1 branes probing the tip of toric CY$_4$ 
cones in type IIB.
It has been shown that in such cases the toric geometry
can be used to obtain the elliptic genus \cite{Franco:2017cjj}.
It would be interesting to further
explore the role of toric geometry in these 2d SCFTs
and look for possible connections, if any, with our results.

\section*{Acknowledgements}

We are grateful to Marcos Crichigno and Domenico Orlando
for useful discussions and comments.
The work of A.A.  is supported by the Swiss National Science Foundation
(snf) under grant number pp00p2$_{-}$157571/1.
 This work has been supported in part by Italian Ministero
dell'Istruzione, Universit\`a e Ricerca (MIUR) and Istituto Nazionale di Fisica Nucleare
(INFN) through the "Gauge Theories, Strings, Supergravity" (GSS) research project.
We thank the Galileo Galilei Institute for Theoretical Physics (GGI) for the hospitality and 
INFN for partial support during the completion of this work, within the program 
"New Developments in AdS$_3$/CFT$_2$ Holography".

\appendix
\section{Mixing parameters for dP$_2$ and dP$_3$}
\label{mixing}

In this appendix we report the value of the mixing parameters for some choices
of fluxes for the dP$_2$ and the dP$_3$ models studied in the body of the paper.
The general results are pretty involved, so we restrict to some simple choices
of fluxes $b_\mathbf{I}$.

In the dP$_2$ case we just show the $\kappa=0$ case and one non vanishing flux
each time.
By following the notations of section \ref{FTside} we refer to $\epsilon$ as the mixing parameters
of the flavor symmetries and to $\eta$ as the mixing parameters of the baryonic symmetries.
We have the following cases
\begin{itemize}
\item{$b_1 \neq 0$}
\begin{eqnarray}
\epsilon _1 &=& \phantom{+} \frac{7 b_1^4+85 b_1^3-1800 b_1^2-4500 b_1+60000}{13 b_1^4+310 b_1^3-6375 b_1^2+7000 b_1+100000}
\nonumber \\
\epsilon _2 &=& \phantom{+} \frac{11 b_1^4+110 b_1^3-225 b_1^2-3250 b_1-10000}{13 b_1^4+310 b_1^3-6375 b_1^2+7000 b_1+100000}
\nonumber \\
\eta_1&=& -\frac{3 b_1^4-20 b_1^3-525 b_1^2+2000 b_1}{13 b_1^4+310 b_1^3-6375 b_1^2+7000 b_1+100000}
\nonumber \\
\eta_2&=& -\frac{18 b_1^4-10 b_1^3-600 b_1^2-500 b_1}{13 b_1^4+310 b_1^3-6375 b_1^2+7000 b_1+100000}
\end{eqnarray}
\item{$b_2 \neq 0$}
\begin{eqnarray}
\epsilon _1 &=& -\frac{8 b_2^4-270 b_2^3+2600 b_2^2-1250 b_2-45000}{180 b_2^3-3825 b_2^2+4500 b_2+75000}
\nonumber \\
\epsilon _2 &=&- \frac{8 b_2^3-175 b_2^2+525 b_2+500}{12 b_2^3-255 b_2^2+300 b_2+5000}
\nonumber \\
\eta_1&=& -\frac{4 b_2^4-135 b_2^3+850 b_2^2-1000 b_2}{180 b_2^3-3825 b_2^2+4500 b_2+75000}
\nonumber \\
\eta_2&=& -\frac{8 b_2^4-270 b_2^3+2000 b_2^2+250 b_2}{180 b_2^3-3825 b_2^2+4500 b_2+75000}
\end{eqnarray}
\item{$b_3 \neq 0$}
\begin{eqnarray}
\epsilon _1 = \frac{3}{5}
, \quad
\epsilon _2 =  -\frac{1}{10}
, \quad
\eta_1= -\frac{b_3}{10}
, \quad
\eta_2= 0
\end{eqnarray}
\item{$b_4 \neq 0$}
\begin{eqnarray}
\epsilon _1 &=& \phantom{+} \frac{95 b_4^3-900 b_4^2+45000}{3 b_4^4-1950 b_4^2+1500 b_4+75000}
\nonumber \\
\epsilon _2 &=& -\frac{4 b_4^4+5 b_4^3-1000 b_4^2+1000 b_4+5000}{2 b_4^4-1300 b_4^2+1000 b_4+50000}
\nonumber \\
\eta_2&=& -\frac{12 b_4^4+175 b_4^3+750 b_4^2-1500 b_4}{6 b_4^4-3900 b_4^2+3000 b_4+150000}
\nonumber \\
\eta_2&=&\phantom{+}  \frac{6 b_4^4+80 b_4^3-1800 b_4^2-6000 b_4}{3 b_4^4-1950 b_4^2+1500 b_4+75000}
\end{eqnarray}
\end{itemize}
It is interesting to observe that in each case all the mixing parameters are non--vanishing, showing the 
general fact that the baryonic symmetries have a non trivial mix with the R--current in 2d.

In the dP$_3$ case we consider the case of generic $\kappa$ ($\kappa^2 = 1$) and again we fix only one non vanishing flux
for each case.
We have the following cases
\begin{itemize}
\item{$b_1 \neq 0$}
\begin{eqnarray}
\epsilon _1 &=&  
\phantom{+} 
\frac{16979 b_1  -6982 b_1^2 \kappa -740 b_1^3+171360 \kappa}
{52998 b_1 -4416 b_1^2 \kappa -1104 b_1^3+235116 \kappa }
\nonumber \\
\epsilon _2 &=&  
-\frac{8165 b_1  -844 b_1^2 \kappa -188 b_1^3+42840 \kappa  }
{26499 b_1  -2208 b_1^2 \kappa -552 b_1^3+117558 \kappa  }
\nonumber \\
\eta_1&=& -\frac{105 b_1 +16 b_1^2 \kappa +4 b_1^3+630 \kappa  }
{8833 b_1-736 b_1^2 \kappa -184 b_1^3+39186 \kappa }
\nonumber \\
\eta_2&=&   
\phantom{+} 
\frac{920 b_1 -1948 b_1^2 \kappa -464 b_1^3-630 \kappa }
{26499 b_1 -2208 b_1^2 \kappa -552 b_1^3+117558 \kappa }
\nonumber \\
\eta_3&=&   
\phantom{+} \frac{605 b_1  -1996 b_1^2 \kappa -476 b_1^3-2520 \kappa}
{52998 b_1 -4416 b_1^2 \kappa -1104 b_1^3+235116 \kappa } 
\end{eqnarray}
\item[$b_2 \neq 0$]
\begin{eqnarray}
\epsilon _1 &=&  
-
\frac
{4 \left(5640 b_2  -41298 b_2^2 \kappa  +840 b_2^3 +737 b_2^4 \kappa -52 b_2^5+514080 \kappa \right)}
{3 \left(43536 b_2+116452 b_2^2 \kappa  -1952 b_2^3  -2053 b_2^4 \kappa +104 b_2^5-940464 \kappa \right)}
\nonumber \\
\epsilon _2 &=&   
\phantom{+} 
\frac
{199452 b_2  -177054 b_2^2 \kappa  +1767 b_2^3  +4138 b_2^4 \kappa -338 b_2^5+1028160 \kappa }
{130608 b_2  +349356 b_2^2 \kappa -5856 b_2^3 -6159 b_2^4 \kappa +312 b_2^5-2821392 \kappa }
\nonumber \\
\eta_1&=&  
\phantom{+} 
\frac
{51420 b_2  +15446 b_2^2 \kappa  -8649 b_2^3  -1289 b_2^4 \kappa +156 b_2^5+15120 \kappa  }
{43536 b_2  +116452 b_2^2 \kappa  -1952 b_2^3  -2053 b_2^4 \kappa +104 b_2^5+940464 \kappa  }
\nonumber \\
\eta_2&=& 
 \phantom{+} 
 \frac
 {960 b_2  +66888 b_2^2 \kappa  -39906 b_2^3  -4054 b_2^4 \kappa +728 b_2^5+15120 \kappa  }
 {130608 b_2  +349356 b_2^2 \kappa -5856 b_2^3 -6159 b_2^4 \kappa +312 b_2^5-2821392 \kappa  }
\nonumber \\
\eta_3&=&     
\phantom{+} 
\frac{79500 b_2  +63198 b_2^2 \kappa -30447 b_2^3 -4835 b_2^4 \kappa +364 b_2^5+30240 \kappa }
{3 \left(43536 b_2 +116452 b_2^2 \kappa -1952 b_2^3  -2053 b_2^4 \kappa +104 b_2^5-940464 \kappa \right)}
\nonumber \\
\end{eqnarray}
\item[$b_3 \neq 0$]
\begin{eqnarray}
\epsilon _1 &=&  
-
\frac
{15660 b_3 \kappa -19524 b_3^2 +1857 b_3^3 \kappa -41 b_3^4+2056320 }
{3 \left(25344 b_3 \kappa +27384 b_3^2 -41 b_3^4-940464 \right)}
\nonumber \\
\epsilon _2 &=&  
\phantom{+} 
\frac
{18180 b_3 \kappa -21588 b_3^2  +3729 b_3^3 \kappa +41 b_3^4+1028160 }
{3 \left(25344 b_3 \kappa +27384 b_3^2  -41 b_3^4-940464  \right)}
\nonumber \\
\eta_1&=& 
\phantom{+} 
\frac
{107484 b_3 \kappa -  3564 b_3^2    - 3245 b_3^3 \kappa +2 b_3^4+15120}
{25344 b_3 \kappa   + 27384 b_3^2  - 41 b_3^4-940464  }
\nonumber \\
\eta_2&=&  
-\frac
{2 \left(11340 b_3 \kappa + 366 b_3^2 +1749 b_3^3 \kappa -164 b_3^4-7560 \right)}
{3 \left(25344 b_3 \kappa + 27384 b_3^2  -41 b_3^4-940464  \right)}
\nonumber \\
\eta_3&=&  
\phantom{+} 
\frac
{4 \left(21690 b_3 \kappa +246 b_3^2 -357 b_3^3 \kappa +44 b_3^4+7560 \right)}
{3 \left(25344 b_3 \kappa +27384 b_3^2 -41 b_3^4-940464 \right)}
\end{eqnarray}
\item[$b_4\neq 0$]
\begin{eqnarray}
\epsilon _1 &=&-   \frac{3450 b_4 \kappa -15095 b_4^2+1028160  }{27288 b_4 \kappa +25014 b_4^2-1410696  }
\nonumber \\
\epsilon _2 &=&  -
 \frac{3450 b_4 \kappa +7645 b_4^2-514080  }{27288 b_4 \kappa +25014 b_4^2-1410696  }
\nonumber \\
\eta_13&=& \phantom{+} 
 \frac{3330 b_4 \kappa -65 b_4^2+7560  }{9096 b_4 \kappa +8338 b_4^2-470232  }
\nonumber \\
\eta_2&=& \phantom{+} 
 \frac{62178 b_4 -1894 b_4^2 \kappa -1137 b_4^3+3780 \kappa }{13644 b_4 \kappa ^2+12507 b_4^2 \kappa -705348 \kappa }
\nonumber \\
\eta_3&=&  \phantom{+} \frac{7410 b_4 \kappa +565 b_4^2+7560  }{13644 b_4 \kappa +12507 b_4^2-705348 }
\end{eqnarray}
\item[$b_5 \neq 0$]
\begin{eqnarray}
\epsilon _1 &=&   -  \frac{3450 b_5 \kappa +15095 b_5^2-1028160}{27288 b_5 \kappa -25014 b_5^2+1410696}
\nonumber \\
\epsilon _2 &=& \phantom{+}  \frac{3450 b_5 \kappa +3725 b_5^2-257040}{13644 b_5 \kappa -12507 b_5^2+705348}
\nonumber \\
\eta_1&=&
-\frac{60 b_5 \kappa -65 b_5^2+3780}{4548 b_5 \kappa -4169 b_5^2+235116}
\nonumber \\
\eta_2&=& 
-\frac{7590 b_5 \kappa -760 b_5^2+3780}{13644 b_5 \kappa -12507 b_5^2+705348}
\nonumber \\
\eta_3&=& 
\phantom{+} 
\frac{3593 b_5^2 \kappa -2274 b_5^3+114366 b_5+15120 \kappa }{25014 b_5^2 \kappa -27288 b_5-1410696 \kappa }
\end{eqnarray}
\end{itemize}
Again we observe that the mixing parameters $\eta_i$, that were vanishing in the 4d case, are
non zero in two dimensions.

\bibliographystyle{JHEP}
\bibliography{References}

\providecommand{\href}[2]{#2}\begingroup\raggedright\begin{thebibliography}{10}

\bibitem{Cardy:1988cwa}
J.~L. Cardy, {\it {Is There a c Theorem in Four-Dimensions?}},  {\em Phys.
  Lett.} {\bf B215} (1988) 749--752.

\bibitem{Komargodski:2011vj}
Z.~Komargodski and A.~Schwimmer, {\it {On Renormalization Group Flows in Four
  Dimensions}},  {\em JHEP} {\bf 12} (2011) 099,
  [\href{http://arxiv.org/abs/1107.3987}{{\tt arXiv:1107.3987}}].

\bibitem{Anselmi:1997am}
D.~Anselmi, D.~Z. Freedman, M.~T. Grisaru, and A.~A. Johansen, {\it
  {Nonperturbative formulas for central functions of supersymmetric gauge
  theories}},  {\em Nucl. Phys.} {\bf B526} (1998) 543--571,
  [\href{http://arxiv.org/abs/hep-th/9708042}{{\tt hep-th/9708042}}].

\bibitem{Intriligator:2003jj}
K.~A. Intriligator and B.~Wecht, {\it {The Exact superconformal R symmetry
  maximizes a}},  {\em Nucl. Phys.} {\bf B667} (2003) 183--200,
  [\href{http://arxiv.org/abs/hep-th/0304128}{{\tt hep-th/0304128}}].

\bibitem{Zamolodchikov:1986gt}
A.~B. Zamolodchikov, {\it {Irreversibility of the Flux of the Renormalization
  Group in a 2D Field Theory}},  {\em JETP Lett.} {\bf 43} (1986) 730--732.
  [Pisma Zh. Eksp. Teor. Fiz.43,565(1986)].

\bibitem{Benini:2012cz}
F.~Benini and N.~Bobev, {\it {Exact two-dimensional superconformal R-symmetry
  and c-extremization}},  {\em Phys. Rev. Lett.} {\bf 110} (2013), no.~6
  061601, [\href{http://arxiv.org/abs/1211.4030}{{\tt arXiv:1211.4030}}].

\bibitem{Witten:1988xj}
E.~Witten, {\it {Topological Sigma Models}},  {\em Commun. Math. Phys.} {\bf
  118} (1988) 411.

\bibitem{Bershadsky:1995vm}
M.~Bershadsky, A.~Johansen, V.~Sadov, and C.~Vafa, {\it {Topological reduction
  of 4-d SYM to 2-d sigma models}},  {\em Nucl. Phys.} {\bf B448} (1995)
  166--186, [\href{http://arxiv.org/abs/hep-th/9501096}{{\tt hep-th/9501096}}].

\bibitem{Festuccia:2011ws}
G.~Festuccia and N.~Seiberg, {\it {Rigid Supersymmetric Theories in Curved
  Superspace}},  {\em JHEP} {\bf 06} (2011) 114,
  [\href{http://arxiv.org/abs/1105.0689}{{\tt arXiv:1105.0689}}].

\bibitem{Kutasov:2013ffl}
D.~Kutasov and J.~Lin, {\it {(0,2) Dynamics From Four Dimensions}},  {\em Phys.
  Rev.} {\bf D89} (2014), no.~8 085025,
  [\href{http://arxiv.org/abs/1310.6032}{{\tt arXiv:1310.6032}}].

\bibitem{Benini:2015bwz}
F.~Benini, N.~Bobev, and P.~M. Crichigno, {\it {Two-dimensional SCFTs from
  D3-branes}},  {\em JHEP} {\bf 07} (2016) 020,
  [\href{http://arxiv.org/abs/1511.09462}{{\tt arXiv:1511.09462}}].

\bibitem{Maldacena:2000mw}
J.~M. Maldacena and C.~Nunez, {\it {Supergravity description of field theories
  on curved manifolds and a no go theorem}},  {\em Int. J. Mod. Phys.} {\bf
  A16} (2001) 822--855, [\href{http://arxiv.org/abs/hep-th/0007018}{{\tt
  hep-th/0007018}}]. [,182(2000)].

\bibitem{Kim:2005ez}
N.~Kim, {\it {AdS(3) solutions of IIB supergravity from D3-branes}},  {\em
  JHEP} {\bf 01} (2006) 094, [\href{http://arxiv.org/abs/hep-th/0511029}{{\tt
  hep-th/0511029}}].

\bibitem{Gauntlett:2007ts}
J.~P. Gauntlett and N.~Kim, {\it {Geometries with Killing Spinors and
  Supersymmetric AdS Solutions}},  {\em Commun. Math. Phys.} {\bf 284} (2008)
  897--918, [\href{http://arxiv.org/abs/0710.2590}{{\tt arXiv:0710.2590}}].

\bibitem{Benvenuti:2004dy}
S.~Benvenuti, S.~Franco, A.~Hanany, D.~Martelli, and J.~Sparks, {\it {An
  Infinite family of superconformal quiver gauge theories with Sasaki-Einstein
  duals}},  {\em JHEP} {\bf 06} (2005) 064,
  [\href{http://arxiv.org/abs/hep-th/0411264}{{\tt hep-th/0411264}}].

\bibitem{Kennaway:2007tq}
K.~D. Kennaway, {\it {Brane Tilings}},  {\em Int. J. Mod. Phys.} {\bf A22}
  (2007) 2977--3038, [\href{http://arxiv.org/abs/0706.1660}{{\tt
  arXiv:0706.1660}}].

\bibitem{Franco:2017jeo}
S.~Franco, Y.-H. He, C.~Sun, and Y.~Xiao, {\it {A Comprehensive Survey of Brane
  Tilings}},  \href{http://arxiv.org/abs/1702.03958}{{\tt arXiv:1702.03958}}.

\bibitem{Martelli:2005tp}
D.~Martelli, J.~Sparks, and S.-T. Yau, {\it {The Geometric dual of
  a-maximisation for Toric Sasaki-Einstein manifolds}},  {\em Commun. Math.
  Phys.} {\bf 268} (2006) 39--65,
  [\href{http://arxiv.org/abs/hep-th/0503183}{{\tt hep-th/0503183}}].

\bibitem{Butti:2005vn}
A.~Butti and A.~Zaffaroni, {\it {R-charges from toric diagrams and the
  equivalence of a-maximization and Z-minimization}},  {\em JHEP} {\bf 11}
  (2005) 019, [\href{http://arxiv.org/abs/hep-th/0506232}{{\tt
  hep-th/0506232}}].

\bibitem{Feng:2000mi}
B.~Feng, A.~Hanany, and Y.-H. He, {\it {D-brane gauge theories from toric
  singularities and toric duality}},  {\em Nucl. Phys.} {\bf B595} (2001)
  165--200, [\href{http://arxiv.org/abs/hep-th/0003085}{{\tt hep-th/0003085}}].

\bibitem{Bertolini:2004xf}
M.~Bertolini, F.~Bigazzi, and A.~L. Cotrone, {\it {New checks and subtleties
  for AdS/CFT and a-maximization}},  {\em JHEP} {\bf 12} (2004) 024,
  [\href{http://arxiv.org/abs/hep-th/0411249}{{\tt hep-th/0411249}}].

\bibitem{Gubser:1998vd}
S.~S. Gubser, {\it {Einstein manifolds and conformal field theories}},  {\em
  Phys. Rev.} {\bf D59} (1999) 025006,
  [\href{http://arxiv.org/abs/hep-th/9807164}{{\tt hep-th/9807164}}].

\bibitem{Gubser:1998fp}
S.~S. Gubser and I.~R. Klebanov, {\it {Baryons and domain walls in an N=1
  superconformal gauge theory}},  {\em Phys. Rev.} {\bf D58} (1998) 125025,
  [\href{http://arxiv.org/abs/hep-th/9808075}{{\tt hep-th/9808075}}].

\bibitem{Martelli:2006yb}
D.~Martelli, J.~Sparks, and S.-T. Yau, {\it {Sasaki-Einstein manifolds and
  volume minimisation}},  {\em Commun. Math. Phys.} {\bf 280} (2008) 611--673,
  [\href{http://arxiv.org/abs/hep-th/0603021}{{\tt hep-th/0603021}}].

\bibitem{Tachikawa:2005tq}
Y.~Tachikawa, {\it {Five-dimensional supergravity dual of a-maximization}},
  {\em Nucl. Phys.} {\bf B733} (2006) 188--203,
  [\href{http://arxiv.org/abs/hep-th/0507057}{{\tt hep-th/0507057}}].

\bibitem{Butti:2005ps}
A.~Butti and A.~Zaffaroni, {\it {From toric geometry to quiver gauge theory:
  The Equivalence of a-maximization and Z-minimization}},  {\em Fortsch. Phys.}
  {\bf 54} (2006) 309--316, [\href{http://arxiv.org/abs/hep-th/0512240}{{\tt
  hep-th/0512240}}].

\bibitem{Butti:2006nk}
A.~Butti, A.~Zaffaroni, and D.~Forcella, {\it {Deformations of conformal
  theories and non-toric quiver gauge theories}},  {\em JHEP} {\bf 02} (2007)
  081, [\href{http://arxiv.org/abs/hep-th/0607147}{{\tt hep-th/0607147}}].

\bibitem{Benvenuti:2006xg}
S.~Benvenuti, L.~A. Pando~Zayas, and Y.~Tachikawa, {\it {Triangle anomalies
  from Einstein manifolds}},  {\em Adv. Theor. Math. Phys.} {\bf 10} (2006),
  no.~3 395--432, [\href{http://arxiv.org/abs/hep-th/0601054}{{\tt
  hep-th/0601054}}].

\bibitem{Lee:2006ru}
S.~Lee and S.-J. Rey, {\it {Comments on anomalies and charges of toric-quiver
  duals}},  {\em JHEP} {\bf 03} (2006) 068,
  [\href{http://arxiv.org/abs/hep-th/0601223}{{\tt hep-th/0601223}}].

\bibitem{Kato:2006vx}
A.~Kato, {\it {Zonotopes and four-dimensional superconformal field theories}},
  {\em JHEP} {\bf 06} (2007) 037,
  [\href{http://arxiv.org/abs/hep-th/0610266}{{\tt hep-th/0610266}}].

\bibitem{Gulotta:2008ef}
D.~R. Gulotta, {\it {Properly ordered dimers, R-charges, and an efficient
  inverse algorithm}},  {\em JHEP} {\bf 10} (2008) 014,
  [\href{http://arxiv.org/abs/0807.3012}{{\tt arXiv:0807.3012}}].

\bibitem{Eager:2010yu}
R.~Eager, {\it {Equivalence of A-Maximization and Volume Minimization}},  {\em
  JHEP} {\bf 01} (2014) 089, [\href{http://arxiv.org/abs/1011.1809}{{\tt
  arXiv:1011.1809}}].

\bibitem{Feng:2005gw}
B.~Feng, Y.-H. He, K.~D. Kennaway, and C.~Vafa, {\it {Dimer models from mirror
  symmetry and quivering amoebae}},  {\em Adv. Theor. Math. Phys.} {\bf 12}
  (2008), no.~3 489--545, [\href{http://arxiv.org/abs/hep-th/0511287}{{\tt
  hep-th/0511287}}].

\bibitem{Hanany:2005ss}
A.~Hanany and D.~Vegh, {\it {Quivers, tilings, branes and rhombi}},  {\em JHEP}
  {\bf 10} (2007) 029, [\href{http://arxiv.org/abs/hep-th/0511063}{{\tt
  hep-th/0511063}}].

\bibitem{Franco:2005sm}
S.~Franco, A.~Hanany, D.~Martelli, J.~Sparks, D.~Vegh, and B.~Wecht, {\it
  {Gauge theories from toric geometry and brane tilings}},  {\em JHEP} {\bf 01}
  (2006) 128, [\href{http://arxiv.org/abs/hep-th/0505211}{{\tt
  hep-th/0505211}}].

\bibitem{Hosseini:2016cyf}
S.~M. Hosseini, A.~Nedelin, and A.~Zaffaroni, {\it {The Cardy limit of the
  topologically twisted index and black strings in AdS$_5$}},
  \href{http://arxiv.org/abs/1611.09374}{{\tt arXiv:1611.09374}}.

\bibitem{Amariti:2017cyd}
A.~Amariti, L.~Cassia, and S.~Penati, {\it {Surveying 4d SCFTs twisted on
  Riemann surfaces}},  \href{http://arxiv.org/abs/1703.08201}{{\tt
  arXiv:1703.08201}}.

\bibitem{Benini:2013cda}
F.~Benini and N.~Bobev, {\it {Two-dimensional SCFTs from wrapped branes and
  c-extremization}},  {\em JHEP} {\bf 06} (2013) 005,
  [\href{http://arxiv.org/abs/1302.4451}{{\tt arXiv:1302.4451}}].

\bibitem{Brown:1986nw}
J.~D. Brown and M.~Henneaux, {\it {Central Charges in the Canonical Realization
  of Asymptotic Symmetries: An Example from Three-Dimensional Gravity}},  {\em
  Commun. Math. Phys.} {\bf 104} (1986) 207--226.

\bibitem{Karndumri:2013iqa}
P.~Karndumri and E.~O~Colgain, {\it {Supergravity dual of $c$-extremization}},
  {\em Phys. Rev.} {\bf D87} (2013), no.~10 101902,
  [\href{http://arxiv.org/abs/1302.6532}{{\tt arXiv:1302.6532}}].

\bibitem{Amariti:2016mnz}
A.~Amariti and C.~Toldo, {\it {Betti multiplets, flows across dimensions and
  c-extremization}},  \href{http://arxiv.org/abs/1610.08858}{{\tt
  arXiv:1610.08858}}.

\bibitem{Benvenuti:2005ja}
S.~Benvenuti and M.~Kruczenski, {\it {From Sasaki-Einstein spaces to quivers
  via BPS geodesics: L**p,q|r}},  {\em JHEP} {\bf 04} (2006) 033,
  [\href{http://arxiv.org/abs/hep-th/0505206}{{\tt hep-th/0505206}}].

\bibitem{Butti:2005sw}
A.~Butti, D.~Forcella, and A.~Zaffaroni, {\it {The Dual superconformal theory
  for L**pqr manifolds}},  {\em JHEP} {\bf 09} (2005) 018,
  [\href{http://arxiv.org/abs/hep-th/0505220}{{\tt hep-th/0505220}}].

\bibitem{Hanany:2005hq}
A.~Hanany, P.~Kazakopoulos, and B.~Wecht, {\it {A New infinite class of quiver
  gauge theories}},  {\em JHEP} {\bf 08} (2005) 054,
  [\href{http://arxiv.org/abs/hep-th/0503177}{{\tt hep-th/0503177}}].

\bibitem{Almuhairi:2011ws}
A.~Almuhairi and J.~Polchinski, {\it {Magnetic AdS$\times R^2$: Supersymmetry
  and stability}},  \href{http://arxiv.org/abs/1108.1213}{{\tt
  arXiv:1108.1213}}.

\bibitem{Gadde:2015wta}
A.~Gadde, S.~S. Razamat, and B.~Willett, {\it {On the reduction of 4d $
  \mathcal{N}=1 $ theories on $ {\mathbb{S}}^2 $}},  {\em JHEP} {\bf 11} (2015)
  163, [\href{http://arxiv.org/abs/1506.08795}{{\tt arXiv:1506.08795}}].

\bibitem{Franco:2015tna}
S.~Franco, D.~Ghim, S.~Lee, R.-K. Seong, and D.~Yokoyama, {\it {2d (0,2) Quiver
  Gauge Theories and D-Branes}},  {\em JHEP} {\bf 09} (2015) 072,
  [\href{http://arxiv.org/abs/1506.03818}{{\tt arXiv:1506.03818}}].

\bibitem{Franco:2015tya}
S.~Franco, S.~Lee, and R.-K. Seong, {\it {Brane Brick Models, Toric Calabi-Yau
  4-Folds and 2d (0,2) Quivers}},  {\em JHEP} {\bf 02} (2016) 047,
  [\href{http://arxiv.org/abs/1510.01744}{{\tt arXiv:1510.01744}}].

\bibitem{Franco:2017cjj}
S.~Franco, D.~Ghim, S.~Lee, and R.-K. Seong, {\it {Elliptic Genera of 2d (0,2)
  Gauge Theories from Brane Brick Models}},
  \href{http://arxiv.org/abs/1702.02948}{{\tt arXiv:1702.02948}}.

\end{thebibliography}\endgroup
\end{document}